\documentclass[useAMS,usenatbib]{mn2e}
\usepackage{graphicx}
\usepackage{longtable}
\usepackage{supertabular}
\usepackage{rotating}
\usepackage{amsmath,amssymb}
\usepackage{color}
\usepackage{here}
\newcommand{\Mo}{M_\odot}
\newcommand{\Mwd}{M_{\rm WD}}
\newcommand{\Rwd}{R_{\rm WD}}

\title[X-ray reflection from cold WDs in mCVs]
{X-ray reflection from cold white dwarfs in magnetic cataclysmic variables}

\author[T. Hayashi, T. Kitaguchi and M.Ishida]{Takayuki Hayashi$^{1, 2}$\thanks{E-mail:thayashi@u.nagoya-u.ac.jp},
Takao Kitaguchi$^{3}$ 
and Manabu Ishida$^{4, 5}$
\\
$^{1}$Division of Particle and Astrophysical Science, Nagoya University, Furo-Cho, Chikusa-ku, Nagoya 464-8602, Japan\\
$^{2}$NASA, Goddard Space Flight Center, Code 662, Greenbelt, MD20771, USA\\
$^{3}$Department of Physical Sciences, Hiroshima University, Higashi-Hiroshima, Hiroshima 739-8526, Japan\\
$^{4}$The Institute of Space and Astronautical Science/JAXA, 3-1-1 Yoshinodai, Chuo-ku, Sagamihara 252-5210\\
$^{5}$Department of Physics, Tokyo Metropolitan University, 1-1 Minami-Osawa, Hachioji, Tokyo 192-0397}

\begin{document}
\pagerange{\pageref{firstpage}--\pageref{lastpage}} \pubyear{2014}
\date{}
\maketitle

\begin{abstract}
We model X-ray reflection from white dwarfs (WD)
in %the 
magnetic cataclysmic variables (mCVs) with a %the 
Monte Carlo simulation.
A point source with a power-law spectrum or a realistic post-shock accretion column (PSAC) source 
 irradiates a %the 
 cool and spherical WD.
 The PSAC source emits thermal spectra of various temperatures 
 stratified along the column according to
 %from their corresponding positions by following 
 the PSAC model.
% The influence of the source height from the WD, the spectral hardness and the abundance 
% on the reflection and, especially, the equivalent width (EW) of the fluorescent iron K$\alpha_{1,2}$ lines and
% their Compton shoulder, and the Compton hump are evaluated with the point source.
 In the point source simulation, we confirm (1) %the 
a source harder and nearer to the WD enhances
 the reflection, %is enhanced by the nearer source to %from 
% the WD %surface 
%lower source 
%and the harder spectrum, 
(2) %the 
higher iron
%larger 
abundance enhances the equivalent widths (EWs) of %the total of 
%the 
fluorescent iron K$\alpha_{1,2}$ lines and their Compton shoulder, and %, on the other hand, 
increases %rises 
%the 
cut-off energy of
%bending energy of the lower energy side of 
%the 
a Compton hump,
and (3) %the 
significant reflection appears 
from an area that is more than 90$^\circ$ apart from the position right under the point X-ray source
%above 90\,deg 
%with the effectively distant %high 
%source 
because of the WD curvature.
The PSAC simulation reveals that (1) a more massive WD basically enhances the intensities of the fluorescent iron K$\alpha_{1,2}$ lines and the Compton hump, except for some specific accretion rate, because the more massive WD makes the hotter PSAC from which higher energy X-rays are preferentially emitted, 
%The PSAC simulation %by parameterizing the WD mass, the specific accretion rate  and the abundance 
%reveals that 
%%The higher %larger 
%%abundance enhances the EW and, on the other hand, 
%%rises the cut-off energy of the lower energy side of the Compton hump.
%% The realistic reflection model %in the mCV 
%% is constructed by parameterizing the WD mass, the specific accretion rate 
%% and the abundance with the PSAC source.
%(1) a more massive WD basically
%enhances the reflection, 
%except for some %limited 
%%cases of a 
%specific accretion rate because the massive WD makes the hotter and taller PSAC, % and a reflecting angle,}
%in which a specific accretion rate is low (a taller PSAC) and a reflecting angle is large,}
%although it can reduce that %e reflection 
%in specific cases of a specific accretion rate and a reflecting angle because it makes the hotter and taller PSAC
%\textcolor{red}{ The more massive WD makes the PSAC hotter and taller,
% and, therefore, reduces the reflection %depending on 
% with some 
% the specific accretion rate and the reflecting angle
% although it enhances the reflection in a basic way.}
%  The more massive WD makes the PSAC hotter and taller,
% and, therefore, reduces or enhances the reflection according to the specific accretion rate.
(2) a larger specific accretion rate monotonically enhances %reduces 
the reflection
 because it makes the hotter %cooler 
 and shorter %taller 
 PSAC,
and (3) the intrinsic thermal component %can 
hardens by occultation of the cool base of the PSAC by the WD.
We quantitatively evaluate %the 
influences of the parameters on the EWs and the Compton hump
with both types of sources.
 We also calculate %construct 
 %the 
 X-ray modulation %profile 
profiles brought about by % in relation to 
the WD spin.
They % The profile 
depend on %the 
angles of %between 
the spin axis from %and 
the line of sight and from %, between the spin axis and 
the PSAC,
 and whether the two PSACs %along the polar axis 
 can be seen. % as well as the WD mass and the specific accretion rate.
 The reflection spectral model and the modulation %profile 
 model involve the fluorescent lines and the Compton hump
and can directly be compared to the data % be applied to data in direct,
 which allow us to evaluate these geometrical %measure the 
 parameters with unprecedented accuracy.

\end{abstract}


\section{Introduction}
A magnetic cataclysmic variable (mCV) is a close binary system 
which is composed of a Roche Lobe-filling late type star
and a highly magnetized ($B >$ 0.1 MG) white dwarf (WD).
The mCV is divided into two subclass, that is, a polar and an intermediate polar (IP). 
In the polar and the IP, the WD magnetic field strength is 
$0.1 < B < 10$\,MG and $10 < B < 230$\,MG, respectively.
In the IP, %only,
accreting matter from the companion star forms an accretion disc by %because of 
its angular momentum.
Because %since 
the magnetic field of the WD is moderately strong,
the accretion disc is disrupted and the accreting matter is caught by the WD magnetic field 
within the Alfv$\acute{\rm e}$n radius.
On the other hand, in the polar, the magnetic field is strong enough to make the WD synchronously rotate
with the orbital motion and make the Alfv$\acute{\rm e}$n radius larger than the inner Lagrangian point, and hence
the accreting %on 
matter overflown from the Lagrangian point is directly %is initially 
caught by the WD magnetic field 
and the accretion disc is not formed.
The accreting matter caught by the magnetic field falls along the line of the magnetic field,
undergoes a strong shock and is highly ionized.
The generated plasma is hot enough to emit X-rays and is cooled %cooling 
via radiation %radiative cooling 
as it approaches the WD surface.
This plasma flow is called a Post-Shock Accretion Column (PSAC).

The PSAC structure and its X-ray spectrum have been %are 
considerably modeled
(e.g. \citealt{1973PThPh..49..776H}, \citealt{1973PThPh..49.1184A}, \citealt{1983ApJ...268..291I},
\citealt{1994ApJ...426..664W}, \citealt{1996A&A...306..232W},
\citealt{1998MNRAS.293..222C}, \citealt{1999MNRAS.306..684C}, \citealt{2005A&A...440..185C},
\citealt{2005MNRAS.360.1091S}, \citealt{2007MNRAS.379..779S} and \citealt{2014MNRAS.438.2267H})
and some of their %the 
X-ray spectral models were applied to observations 
(e.g. \citealt{2000MNRAS.316..225R}, \citealt{2005A&A...435..191S}, 
\citealt{2009A&A...496..121B}, \citealt{2010A&A...520A..25Y} and \citealt{2014MNRAS.441.3718H}).
%by a lot of works.
The PSAC structure models were constructed by %solving 
hydrodynamics.
%hydrodynamical equations,
%that is, momentum, energy and mass continuity equations with the aid of 
%the equation of the state of ideal gas.
Physical effects such as %of 
release of the gravitational potential, 
non-equipartition between electrons and ions, ionization non-equilibrium,
cyclotron cooling, %or 
energy conversion between thermal and kinetic energies by %because of 
variation %change 
of the PSAC cross section,
or variation of accretion rate per unit area 
were taken into account. % in the solution.
We note that the accretion rate per unit area is often 
called 'specific accretion rate' or '$a$' in g\,cm$^{-2}$\,s$^{-1}$.
The X-ray spectral models of the PSAC were constructed with 
distributions of physical quantities of such as temperature 
calculated by the corresponding PSAC models.
%Physical effects of release of the gravity potential, 
%non-equipartition between electron and ion, ionization non-equilibrium,
%cyclotron cooling or energy conversion between thermal and because of variation %change 
%of the PSAC cross section 
%were taken in to account. % in the solution.
%The X-ray spectral models were constructed by integrating 
%existing single-temperature plasma emission model along the PSAC
%based on the PSAC structure models.

%On the other hand, 
Although, by contrast, X-rays reflected by the WD surface 
are %which is 
another important component %in spectrum 
of the mCV spectrum, it is insufficiently modeled.
%where the source illuminating the WD is close to 
%A part of the X-ray photons emitted from the PSAC reaches the WD surface 
%and is absorbed or escapes in a sequence of interaction of scattering, 
%photoelectric absorption and re-emission by atoms or electron covering the WD surface.
%The existence of the reflection X-ray was proved by detection of 
%the fluorescent Fe-K$\alpha$ line.
The reflected X-rays contaminate the intrinsic thermal spectrum,
%\textcolor{blue}{with a non-negligible fraction, having a characteristic spectrum (see\,\S\ref{sec:result_po} and \S\ref{sec:result_psac}})
%with a non-negligible fraction, 
having a spectrum experiencing an energy-dependent modification from the intrinsic one,
 % emitted from the PSAC
and make it difficult to extract physical parameters, %which can be estimated with the thermal spectrum,
for example the WD mass.
Moreover,
%On the other hand, 
the reflection contains important information
of the PSAC geometry such as the height of the PSAC
%which constrains the physical parameters 
and the viewing angle from the PSAC
(\citealt{2011PASJ...63S.739H}, the angle $i$ in figure\,\ref{fig:ref_view}).
The reflection especially forms the fluorescent iron K$\alpha_{1,2}$ lines and their Compton shoulder around 6\,keV
and Compton hump around 10--50\,keV (e.g. \citealt{1991MNRAS.249..352G}).
The features are %more 
intense and become comparable to %directly observed 
the intrinsic thermal component. % at the specified %individual 
%energies.
%Since the intensity and the shape of the reflection spectral features depend on the PSAC structure %geometry 
%and the viewing angle,
%we can extract their %these 
%information by careful spectral modeling of the reflection.
We can extract information of the PSAC structure and the viewing angle by careful spectral modeling of the reflection
because the intensity and the spectral shape of the reflection %spectral features 
depend on them.

We therefore construct a spectral model of the reflection using a %the 
Monte Carlo simulation.
We assume a point source with a power-law spectrum and  
%the %more 
a realistic PASC source which has finite height %length 
and distributions of physical quantities
calculated by \cite{2014MNRAS.438.2267H}.
%Sphere is adopted for the 
A cool and spherical WD %shape 
and occultation by the limb of the WD is also considered.

This paper is organized as follows.
We define the framework %assumptions 
of our Monte Carlo simulation %for the reflection 
in section\,\ref{sec:model}.
In section\,\ref{sec:result_po} and \ref{sec:result_psac}, calculation results are shown
with the point and the PSAC %column 
sources, respectively.
We discuss application %utility 
of the reflection to %on 
observations using our results in section\,\ref{sec:discussion}.
Finally, we summarize our results in section\,\ref{sec:summary}.

 \section{Modeling and Calculation}\label{sec:model}

In this section, we describe the framework %assumptions %and geometry 
%which adopted 
for our Monte Carlo simulation to model the X-ray spectrum 
reflected by the WD surface in the mCVs.

\subsection{X-ray source}\label{sec:source}

\begin{figure}
\includegraphics[width=80mm]{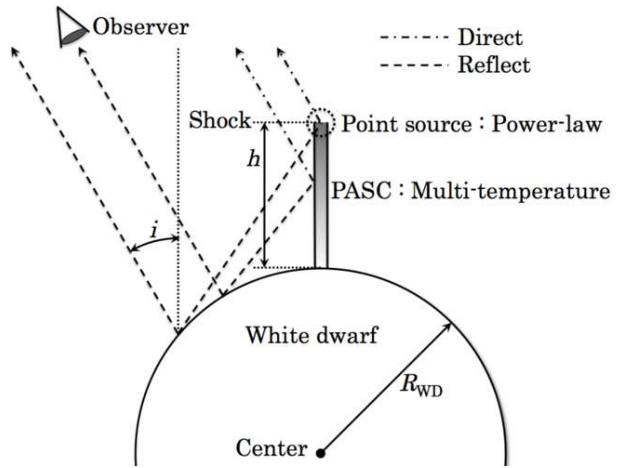}
\caption{Cross sectional view of a %the 
mass accreting magnetic WDs in the mCVs. % assumed in our simulation.
We adopted two types of X-ray sources: one is point source (dotted circle) whose spectrum 
follows a power-law function 
and the other is the PSAC source (gradated gray column) 
which has finite length and emits multi-temperature thermal plasma spectrum 
based on the calculation %depending on calculated 
by \citet{2014MNRAS.438.2267H}.
Dash-dotted and dashed lines show the X-ray paths toward the observer directly from the sources 
and reflected by the WD surface, respectively.
$h$ is distance from the WD surface for the point source and
the shock height for the PSAC source.
$i$ is called viewing angle.
\label{fig:ref_view}}
\end{figure}

Two types of X-ray sources are %were 
assumed in our simulation,
that is, the point source (\S\ref{sec:result_po}) and the %a more 
realistic PSAC source (\S\ref{sec:result_psac}).
The energy range of the %generating 
X-ray photons used in the simulation is 1--500\,keV. % and common in the two types.
%A part of X-rays from the sources arrive at the WD surface.
The photon number %energy of the emitted X-rays 
follows a power law %distribution 
%function $N_0$
of the energy $E$ with the power %an arbitrary photon 
index ${\it \Gamma}$,
\begin{eqnarray}
N_0 (E) \propto E^{-{\it \Gamma}}\,{\rm photons\,keV}^{-1}.
\label{eq:pl}
\end{eqnarray}
The point source is away from the WD surface by $h$ as shown in figure\,\ref{fig:ref_view}
and isotropically emits X-rays. 

The PSAC source is a pole of the thin thermal plasma 
which has finite height of $h$ (figure\,\ref{fig:ref_view}) and negligible cross section.
The %distributions of the 
temperature and the emission measure 
vary %change 
along the pole %direction normal to the WD surface
following the hydrodynamical calculation of the dipolar PSAC calculated by \citet{2014MNRAS.438.2267H}.
%In our simulation, 
The PSAC is divided into one hundred segments of %point sources by 
logarithmically even interval.
Each piece irradiates the WD surface %from the corresponding height
with one-temperature thin thermal plasma emission spectrum 
weighted by its
%of its %individual 
%temperature and 
emission measure. 
%from %corresponding to 
%its height.
The one-temperature spectrum is calculated by {\sc SPEX} package ver\,2.05.04 \citep{1996uxsa.conf..411K}.
%with the corresponding temperature and the emission measure.
%Note that this source is sometimes partially occulted by the WD limb 
%because of its finite size %of the source 
%unlike the power-law point source.

%Reflecting angle $i$ is defined as angle between reflected direction and 
%perpendicular line from the point source to the WD surface or the line along the PSAC source.

\subsection{Interaction on WD surface}\label{sec:interact}

%A part of the 
X-rays arrived at the WD surface experience either
%gets out after undergoing 
scattering or %, 
absorption. % and/or re-emit.
The WD is assumed to be spherical
and its surface is covered %by 
with optically thick cool gas composed of various elements.
%Therefore, the arrived X-rays have to interacts once at least on the WD surface.
A %The 
sequence of the interactions %Our calculation 
continues %to perform for the sequence of the interactions
until the tracing photon escapes from the WD,
is absorbed or its energy %of the tracing photon 
goes down below 1\,keV.
A viewing angle $i$ is defined as the angle between the reflected direction and 
%perpendicular 
the line connecting %from 
the point source to the center of the WD %to the WD surface 
or the line along the pole of the PSAC source 
(figure\,\ref{fig:ref_view}).

\begin{figure}
\includegraphics[width=80mm]{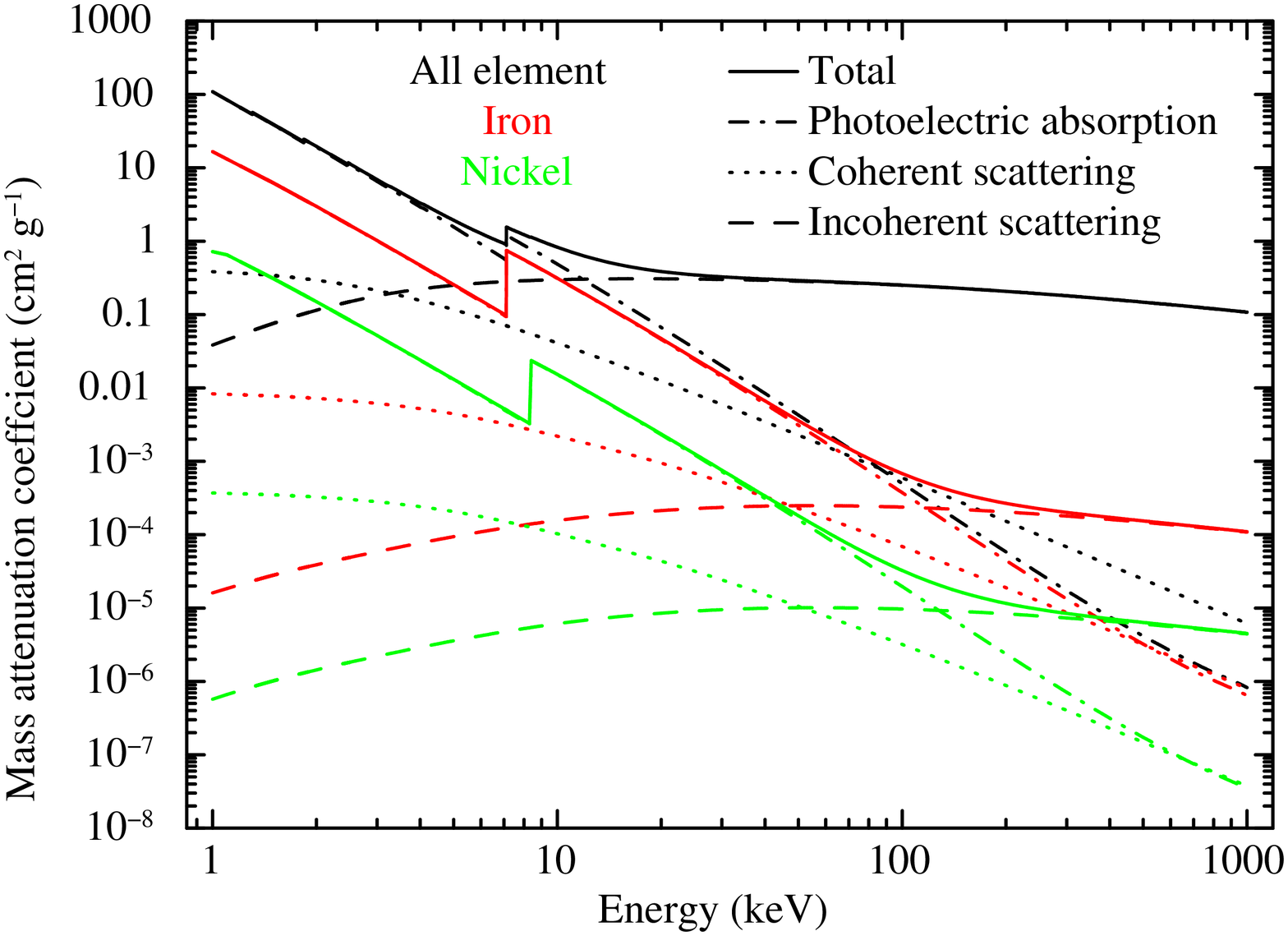}
\caption{Mass attenuation coefficient of one solar abundance.
%Dashed-dotted is photoelectric absorption, dotted is coherent scattering and dashed is incoherent scattering.
The dashed-dotted, dotted, dashed curves are the photoelectric absorption, the coherent and incoherent scatterings, respectively.
Solid line is the total of them. %photoelectric absorption (dashed-dotted), coherent scattering (dotted) and incoherent scattering (dashed).
%Black, red and green show the all element, iron and nickel, respectively. 
Black lines represent the mass attenuation coefficient of the all element,
while red and green lines show fractional ones of iron and nickel, respectively.
\label{fig:mass_att_co}}
\end{figure}

Figure\,\ref{fig:mass_att_co} shows mass attenuation coefficients calculated by NIST XCOM
based on \cite{1975JPCRD...4..471H}
of photoelectric absorption, % (dash-dot), 
coherent scattering, % (dotted), 
incoherent scattering % (dashed) 
and total of them %these interactions (solid) 
of all elements, %(black), 
iron %(red) 
and Nickel %(green) 
of one solar abundance,
where solar abundance of \cite{1989GeCoA..53..197A} is adopted.
We remark that both the coherent and incoherent scatterings are the scattering processes of electrons bound in an atom. 
The incoherent scattering is accompanied by photo-ionization %\textcolor{blue}{photoexcitation} %ionization 
of the atom. 
In the coherent scattering, on the other hand, the initial and final states of the bound electron are the same. 
Since the associated electrons are in bound states, 
the cross sections of the coherent and incoherent scatterings 
are somewhat different from those of the Thomson and Compton scatterings
with a free electron at rest, respectively.
In the lower and higher energies, the photoelectric absorption and the incoherent scattering 
are dominant, respectively, although the energy %etic 
domains are different for each element.
On the other hand, the coherent scattering is relatively minor interaction at any energy.
 
%As shown above, scattering by not a free electron but a electron bound in a atom 
%is divided into two categories of the coherent and the incoherent  scatterings. %, %.
%%Scattering by not free electron but atom is divided into two categories of the coherent and the incoherent 
%%by bounding energy of a electron responsible to the scattering
%%relative to the incident X-ray energy 
%whose cross sections against the incident energy are %as 
%shown in figure\,\ref{fig:mass_att_co}.
%%Figure \ref{fig:dif_cross_section} shows differential cross section of the coherent and the incoherent scatterings
%%for photons of 6.4 and 50\,keV with one solar abundance.
In the coherent scattering, the scattered %ing 
X-ray does not lose its energy and
the differential cross section can be represented as
\begin{eqnarray}
%\frac{{\rm d} \sigma_{\rm coh}}{{\rm d} \theta} = \frac{{\rm d} \sigma_{\rm TM}}{{\rm d} \theta} F_m^2 (x,Z),
\frac{{\rm d} \sigma_{\rm coh}}{{\rm d} \Omega} %= \frac{{\rm d} \sigma_{\rm TM}}{{\rm d} \Omega} 
= \frac{r_{\rm e}^2}{2} (1-\cos^2{\theta})F_m^2 (x,Z)
= \frac{{\rm d} \sigma_{\rm TM}}{{\rm d} \Omega} F_m^2 (x,Z),
\end{eqnarray}
where $r_{\rm e}$ is the classical electron radius, $\theta$ is a scattering angle,
%${\rm d}
$\sigma_{\rm TM}$
% / {{\rm d} \theta}$ 
 is the Thomson %differential 
cross section and 
$F_m (x,Z)$ is an atomic form factor as a %, which is 
function of an atomic number $Z$ and a momentum transfer 
\begin{eqnarray}
x = \frac{\sin{(\theta/2)}}{\lambda},
\end{eqnarray}
where $\lambda$ is the wavelength of the incident X-ray in angstrom. %, and atomic number $Z$.
We adopted %$S_m$ 
$F_m (x,Z)$ evaluated by \cite{1975JPCRD...4..471H}. % in this work.

By contrast, in the incoherent scattering, 
the incident X-ray gives energy to an electron and loses a part of its energy.
The transferred energy is used for ionization, %\textcolor{blue}{excitation}
and hence is %has to be 
larger than the binding energy of the interacting electron.
Moreover, the energy of the scattered photon %s scattered by the bound electrons %in atoms 
%is 
shifts from that scattered by a free %an 
electron %s 
 at rest because of the Doppler effect
 due to electron orbital motion. % in the atom.
A double differential cross section is used for the incoherent scattering and can be represented by \cite{Ribberfors&Berggren} as
\begin{eqnarray}
%\frac{{\rm d^2} \sigma_{\rm incoh}}{{\rm d} \theta {\rm d}} = \frac{{\rm d} \sigma_{\rm KN}}{{\rm d} \theta} S_m (x),
\frac{{\rm d^2} \sigma_{\rm incoh}}{{\rm d} E{\rm d} \Omega} = %\frac{{\rm d} \sigma_{\rm KN}}{{\rm d} \theta} S_m (x),
\frac{r_{\rm e}^2}{2} \frac{Em_{\rm e}c}{E_0cq} \left[1+\left(\frac{p_{\rm z}}{m_{\rm e}c}\right)^2\right]^{-1/2} X J(p_{\rm z})\label{eq:ddcs}
\end{eqnarray}
%where ${\rm d}\sigma_{\rm KN} / {{\rm d} \theta}$ is the Klein-Nishina differential cross section
%and $S_m (x)$ is incoherent scattering function, which is also function of $x$.
with the impulse approximation, where $E_0$ and $E$ are energies of incident and scattered photon, %s, 
respectively,
$m_{\rm e}$ is the electron mass, $c$ is the velocity of light
and $q$ is the modulus of the momentum transfer vector, ${\boldsymbol q} = {\boldsymbol k_0} - {\boldsymbol k}$,
where ${\boldsymbol k_0}$ and ${\boldsymbol k}$ are momenta of incident and scattered photons, respectively.
%$F_m$ and $S_m$ depend on element, we adopted those evaluated by \cite{1975JPCRD...4..471H}
%in this work.
$p_{\rm z}$ is the projection of the initial momentum ${\boldsymbol p}$ of the electron on the direction of 
the scattering vector ${\boldsymbol k} - {\boldsymbol k_0} = -{\boldsymbol q}$, that is,
\begin{eqnarray}
p_{\rm z} \equiv -\frac{{\boldsymbol p}\cdot {\boldsymbol q}}{q}
= \frac{E_0E(1-\cos{\theta})-m_{\rm e}c^2(E_0-E)}{c^2q}
\end{eqnarray}
or 
\begin{eqnarray}
\frac{p_{\rm z}}{m_{\rm e}c} = \frac{E_0(E-E_{\rm C})}{E_{\rm C}cq},
\end{eqnarray}
where 
\begin{eqnarray}
E_{\rm C} \equiv \frac{m_{\rm e}c^2E_0}{m_{\rm e}c^2+E_0(1-\cos{\theta})} 
\end{eqnarray}
is the energy of a photon %s 
scattered by a free electron at rest (Compton scattering).
$X$ in equation\,\ref{eq:ddcs} is defined as 
\begin{eqnarray}
X \equiv \frac{R}{R'} + \frac{R'}{R} + 2 \left(\frac{1}{R}-\frac{1}{R'}\right) + \left(\frac{1}{R}-\frac{1}{R'}\right)^2,\label{eq:x}
\end{eqnarray}
with 
\begin{eqnarray}
R = \frac{E}{m_e c^2}\left\{\left[1+\left(\frac{p_{\rm z}}{m_{\rm e}c}\right)^2\right]^{1/2}+\frac{E_0-E \cos{\theta}}{cq} \frac{p_{\rm z}}{m_e c}\right\},\label{eq:r}
\end{eqnarray}
and
\begin{eqnarray}
R' = R - \frac{E_0E}{m_{\rm e}^2c^4}(1-\cos{\theta}) = R - \frac{E}{m_{\rm e}c^2}\left(\frac{E_0}{E_{\rm C}}-1\right)\label{eq:r-d}
\end{eqnarray}
$J(p_{\rm z})$ in equation\,\ref{eq:ddcs} is the Compton profile,
\begin{eqnarray}
J(p_{\rm z}) \equiv \int\!\!\!\int\rho({\boldsymbol p}) {\rm d}p_{\rm x} {\rm d}p_{\rm y}, 
\end{eqnarray}
where $\rho({\boldsymbol p})$ is the electron momentum distribution,
which can be determined by the state %sate 
wave function.

We follow a method described by \cite{Brusa1996} to reproduce the incoherent scattering.
They set $p_{\rm z}/m_{\rm e} c = 0$ and $E = E_{\rm C}$ in equation \ref{eq:r} and \ref{eq:r-d}
because %the most probable of 
$|p_{\rm z}| \sim 0$ is most probable 
and, therefore, equation\,\ref{eq:x} reduces to the Klein-Nishina $X$-factor,
\begin{eqnarray}
X_{\rm KN} = \frac{E_{\rm C}}{E_0} + \frac{E_0}{E_{\rm C}}-\sin^2{\theta}.
\end{eqnarray}
Moreover, they reproduced the one-electron Compton profiles in the $i$-th shell, $J_i(p_{\rm z})$ by an analytical function
and obtained $J(p_{\rm z})$ by summing up the one-electron Compton profiles with number of electrons in the $i$-th shell ($Z_i$),
\begin{eqnarray}
J(p_{\rm z}) = \sum_i Z_i J_i(p_{\rm z}).
\end{eqnarray}

\begin{figure}
\includegraphics[width=70mm]{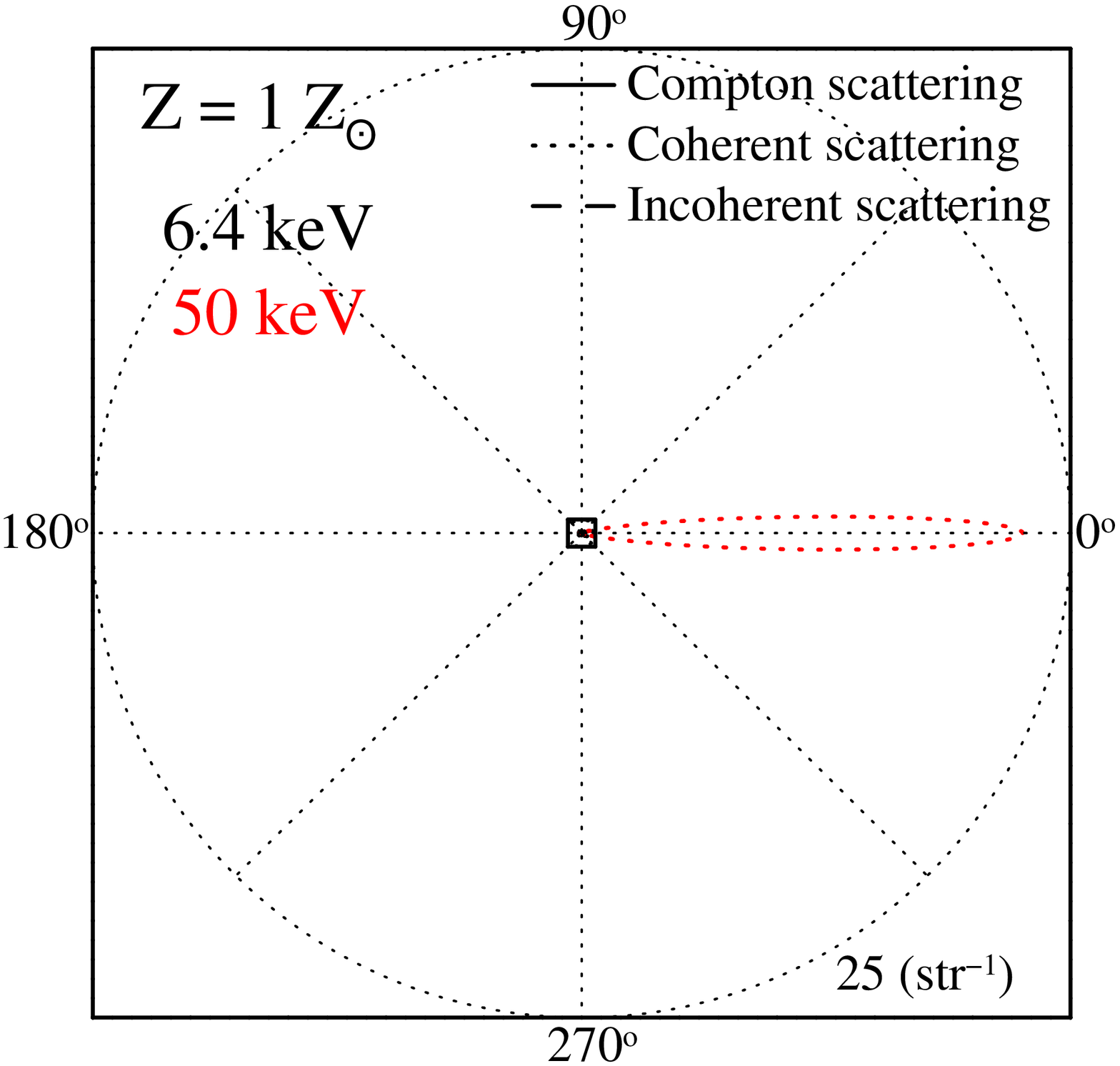}
\includegraphics[width=70mm]{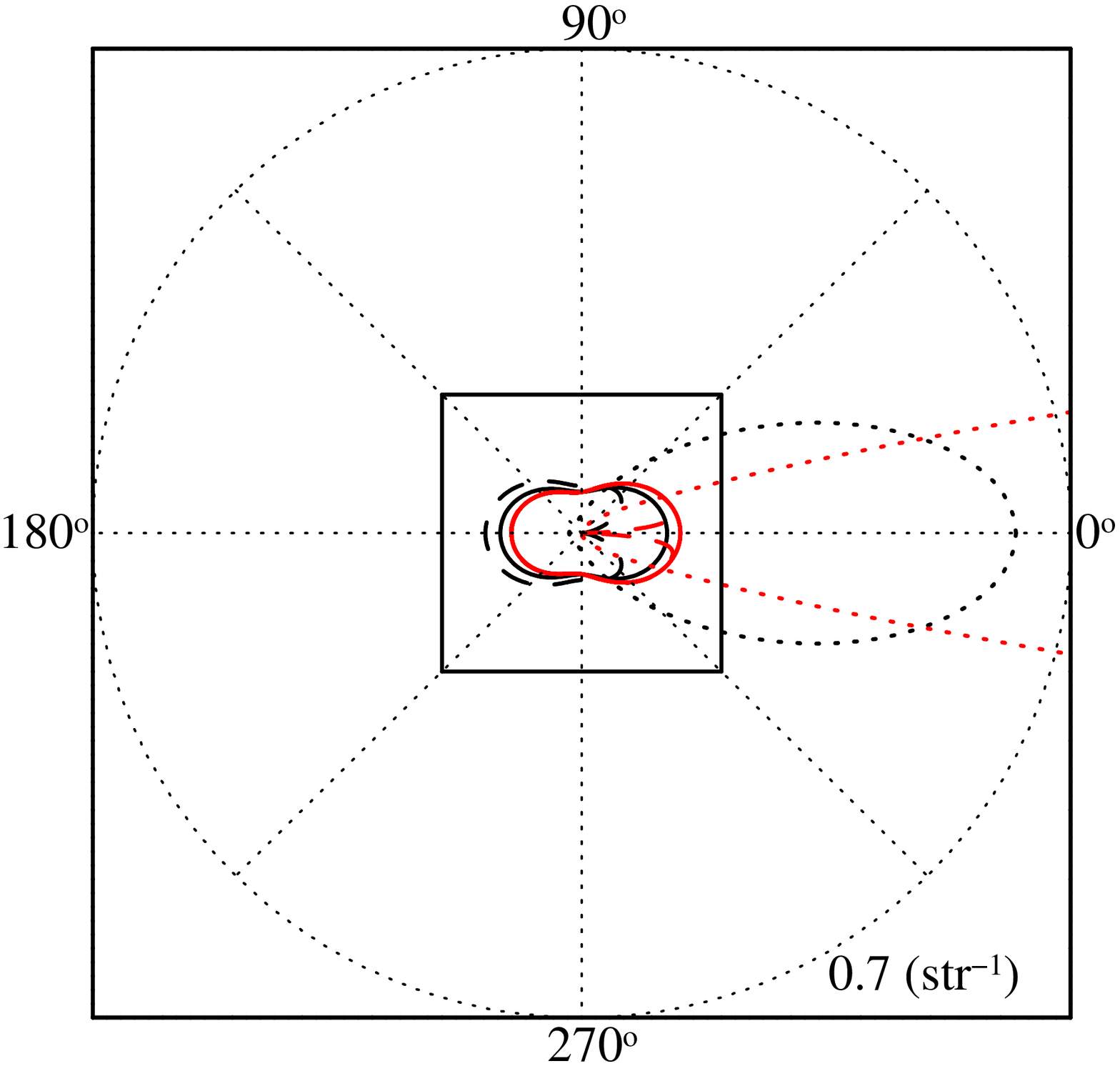}
\includegraphics[width=70mm]{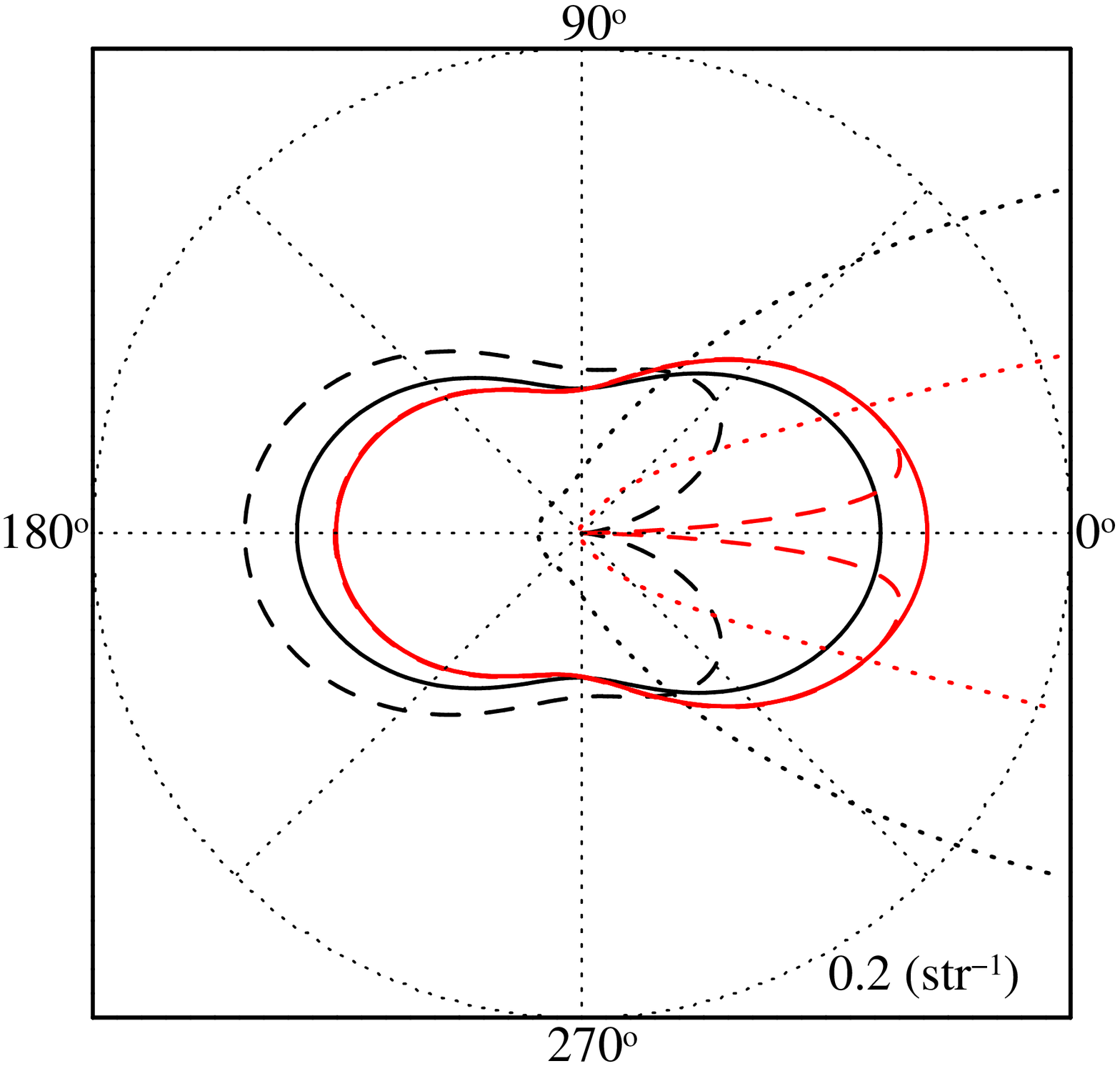}
\caption{Differential cross sections for an incident X-ray of 6.4\,keV (black) and 50\,keV (red) 
scattered off %averaged over 
by one solar abundance material.
Solid, dotted and dashed lines show the differential cross sections of Compton scattering of a free electron at rest (Klein-Nishina),
coherent scattering and incoherent scattering by bound %with a bounding 
electrons, respectively.
These differential cross sections are normalized so that each total cross section becomes unity.
In the top panel, %From top to bottom panel, showing range becomes narrower and 
the radius %i 
of the dotted circle is %are 
25%, 0.7 and 0.2 normalized 
\,str$^{-1}$.
%cm$^{-2}$\,str$^{-1}$. %, respectively.
%The solid squares of top and middle panels correspond to showing ranges of the middle and bottom panels, respectively.
The middle panel is a blowup of the central square in the top panel. The bottom panel is again a blowup of the central square in the middle panel. The radii of the dotted circles are 0.7 and 0.2\,str$^{-1}$% cm$^{-2}$ str$^{-1}$, respectively.
\label{fig:dif_cross_section}}
\end{figure}

Figure \ref{fig:dif_cross_section} shows the averaged 
differential cross sections 
%at 6.4\,keV % (black) 
%and 50\,keV %(red) 
%for one solar abundance 
of 
the 
Compton, %(free electron at rest, solid), 
%the 
coherent %(dotted) 
and %the 
incoherent %(dashed) 
scatterings at 6.4 keV and 50 keV. 
The scattering %reflecting 
material is of one solar abundance.
Each differential cross section in this figure is normalized %divided 
by its total cross section,
which is
%that is, surface integral of 
each differential cross section 
integrated over the 4$\pi$ solid angle.
%is normalized to unity.
%whose total cross sections are normalized to unity.}
The incident X-ray is injected %moves 
from the direction of 180$^{\circ}$ %to 0$^{\circ}$ 
and collides with the electron
at the center. 
% bound in the atom at center of each panel. 
%in this figure.
In the coherent scattering, the differential cross section is larger 
for forward (to 0$^{\circ}$) scattering than backward (to 180$^{\circ}$) scattering
which becomes more conspicuous for higher energy incident X-rays.
Below about 1\,keV, it %the differential cross section of the coherent scattering 
almost converges to %on 
that of the Thomson scattering.
In contrast, in the incoherent scattering, the differential cross section is larger for backward scattering
for lower energy %etic 
incident X-rays.
For a higher energy %etic 
incident X-ray, it %the differential cross section of the incoherent scattering 
converges to %on 
that of the Compton scattering except for the forward scattering.
The differential cross section of the incoherent scattering must drop for the forward scattering 
at %for
 any %high 
energy of the incident X-ray
although the range of the scattering angle within which the %corresponding to %of 
%the limited 
forward scattering is reduced, % becomes narrower,
where the coherent scattering is dominant.

In this work, re-emission of iron K$\alpha_1$ and $\alpha_2$ (6.404 and 6.391\,keV, respectively), iron K$\beta$ (7.058\,keV), 
nickel K$\alpha_1$ and $\alpha_2$ (7.478  and 7.461\,keV, respectively) and nickel K$\beta$ (8.265\,keV) are considered.
%In this work, re-emission of Fe-K$\alpha_1$ and $\alpha_2$ (6.40\,keV), Fe-K$\beta$ (7.06\,keV), 
%Ni-K$\alpha$ (7.47\,keV) and Ni-K$\beta$ (8.26\,keV) are considered.
The photoelectric absorption by iron %of photons whose energy is 
in a higher energy %region higher 
than %that of 
its %the iron 
K-edge (7.112\,keV)
can be %are 
undertaken by %responsible for 
the K-shell electrons. 
The K-shell absorption of iron occupies a fraction of 0.876 of the total photoelectric absorption cross section
%whose %with a 
%fraction is %of 
%0.876 
\citep{1970NDT.....7..565S}.
The fluorescence yield of the iron K-shell is 0.34 
%absorbing iron atom with its K-shell re-emits fluorescent K line following the fluorescence yield of 0.34
and the line relative intensity of K$\alpha_1$, $\alpha_2$ and K$\beta$ is known to be %determined by a ratio of 
100:50:17. %\citep{Kikon}.
Note that the fraction of the K-shell electron absorption 
to the total photoelectric absorption is assumed to be energy independent % in this work, 
although it slightly varies with incident X-ray energy.
For Ni, %procedure is common with iron although 
the energy of K-edge is 8.333\,keV, the fluorescence yield is 0.38, absorbing faction of K-shell is 0.872
and K$\alpha_1$, $\alpha_2$ and K$\beta$ line ratio is 100:51:17.
%7.112
%0.34
%Alpha:Beta = 150:17 (Kikoin 1976 used in Ikeda+ 2009)
%
% static double NiK_edge = 8.332;
%  static double K_fluo = 0.3780; --> 0.38 Fink 1966
%  //print (100+51)/(100+51+17.)
%  //0.898809523809524
%  static double KL_Rint = 0.898809523809524; //Alpha:Beta = 151:17 (xdb)

We note that the general relativistic effect is not considered in this work. 
The energies of the fluorescent lines can be shifted
%although
%it may shift the X-ray energy 
to lower direction by a few\,eV
with the massive WD
because these lines are emitted from the surface of the WD where the gravitational redshift is not negligible. The general relativistic effect may work for thermal emission lines as well.
% and the low viewing angle.
We will consider them
%include this effect into both of the reflection and the thermal emission
as well as Doppler effect 
associated with the plasma flow in the PSAC
%into the thermal emission
to apply our model to high energy resolution X-ray data in the future.

 \section{Results of point source}\label{sec:result_po}
In this section, results of the Monte Carlo simulation with the point source of %with %whose spectrum is 
a power-law spectrum are summarized.
%${\it \Gamma}$ in equation\,\ref{eq:pl} are varied from 1.0 to 2.4.
  
 \subsection{Reflected X-ray image}
\begin{figure}
\includegraphics[width=180mm]{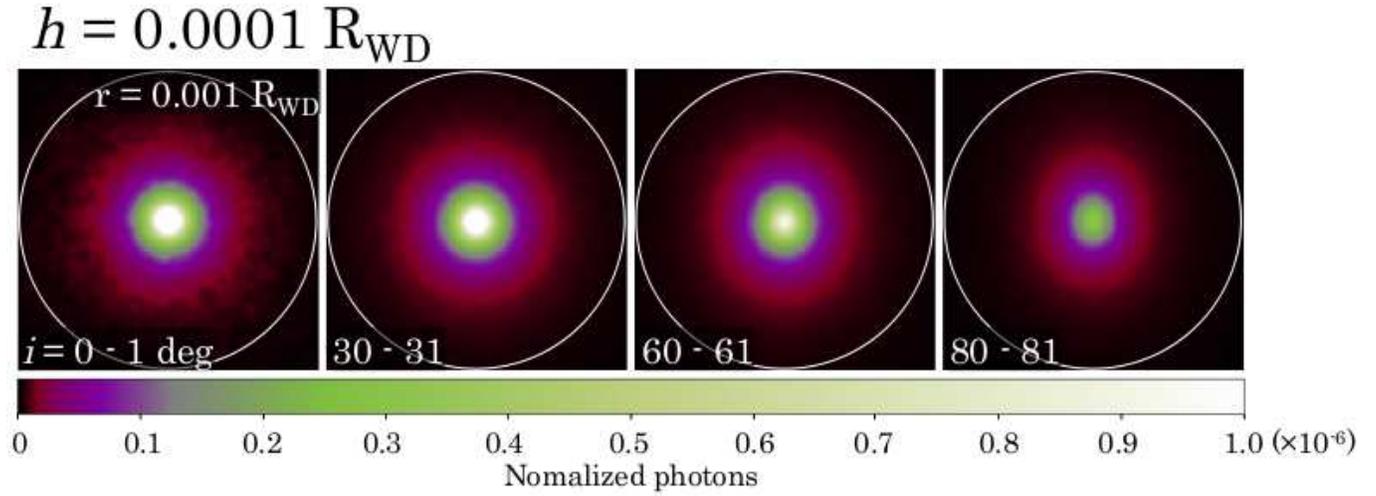}
\vspace{1mm}
 \caption{Reflection X-ray images with a point %X-ray 
 source located at 0.0001 $\Rwd$ above the WD surface.
Incident %Generated 
X-rays have a 
% surface
% which generates X-ray of %following 
 power-law spectrum with ${\it \Gamma}$ = 2.0 in the band %between 
 1--500\,keV.
A white circle in each panel is %White solid line is a circle 
centered on the position on the WD surface right down the point source
%the foot of a perpendicular line from the source to the WD surface 
with a radius of 0.001\,$\Rwd$.
The scattering direction is contained in a plane defined by %the perpendicular line and vertical line in this figure.
an in-plane vertical line and a normal to the page.
Range of %reflecting 
viewing angle ($i$, see Fig.\,\ref{fig:ref_view}) %}) between the perpendicular line and the scattering direction
is 0--1, 30--31, 60--61and 80--81 degrees from left to right.
These images are normalized with generated photon number  
%(5$\times$10$^8$ for $i$ = 0-1\,deg
%and 10$^8$ for the others) 
and smoothed by Gaussian function with $\sigma$ of 3$\times$10$^{-5}$\,$\Rwd$.
%Reflection images with a point X-ray source 0.0001 $\Rwd$ above the WD surface
% which generates X-ray of power-law spectrum with  ${\it \Gamma}$ = 2 between 1-500\,keV.
% White solid circles show a circle centered on the foot of 
% a perpendicular line from the source to the WD surface with a radius of 0.001\,$\Rwd$.
%The scattering directional vector is contained in a plane defined 
%by the perpendicular line and vertical line in this figure.
%Range of reflecting angle ($i$) between the perpendicular line and the scattering direction
%is 0 - 1 degree for of the left top panel and increases to right bottom by 10 degrees.
%These images are normalized with irradiated photon number (10$^8$) and multiplied by 10$^8$
%after smoothing by Gaussian function with $\sigma$ of 10$^{-6}$\,$\Rwd$.
Note that the fluctuation appeared in the viewing angle of $i = 0-1$\,deg %in figure\,\ref{fig:images_h0p0001}
is caused by Poisson noise 
because the solid angle %corresponding that reflecting angle 
is extremely small 
compared to the other viewing angle. % shown in figure\,\ref{fig:images_h0p0001}.
 \label{fig:images_h0p0001}}
\end{figure}

\begin{figure}
\includegraphics[width=180mm]{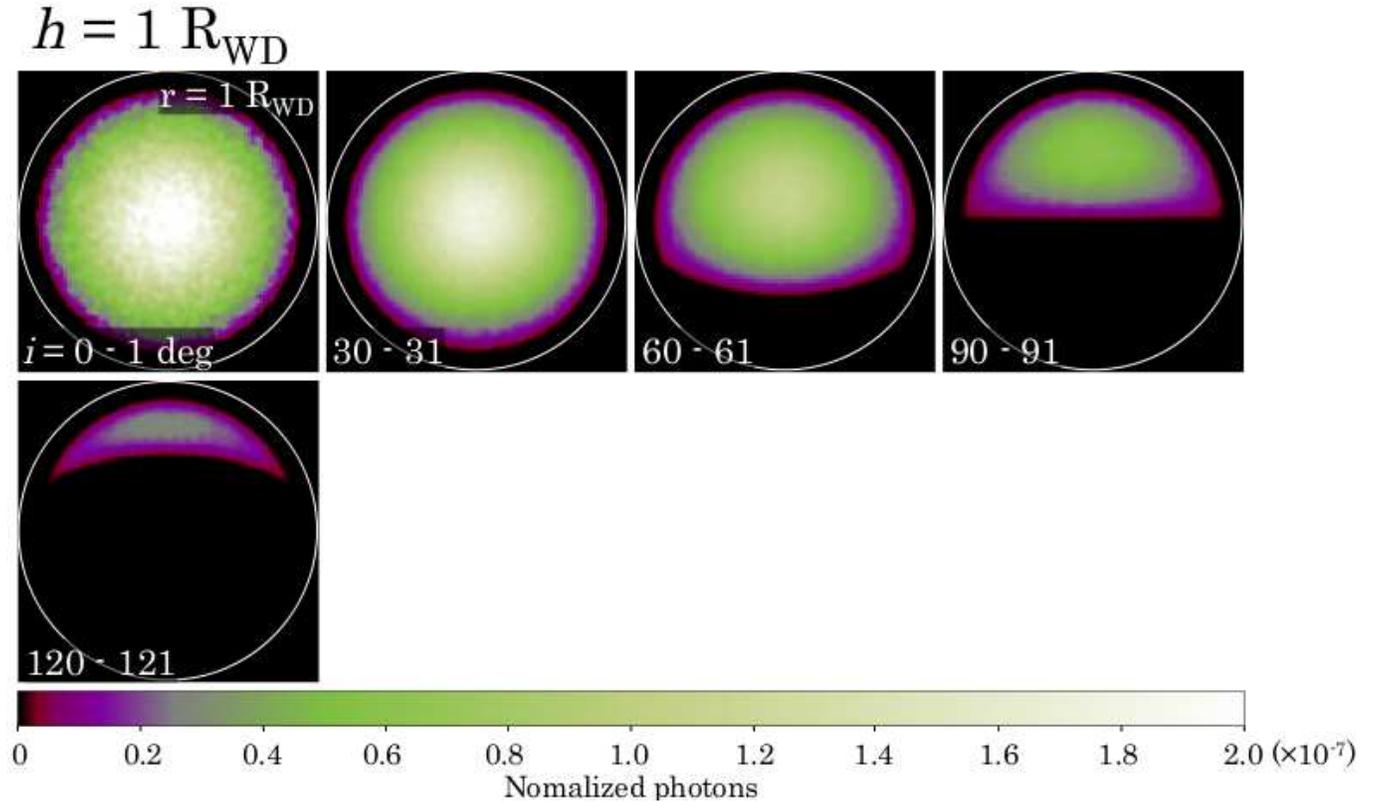}
\vspace{1mm}
 \caption{Same as Figure\,\ref{fig:images_h0p0001}, but a point X-ray source is located at 1\,$\Rwd$ above the WD, where
%Reflection images with a point X-ray source of 1 $\Rwd$ above the WD surface.
%The spectrum and energy range of the source, and the definition of the coordinate are common to 
%Fig.\,\ref{fig:images_h0p0001}.
%Radius of the white circle is 
1\,$\Rwd$ is the radius of the WD. % whose center stands on same place as Fig.\,\ref{fig:images_h0p0001}.
Range of reflecting %viewing 
angle ($i$) %between the perpendicular line and the scattering direction
is 0--1, 30--31, 60--61, 90--91 and 120--121 degrees from left to right and top to bottom. %upper left to lower right.
These images are normalized by generated photon number  
and smoothed by Gaussian function with $\sigma$ of 3$\times$10$^{-2}$\,$\Rwd$.
 \label{fig:images_h1p0}}
\end{figure}

Figures\,\ref{fig:images_h0p0001} and \ref{fig:images_h1p0} are reflected X-ray images 
%with the 
formed by a
point source emitting the power-law spectrum of ${\it \Gamma} = 2.0$
from $h$ = 0.0001 and 1\,$\Rwd$, % distant from the WD surface, 
respectively.
%Note that the fluctuation appeared in the reflecting angle of $i = 0-1$\,deg in figure\,\ref{fig:images_h0p0001}
%is caused by Poisson noise 
%because the solid angle %corresponding that reflecting angle 
%is extremely small 
%compared to the other reflecting angle shown in figure\,\ref{fig:images_h0p0001}.
%In these figure, the point sources  lie in the center of each panel
%and scattering direction is contained in a plane defined 
%by a perpendicular line from the source to the WD surface and vertical line in these figure.
%Some reflecting angles are shown in each $h$.

In the case of $h$ = 0.0001 (figure\,\ref{fig:images_h0p0001}), the curvature of the %reflecting 
WD surface in the reflecting area is negligible and the reflecting surface can be approximated as 
%reflection from 
an infinite plane. % above which the point source lies.
Because  
$h$ is sufficiently small compared %ing 
with the WD radius,
%and 
the reflecting area on the WD is sufficiently narrow.
Most reflection occurs %is occurred 
in a circle of %the whose 
the radius of %is 
0.001\,$\Rwd$ (see figure\,\ref{fig:images_h0p0001}).
For a %the 
larger viewing angle, the image is elongated %extends 
along the reflected direction.
This feature is caused by the anisotropy of the differential cross sections of the scattering %s
as %and 
previously reported by \cite{1991MNRAS.249..352G}.
Incident X-rays are more easily scattered to a direction close to %the parallel to 
the incident direction.
%And 
The incident direction and reflecting direction is closer 
when reflection position is more distant from the position right down %foot of the perpendicular line from %print of 
the point source. % in reflected direction.
Consequently %Therefore, 
when the viewing angle is larger, the image extends more in the reflecting direction.

In the case of %When 
$h = \Rwd$ in figure\,\ref{fig:images_h1p0}, %the distance of the point source from the WD surface of 
%$h$ = 1\,$\Rwd$ 
%is comparable to the WD radius and 
the WD curvature is not negligible.
Because of the curvature, X-rays are reflected to the viewing angle over %above 
90\,deg
and occultation by the WD limb occurs unlike the case of $h$ = 0.0001\,$\Rwd$.
Moreover, the image extension appearing in the large viewing angle of the %case of 
$h$ = 0.0001\,$\Rwd$
case does %
%dose 
not appear because the curvature 
radius of the reflector (WD) comparable to the height of the X-ray source precludes 
%makes 
the incident direction and reflected direction from getting closer, 
unlike the case of $h = 0.0001\,R_{\rm WD}$.
%more separate.
%which reduces the reflectivity % following the differential cross sections %of the scattering
%when the reflecting position from the point source footprint is more distant.
%These effects become more important for larger $h$.

 \onecolumn  
%
%\begin{figure}
%\includegraphics[width=180mm]{calc_ref_images_h0p01.eps}
%\vspace{1mm}
% \caption{Reflection images with a point X-ray source of 0.01 $\Rwd$ above the WD surface.
%The spectrum and energy range of the source, and the definition of the coordinate are common to 
%Fig.\,\ref{fig:images_h0p0001}.
%Radius of the white circle is 0.1\,$\Rwd$ whose center stands on same place as Fig.\,\ref{fig:images_h0p0001}.
%Range of reflecting angle ($i$) between the perpendicular line and the scattering direction
%is 0-1, 30-31, 60-61and 90-91 degrees from left to right.
%These images are normalized with generated photon number  
%and smoothed by Gaussian function with $\sigma$ of 3$\times$10$^{-3}$\,$\Rwd$.
% \label{fig:images_h0p01}}
%\end{figure}
%
%\begin{figure}
%\includegraphics[width=180mm]{calc_ref_images_h0p10.eps}
%\vspace{1mm}
% \caption{Reflection images with a point X-ray source of 0.1 $\Rwd$ above the WD surface.
%The spectrum and energy range of the source, and the definition of the coordinate are common to 
%Fig.\,\ref{fig:images_h0p0001}.
%Radius of the white circle is 1\,$\Rwd$ whose center stands on same place as Fig.\,\ref{fig:images_h0p0001}.
%Range of reflecting angle ($i$) between the perpendicular line and the scattering direction
%is 0-1, 30-31, 60-61and 90-91 degrees from left to right.
%These images are normalized with generated photon number  
%and smoothed by Gaussian function with $\sigma$ of 3$\times$10$^{-2}$\,$\Rwd$.
% \label{fig:images_h0p1}}
%\end{figure}
%

\twocolumn

\subsection{Spectrum}\label{subsec:po_spe}

\begin{figure}
\includegraphics[width=80mm]{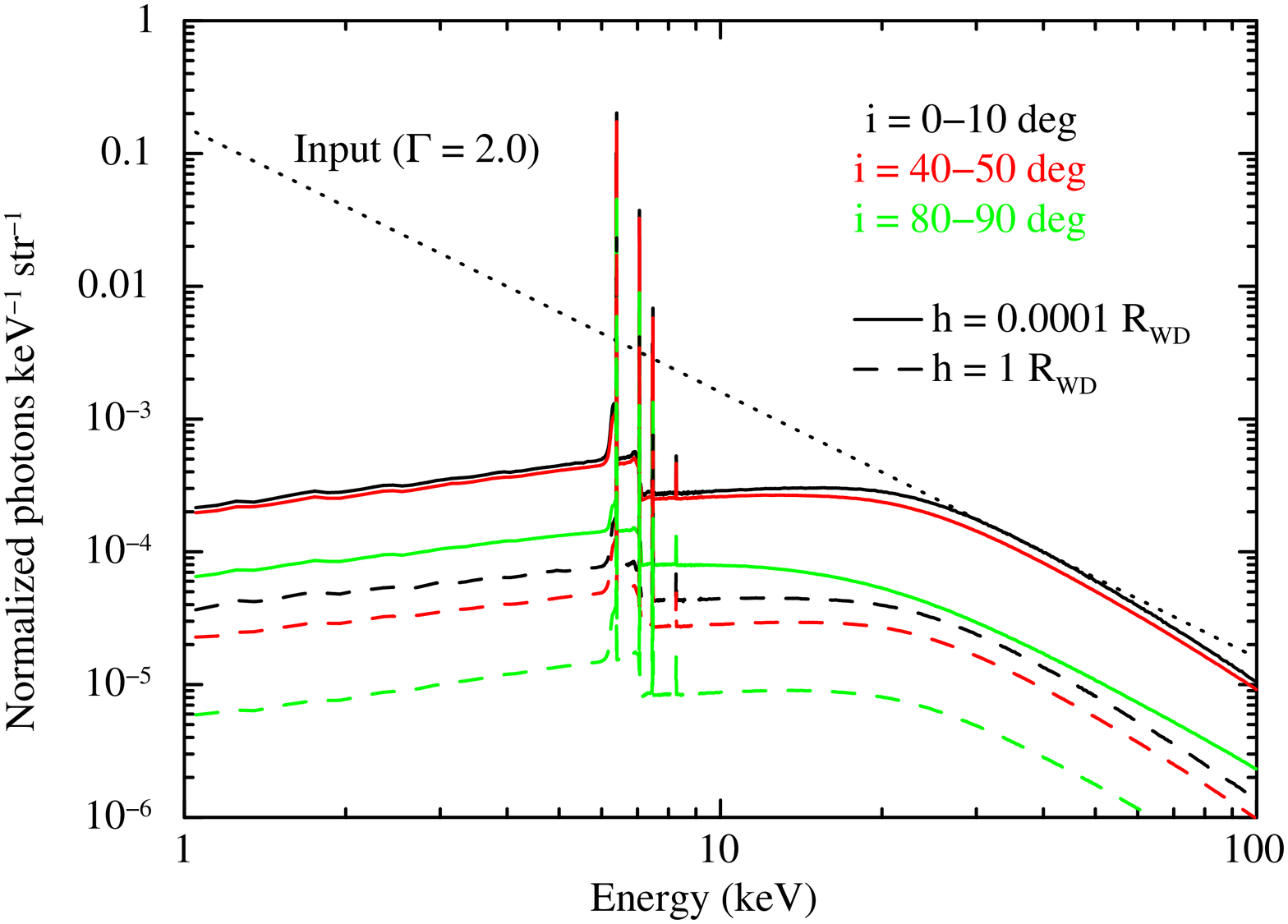}
\includegraphics[width=80mm]{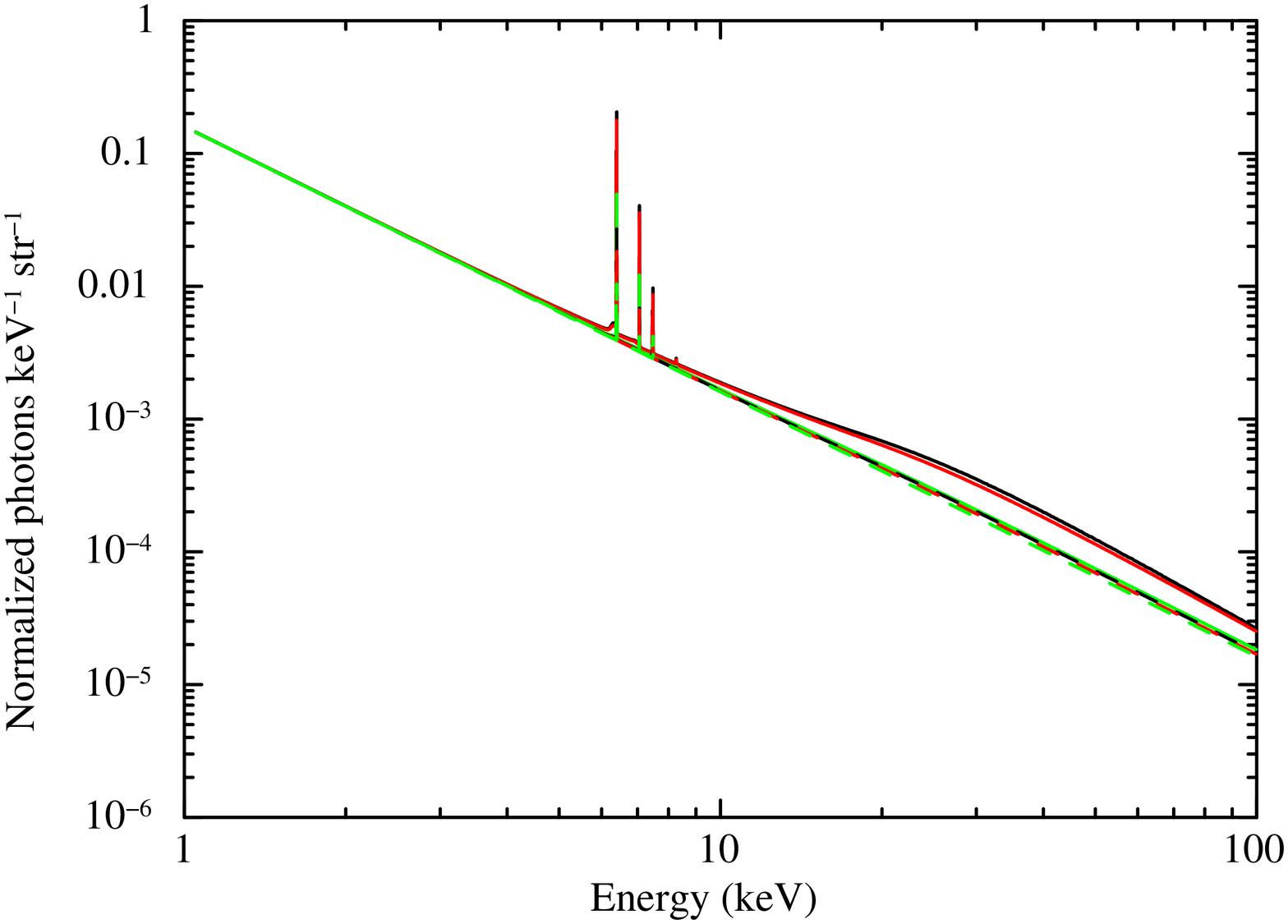}
\includegraphics[width=80mm]
{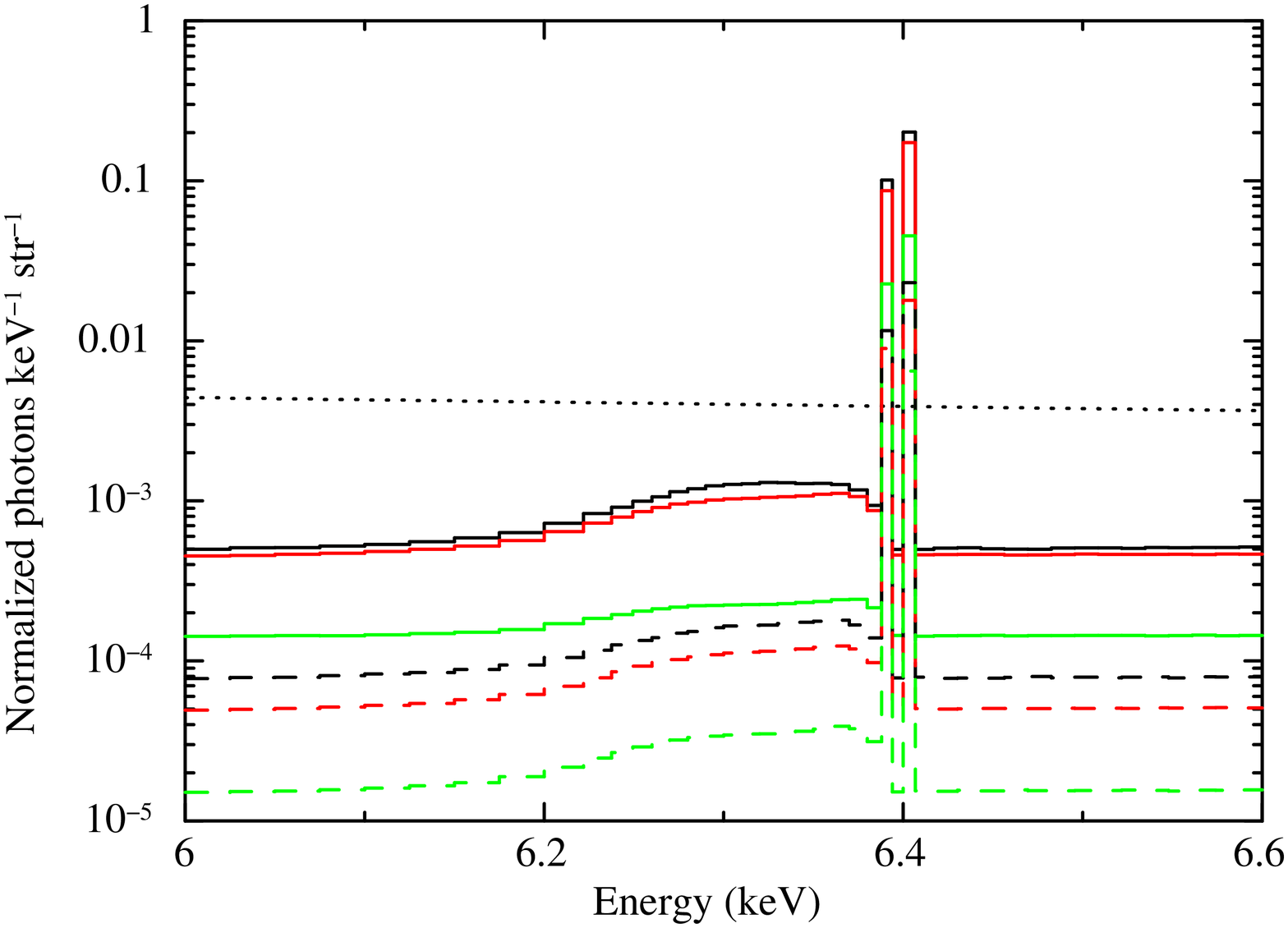}
\vspace{1mm}
 \caption{Reflected X-ray spectra from a %the 
 cool and spherical WD. 
 The spectrum of the irradiating point source is a power law with ${\it \Gamma}$ = 2.0 (dotted line).
 The height of the point source is assumed at $h$ = 0.0001\,$\Rwd$ (solid line) 
 and 1\,$\Rwd$ (dashed line). %\,$\Rwd$. 
 The viewing %reflecting 
 angle $i$ is 0--10\,(black), 40--50\,(red) and 80--90\,deg (green). %, respectively.
The middle panel shows the total spectra of the intrinsic and reflect%ion
ed emission components.
 The bottom panel is magnification of part of the iron K$\alpha_{1,2}$ % and Ni fluorescent 
 lines %part 
 of the top panel.
 Note that horizontal axis of the bottom panel is linear scale unlike the other panels.
 \label{fig:spectra}}
\end{figure}

Spectral structures of the %associated with the %The 
reflection %spectrum 
manifest themselves over
%are %is 
%comparable to %against 
the intrinsic spectrum %component 
in the following two energy domains. One is
%at two domains of %such as 
the fluorescent emission line in the band of 6--9\,keV, and the other is the Compton hump
%The Compton hump is 
formed around 10-50\,keV 
by a sequence of energy loss %down scattering 
by the incoherent scattering and photoelectric absorption %in the WD
(figure\,\ref{fig:mass_att_co}).
Figure \ref{fig:spectra} shows the reflection spectra %assumed 
with
${\it \Gamma }= 2.0$ for intrinsic power-law spectrum, % (dotted line), 
$h$ = 0.0001 %(solid line) 
and 1% (dashed line)
\,$\Rwd$ 
and, $i$ = 0--10, % (black), 
30--40 %(red) 
and 80--90% (green)
\,deg. 
This figure clearly shows the fluorescent lines, % and 
their Compton shoulder and the Compton hump.
%Note that 
The Compton shoulder is formed by the incoherent %Compton 
scattering of the fluorescent %reemitted 
photons.
The reflection containing %the 
all of the spectral features mentioned above is %are 
reduced as $h$ increases
because a solid angle of the WD surface viewed %ing 
from the point source becomes small. %is reduced. % for the higher $h$.
The reflection %also depends on $i$ and 
becomes less
%more 
intense for larger %smaller 
$i$
because a longer optical path is needed to %or the 
escape out from the WD. % with larger $i$.
%and the differential cross section of the scattering.
The details of the line emissions and the Compton hump are %is %are 
described %presented 
below.

\subsubsection{Iron K$\alpha_{1,2}$ %and nickel fluorescent 
lines and their Compton shoulders}\label{sec:fe_line}

The bottom panel of %the 
figure\,\ref{fig:spectra} is
magnification of the reflected X-ray spectra in %between 
6--8.5\,keV
where the iron K$\alpha_{1, 2}$ %, iron K$\beta$, nickel K$\alpha_{1, 2}$ and nickel K$\beta$ fluorescent 
lines appear with their Compton shoulder.
%The two iron fluorescent K$\alpha$ lines \textcolor{red}{out} of the six are especially intense 
%and their Compton shoulders are clearly seen 
%because of %its 
%large %the 
%\textcolor{red}{iron}
%abundance. % of the iron element.
A valley between the fluorescent iron K$\alpha_{1, 2}$ lines and their %its 
Compton shoulder is due to
the small cross section of incoherent forward scattering (see figure\,\ref{fig:dif_cross_section}).
%\textcolor{red}{A valley between the fluorescent line and its Compton shoulder 
%is due to small fraction of thin-angle incoherent scattering.
%%must exists because the coherent scattering dominates around scattering angle $\sim$ 0
%%(see figure \ref{fig:dif_cross_section})}.
Moreover, shape of the Compton shoulder is not double peak, unlike the case of the %as that of 
Compton scattering by free electrons at rest,
because the double peak is blurred by the Doppler effect caused by the momentum of the bound electrons.
%Note that fluctuation by $<$ 20\,\% is caused by the Poisson error.
We note that the lines at the original energies contain %reemitted 
fluorescent
photons directly arriving at the observer and those undergoing the coherent scattering(s).
We, however, do not discriminate them and just call their original name %,
%however, we call these lines original name of the reemission line 
such as fluorescent iron K$\alpha_1$ line.

Hereafter, we concentrate on the iron K$\alpha_{1, 2}$ lines.
%which is the most intense fluorescent line in the X-ray astrophysics in general.
We assume that the abundance is 1\,$Z_{\odot}$ %hereafter 
unless otherwise noted.
%in this section of \S\ref{sec:fe_line} except for \S\ref{sec:influ_abund}.
We define an energy centroid of the iron K$\alpha_{1, 2}$ fluorescent lines with their Compton shoulder
as
\begin{eqnarray}
%EW = \int^{E_1}_{E_0}{\rm d}E\frac{F_{\rm ref}(E)-F_{\rm ref}(6.41\,{\rm keV})}{((F_{\rm pl}(E_1)-F_{\rm pl}(E_0))/2)+F_{\rm ref}(6.41\,{\rm keV})}\label{eq:ew}
%EW = \int^{E_1}_{E_0}{\rm d}E\frac{F_{\rm ref}(E)-F_{\rm ref\mathchar`-cont}(E)}{F_{\rm pl}(E_{\rm c})+F_{\rm ref\mathchar`-cont}(E_{\rm c})}\label{eq:ew}
E_{\rm cen} = \frac{\int^{E_1}_{E_0}{\rm d}E E \left\{I_{\rm ref}(E)-I_{\rm ref\mathchar`-cont}(E)\right\}}{\int^{E_1}_{E_0}{\rm d}E \left\{I_{\rm ref}(E)-I_{\rm ref\mathchar`-cont}(E)\right\}} \label{eq:ec},
\end{eqnarray}
where $I_{\rm ref}(E)$ and $I_{\rm ref\mathchar`-cont}(E)$
are the intensity %ies 
of the reflection in total
and that of the continuum %component of the reflection
(excluding fluorescent %reemitted 
photons), respectively, 
at an energy of $E$ in unit of keV. %, respectively.
We adopted (6.100, 6.405) %(6.100, 6.404)
 as the set of ($E_0$, $E_1$).
%In equation\,\ref{eq:ec},  $I_{\rm ref\mathchar`-cont}(E)$ %the reflection continuum flux %at 6.41\,keV 
%is subtracted from $I_{\rm ref}(E)$
%in order to remove influence of the reflection continuum on the estimation of the energy centroid.
%$I_{\rm ref\mathchar`-cont}(E)$ is obtained by a %nother 
%supplemental simulation
%in which the iron K$\alpha_{1, 2}$ lines are not reemitted and the other calculating processes and inputted parameters are identical to the main simulation.
We remark that 
$I_{\rm ref}(E) - I_{\rm ref\mathchar`-cont}(E)$ is the spectrum of the iron K$\alpha$ lines and their reprocessed emission.

\begin{figure}
\includegraphics[width=80mm]{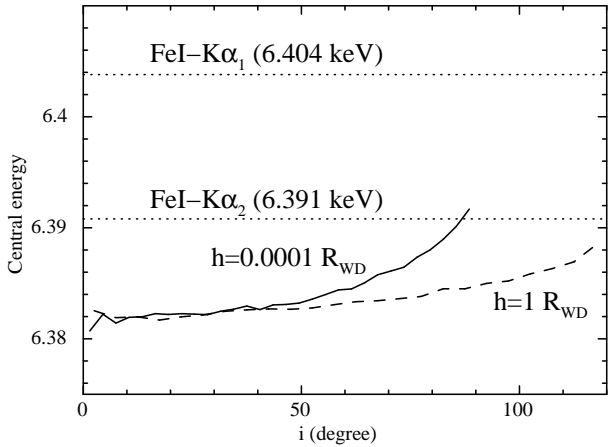}
\vspace{1mm}
 \caption{Relation between the %reflecting 
 viewing 
 angle $i$ and the energy centroid averaged over %of the total of 
 the fluorescent iron K$\alpha_{1, 2}$ lines 
and %with 
their %its 
 Compton shoulders.  
Solid and dashed lines assume $h$ = 0.0001 and 1\,$\Rwd$ for the irradiating point source.
%Abundance in the WD surface and 
Photon index of the source
is %are %commonly 
assumed to be %at 
%$Z$ = 1\,$Z_{\odot}$ and 
${\it \Gamma}$ = 2.0. %, respectively.
%Note that data is plotted only when the EW of the fluorescent Fe-K$\alpha_{1, 2}$ line no less than 1\,eV
%(see figure\,\ref{fig:i_ew}).
%\textcolor{red}{This figure will be replaced by results of 10$^{10}$ of the escapes.}
\label{fig:i_ecen}}
\end{figure}

Figure \ref{fig:i_ecen} shows the energy centroid %in the energy band 
%of the Fe-K$\alpha_{1, 2}$ fluorescent lines with their Compton shoulder 
slightly increases by $\sim$ 10\,eV as $i$ increases.
%
%settles down around 6.38\,keV
%and also varies within 40\,eV.
This is because the %The variation mainly anti-correlates to 
intensity ratio of the Compton shoulders to that of the %lines at 
original lines (see below) %, that is, the energy centroid 
is smaller with increasing $i$.
%lower when the Compton shoulders are more intense relative to the original lines. % ratio is higher.
We note that the difference of the centroid from 6.4\,keV is no more than 20\,eV.
%centroid is different from 6.4\,keV by only 6\,eV
%Energy centroid of of Fe-K$\alpha_{1, 2}$ with their Compton shoulders depend on $i$
%because relative intensity of Compton shoulder to the lines at original energies and shape of the Compton shoulder varies with $i$.

%Figure\,\ref{fig:i_ew} shows equivalent width (EW) of the fluorescent Fe-K$\alpha_{1, 2}$ lines and those with their Compton shoulder 
%against the reflecting angle $i$
%with the source height of $h$ = 0.0001, 0.01, 0.1 and 1\,$\Rwd$.
We define a line equivalent width (EW) as 
\begin{eqnarray}
%EW = \int^{E_1}_{E_0}{\rm d}E\frac{F_{\rm ref}(E)-F_{\rm ref}(6.41\,{\rm keV})}{((F_{\rm pl}(E_1)-F_{\rm pl}(E_0))/2)+F_{\rm ref}(6.41\,{\rm keV})}\label{eq:ew}
%EW = \int^{E_1}_{E_0}{\rm d}E\frac{F_{\rm ref}(E)-F_{\rm ref\mathchar`-cont}(E)}{F_{\rm pl}(E_{\rm c})+F_{\rm ref\mathchar`-cont}(E_{\rm c})}\label{eq:ew}
EW = \int^{E_1}_{E_0}{\rm d}E\frac{I_{\rm ref}(E)-I_{\rm ref\mathchar`-cont}(E)}{I_{\rm pl}(E_{\rm cen})+I_{\rm ref\mathchar`-cont}(E_{\rm cen})}\label{eq:ew}
\end{eqnarray}
where %$I_{\rm ref}(E)$, $I_{\rm ref\mathchar`-cont}(E)$ and 
$I_{\rm pl}(E)$
is intensity of %are intensities of the reflection, continuum of the reflection (excluding reemitted photons) and 
the intrinsic power law at a energy of $E$ % in unit of keV. %, respectively,
and $E_{\rm cen}$ is the energy centroid of the line in question. %objective line. % shown in figure\,\ref{fig:i_ecen}.
In equation\,\ref{eq:ew},  $I_{\rm ref\mathchar`-cont}(E)$ %the reflection continuum flux %at 6.41\,keV 
is subtracted from $I_{\rm ref}(E)$ to extract the line photons only in the numerator,
%in order to remove contribution of the reflection continuum to the objective line
and add the reflection continuum to the power law continuum to calculate 
total continuum in the denominator. %the EW.
%$I_{\rm ref\mathchar`-cont}(E)$ is obtained by another supplemental simulation
%in which Fe-K$\alpha_{1, 2}$ are not reemitted and the other calculating processes and inputted parameters are identical to the main simulation.
%Abundance of the WD surface and 
%Cases of the point source height of $h$ = 0.0001, 0.01, 0.1 and 1\,$\Rwd$ are shown in this figure
%with solid, dashed, chain and dotted lines, respectively.
Sets of ($E_0$, $E_1$) for K$\alpha_1$ and K$\alpha_2$  
%(which includes fluorescent Fe-K$\alpha$ photon escaped by direct or by coherent scattering(s))
are (6.403, 6.405) %(6.402, 6.404) 
and (6.390, 6.392), respectively, 
%for the Compton shoulder of the K$\alpha_{1, 2}$ lines is (6.100, 6.390)
and for the total of the K$\alpha_{1, 2}$  lines with their Compton shoulder  is (6.100, 6.405). %(6.100, 6.404).
%and its corresponding Compton shoulder are $E_1=6.240$\,keV and $E_2=6.402$\,keV, respectively.
%The lower end of the energy range of the Compton shoulder is determined 
%to be the minimum energy of fluorescent Fe-K$\alpha$ photon scattered once. % whose original energy is 6.4 keV.
$E_{\rm cen}$ for the K$\alpha_{1, 2}$ lines are their nominal energies (see \S\ref{sec:interact})
and that for total of the K$\alpha_{1, 2}$ lines with their Compton shoulder is %$E_{\rm cen}$ 
shown in figure\,\ref{fig:i_ecen}.
%The photon index of the point source
%is assumed to be %1\,solar abundance and 
%${\it \Gamma}$ = 2. %, respectively.

\begin{figure}
\includegraphics[width=80mm]{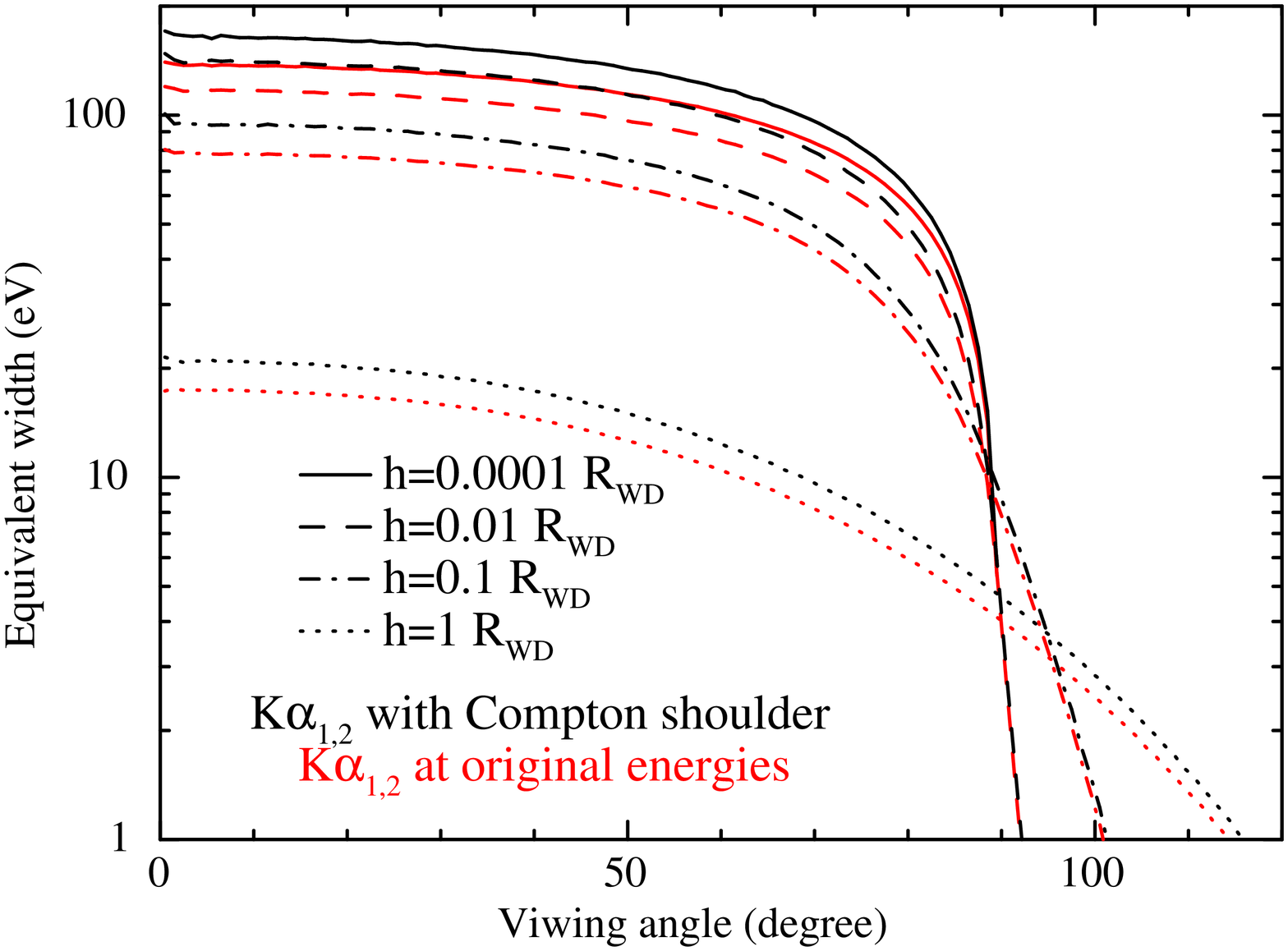}
\vspace{1mm}
 \caption{Relation between the viewing angle $i$ and the EWs of the complex of 
 the fluorescent iron K$\alpha$ line.
Solid, dashed, dashed-dotted
%chain 
and dotted lines are the cases of $h$ %assuming h 
= 0.0001, 0.01, 0.1 and 1\,$\Rwd$ 
for the point source, respectively.
The black and red lines show the
%Black %Thick 
%line shows 
EW of sum of the iron K$\alpha_{1, 2}$ %6.4\,keV 
lines with and without their Compton shoulder, respectively.
%and  red %thin 
%line shows the sum of the iron K$\alpha_{1, 2}$. %EW of 6.4\,keV line.
Photon index of the source
is %are %commonly 
assumed to be %at 
%$Z$ = 1\,$Z_{\odot}$ and 
${\it \Gamma}$ = 2.0. %, respectively.
%Abundance in the WD surface and photon index of the source
%are %commonly 
%assumed at $Z$ = 1\,$Z_{\odot}$ and ${\it \Gamma}$ = 2.0, respectively.
%%\textcolor{red}{This figure will be replaced by results of 10$^{10}$ of the escapes. Total and direct component should be appear.}
\label{fig:i_ew}}
\end{figure}

Figure\,\ref{fig:i_ew} %This figure 
%is relation 
clearly exhibits that monotonic decrease of the EWs against %the 
increase of the viewing angle $i$
irrespective of $h$. % as referred above.
This is because the optical path is longer in larger $i$, which makes %so that 
the line photon difficult to %can 
escape from the WD.
The EW of the K$\alpha_{1, 2}$ lines 
in the case of the $h = 0.0001\,\Rwd$ is consistent 
with the calculation of \cite{1991MNRAS.249..352G}
although %However, 
the authors assumed the Compton scattering 
with free and stationary electrons which are distributed on a flat disc 
with a radius of 100 times larger than a point source height.
%The authors assumed that a flat disc whose radius is 100 times length of a point source height 
%and the scattering is attributed to the free electrons at rest (Compton scattering).
In higher $h,$ the solid angle of the WD viewed %ing 
from the point source becomes smaller
and therefore, the EW becomes smaller.
Moreover, significant EW appears in the %larger 
viewing angle %more above %than 
%above 
larger than 90\,deg 
in higher $h$
because of the spherical shape of the %reflecting 
WD 
%and $i$ of the tangential line from the point source to the WD surface over 90\,deg
%which corresponds to the maximum $i$,
although the EW is %somewhat 
small ($<$ 10\,eV). % to present practical observations.
We note that EW ratio of the K$\alpha_{1, 2}$ is 2:1 
with any $i$ and $h$
according to their relative intensity %we adopted 
(\S\ref{sec:interact}). % with any $i$ and $h$.

\begin{figure}
\includegraphics[width=80mm]{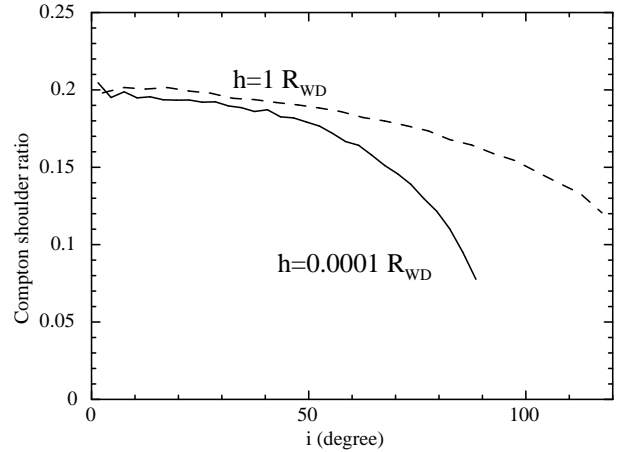}
\vspace{1mm}
 \caption{%Ratio of EW of the Compton shoulder of the fluorescent Fe-K$\alpha_{1, 2}$ to sum of the EWs of 
 EW ratio of the Compton shoulder of the fluorescent iron K$\alpha_{1,2}$ lines to the sum of the EWs of %the EWs for
% \textcolor{blue}{the intrinsic} 
iron K$\alpha_{1, 2}$ lines at respective original energies. 
 % at their original energy. %directly escaped fluorescent Fe-K$\alpha$ line.
 Solid and dashed lines are the cases of $h$ %assume h 
 = 0.0001 and 1\,$\Rwd$ for the irradiating point source.
% Note that data is plotted only when the sum of the EWs of the iron K$\alpha_{1, 2}$ lines no less than 1\,eV
%(see figure\,\ref{fig:i_ew}).
%\textcolor{red}{This figure will be replaced by results of 10$^{10}$ of the escapes.}
\label{fig:i_intensity_ratio}}
\end{figure}

%The Compton shoulder varies with associating $i$ and $h$ in its intensity and shape
%as shown in the bottom panel of figure \ref{fig:spectra}.
%Although the EW of the Compton shoulder is roughly proportional to that of directly escaped fluorescent line,
The EW ratio
%R%A r
%atio of the EW 
of the Compton shoulder to %that of %summed EW of 
sum of the fluorescent iron K$\alpha_{1, 2}$ lines %at their original energies %directly escaped fluorescence
%and 
decreases as $i$ increases as shown in figure \ref{fig:i_intensity_ratio}.
The EW of the Compton shoulder is obtained by equations\,\ref{eq:ec} and \ref{eq:ew}
%the common procedure to that of the K$\alpha_{1, 2}$ lines with their Compton shoulder 
using ($E_0$, $E_1$) of (6.100, 6.390). % for equation\,\ref{eq:ec} and \ref{eq:ew}.
%With lower $h$ and with higher $h$ in lower $i$, 
%the intensity of the Compton shoulder reduces against that of the fluorescence line in higher $i$.
This drop %as $i$ %variation 
of the intensity ratio with increasing $i$ is caused by the following two effects.
One is 
the fact that a fluorescent line photon generated deeper in the WD can experience a larger number of scatterings before escaping from the WD surface. 
This means that a line spectrum from a deeper layer possesses a more prominent Compton shoulder.
%that interaction occurs %average of interacting 
%at deeper inside of the WD %increases
%with larger number of interactions of the tracing photon
%because a part of interacting photons moving to the WD surface escapes from the WD.
The other is that in larger $i$, only photon whose last interaction occurs at a shallower layer can escape out from the WD surface
because the optical path to escape from the WD
increases by 1/$\sin(90-i)$.
One can observe a reflection spectrum from a shallower layer of the WD in a larger $i$ configuration, 
which reduces the EW of the Compton shoulder relative to the seed iron K$\alpha_{1,2}$ lines.
%That is to say, escaped photons undergo less interactions in larger $i$,
%which means escaped photon hardly %dose not 
%undergo scattering after reemission 
%and Compton shoulder becomes less intense as $i$ increases.
%The EW ratio is lower than 0.1 at any $i$. 
%By contrast, in higher $i$ with higher $h$, 
%the intensity ratio of the Compton shoulder increases with higher $i$.
%In reality, the increase caused by the scattering of not the reemitted photon
%but the incident photon directly emitted from the point source.
%In this situation, the scattering of the incident photon becomes significant 
%since the fluorescent line is weak.
%Moreover, scattering efficiency of escaping photon who undergoes only one interaction of scattering is high
%because the reflecting region on the WD surface is narrow because of occultation with the limb 
%(see the bottom panel in figure \ref{fig:images_h1p0})
%and, therefore, incident directions of those photons are aligned and incident to the WD with a thiner angle.

\begin{figure}
\includegraphics[width=80mm]{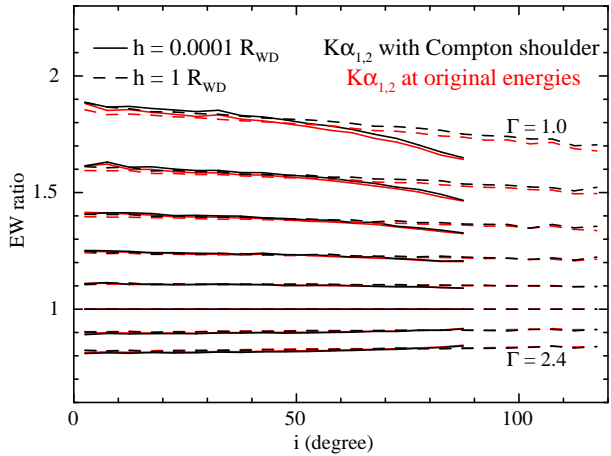}
\vspace{1mm}
 \caption{EW ratio of the sum of the fluorescent iron K$\alpha_{1, 2}$ lines of various  ${\it \Gamma}$ to that of ${\it \Gamma} = 2.0$
(see figure\,\ref{fig:i_ew}) versus %in 
the viewing %reflecting 
angle $i$.
 Solid and dashed lines are cases of $h$ = %assume the source height of 
 0.0001 and 1\,R$_{\rm WD}$, respectively.
 Black %Thick 
 and red %thin 
 lines contain the iron K$\alpha_{1, 2}$ lines with and without their Compton shoulder, % and the %sum of 
 %only the iron K$\alpha_{1, 2}$,
 %lines at their original energies, 
 respectively.
 These four cases are calculated for
 ${\it \Gamma}$ of 1.0 to 2.4 with an interval %at intervals 
 of 0.2. % are used for the four-line groups from top to bottom.
% \textcolor{red}{This figure will be replaced by results of 10$^9$ of the escapes.}
 \label{fig:ew_gamma_ratio}}
\end{figure}

Figure\,\ref{fig:ew_gamma_ratio} shows ratio of the EW calculated for various intrinsic spectra
to that for the spectrum with ${\it \Gamma}=2.0$.
The
EW ratio increases with decreasing ${\it \Gamma}$.
%, but is insensitive to 
%$h$ especially in high ${\it \Gamma}$.
%The EW increases as ${\it \Gamma}$ decreases irrespective of $h$ (figure\,\ref{fig:ew_gamma_ratio}).
This is because the number of photons with the energy
%photons whose energy is 
higher than the absorption edge (7.11\,keV) 
increases when %are more %than those 
%with larger 
 ${\it \Gamma}$ decreases.
Note that the EW ratio is insensitive to $h$, especially in high ${\it \Gamma}$.
The larger $i$ suppresses the number of %interactions of 
the escaping %ed 
photons from deeper layer,  as explained %referred 
above.
A smaller ${\it \Gamma}$ spectrum contains a larger number of more energetic photons that can produce the iron K$\alpha_{1,2}$ lines in a deeper layer of the WD. Such iron lines, however, do not escape from the WD in a higher $i$ case, because of a longer absorption path within the WD.
%Energetic photons such as 100\,keV should undergo a number of %sequence of 
%the incoherent scattering to be absorbed by iron because of its small cross section in the higher energy.
%The sequence of the incoherent scattering makes the photon go down to the deeper layer.
%%Energetic photons should undergo a sequence of the incoherent scattering to be absorbed.
Consequently, %Therefore,
larger $i$ reduces the difference of the EW caused by  ${\it \Gamma}$.

\subsubsection{Compton hump}\label{sec:comp_hump}

\begin{figure}
\includegraphics[width=80mm]{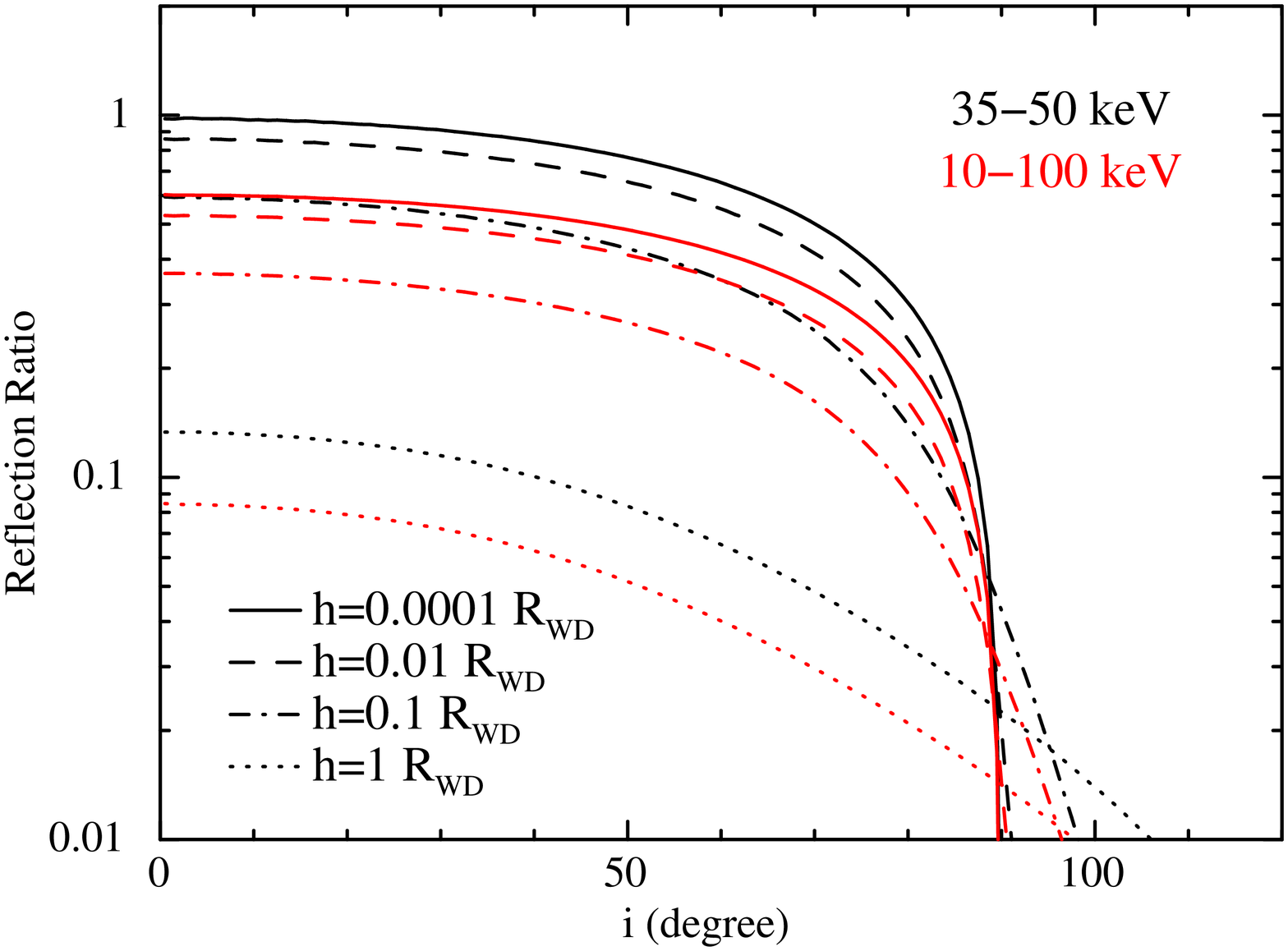}
\vspace{1mm}
 \caption{Ratio of the X-ray reflection to the intrinsic component in the viewing angle $i$.
Black %Thick 
and red %thin 
lines calculated in % assume 
energy bands of 
35--50\,keV and 10--100\,keV, respectively.
Solid, dashed, dashed-dotted
%chain 
and dotted lines are cases of %assume 
$h$ of 0.0001, 0.01, 0.1 and 1\,$R_{\rm WD}$.
 ${\it \Gamma}$ = 2.0 is assumed.
 \label{fig:albedo_ratio}}
\end{figure}

Compton hump varies %associated 
with $i$ and $h$ in its intensity and shape (see figure\,\ref{fig:spectra}). 
%as reemission lines.
%We note that abundance is assumed to be 1\,$Z_{\odot}$ in this \S\ref{sec:comp_hump} as well as \S\ref{sec:fe_line}.
Figure\,\ref{fig:albedo_ratio} shows intensity ratios of the reflection %component 
to the intrinsic power law %component 
%against 
versus the viewing angle
in 35--50\,keV %(thick line) 
and 10--100\,keV %(thin line) 
for various $h$. %with some $h$s.
%$h$ is assumed to be 0.0001\,(solid), 0.01\,(dash), 0.1\,(chain) and 1\,(dotted)\,$R_{\rm WD}$.
The relation between the ratio and $i$ %is determined by the optical depth relating $i$.
%In fact, the relation 
is similar to the relation between the EW and $i$ %the reflecting angle 
(figure\,\ref{fig:i_ew}).

%Shape of the Compton hump also varies relating $i$ and $h$.
The top panel of figure\,\ref{fig:spectra} shows that the 
spectrum scattered off in 0--10 deg has a well defined
Compton hump%in 0--10\,deg is sharper than that 
, whose %which 
peak is less prominent in larger $i$,
especially in 80--90\,deg.
As previously mentioned, the larger $i$ suppresses the number of interactions
(incoherent scattering is dominant above 10\,keV, see figure\,\ref{fig:mass_att_co}), therefore, %that is,
% of the escaped photon, 
the less %larger 
number of interactions
%which 
makes the shape of Compton hump %reflection spectrum 
close to that of the intrinsic spectrum.
 
\begin{figure}
\includegraphics[width=80mm]{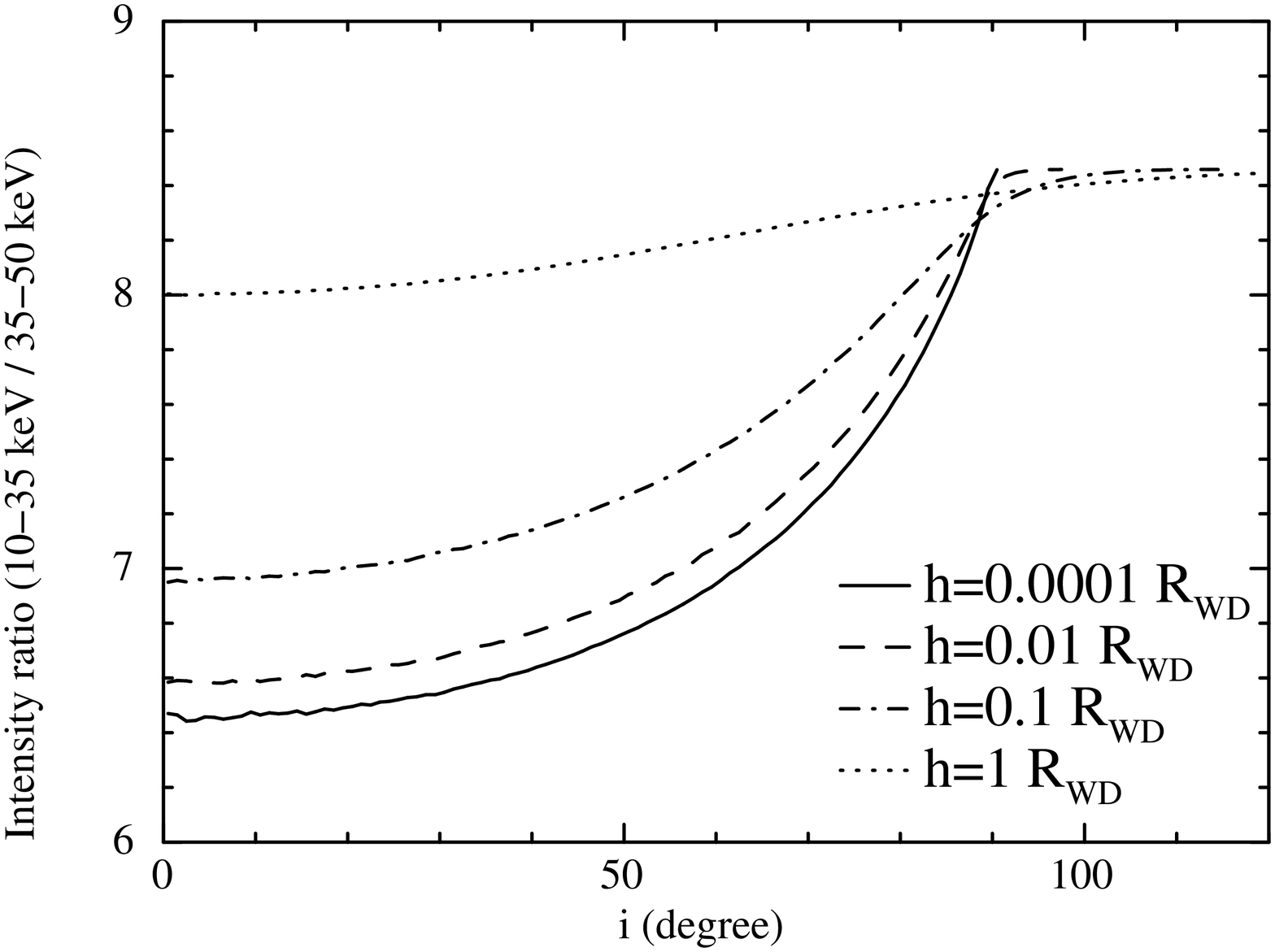}
\includegraphics[width=80mm]{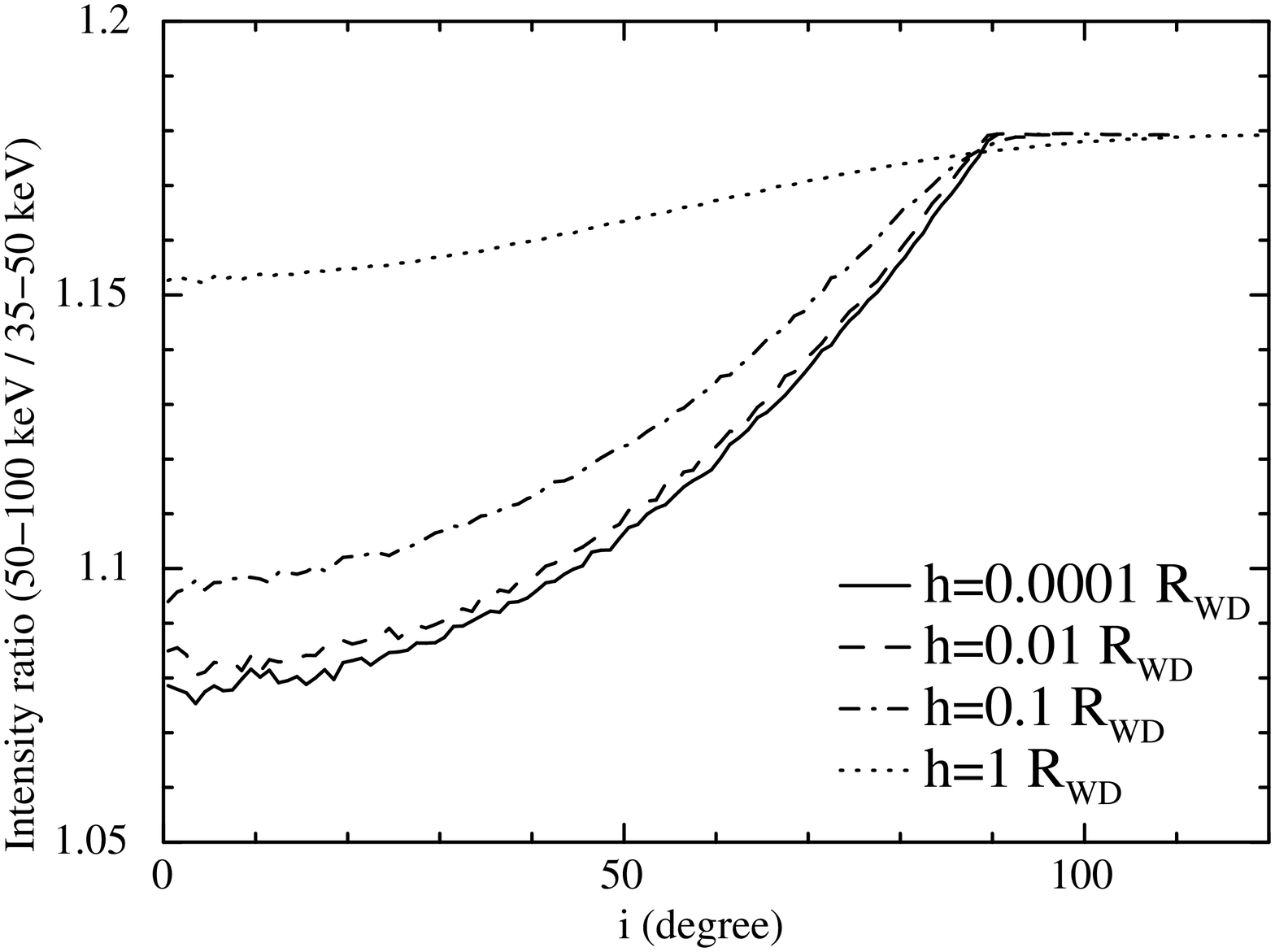}
\vspace{1mm}
 \caption{Ratio of intensity of the total (intrinsic + reflected) spectrum
 in 10--35\,keV (top) and 50--100\,keV (bottom) to that in 35--50\,keV 
versus %in 
the viewing %reflecting 
angle $i$. 
 Solid, dashed, dashed-dotted
 %chain 
 and dotted lines are cases of %assume 
 $h$ of 0.0001, 0.01, 0.1 and 1\,$R_{\rm WD}$.
 ${\it \Gamma}$ = 2.0 is assumed.
%\textcolor{red}{This figure will be replaced by results of 10$^{10}$ of the escapes.}
 \label{fig:intensity_ratio}}
\end{figure}

%By contrast, 
Unlike the emission line, 
the Compton hump can hardly be %hardly 
separated from the intrinsic continuum %component 
in actual data.
Nevertheless, %Therefore, 
we should evaluate the shape of the Compton hump 
with the total spectrum which includes the intrinsic power law and the reflection. % components.
%Therefore, the shape of observable total spectrum is dominated by not shape but intensity of the reflection
% (middle panel of figure\,\ref{fig:spectra}).
Figure\,\ref{fig:intensity_ratio} shows ratios
of intensity of the total spectrum in 10--35\,keV % (top) 
and 50--100\,keV %(bottom) 
to 35--50\,keV.
%with the total spectrum.
The ratios drop toward a smaller $i$, which shows that the reflection component in 35--50~keV is stronger than those in the other bands, especially in small $i$. The ratios with $i>90$~deg are almost constant because they are determined by the intrinsic power law.
%In smaller $i$, the reflection component is stronger especially in 35--50\,keV. % where the denominator is integrated. %calcuculated.
%Thus, these intensity ratios are smaller in smaller $i$ in %with
%%the 
%both energy bands of the numerator. %in figure\,\ref{fig:intensity_ratio}.
%The reflection %component 
%decreases as the $i$ increases
%which increases these intensity ratios up to a value determined by only the intrinsic power law. %component.

\begin{figure}
\includegraphics[width=80mm]{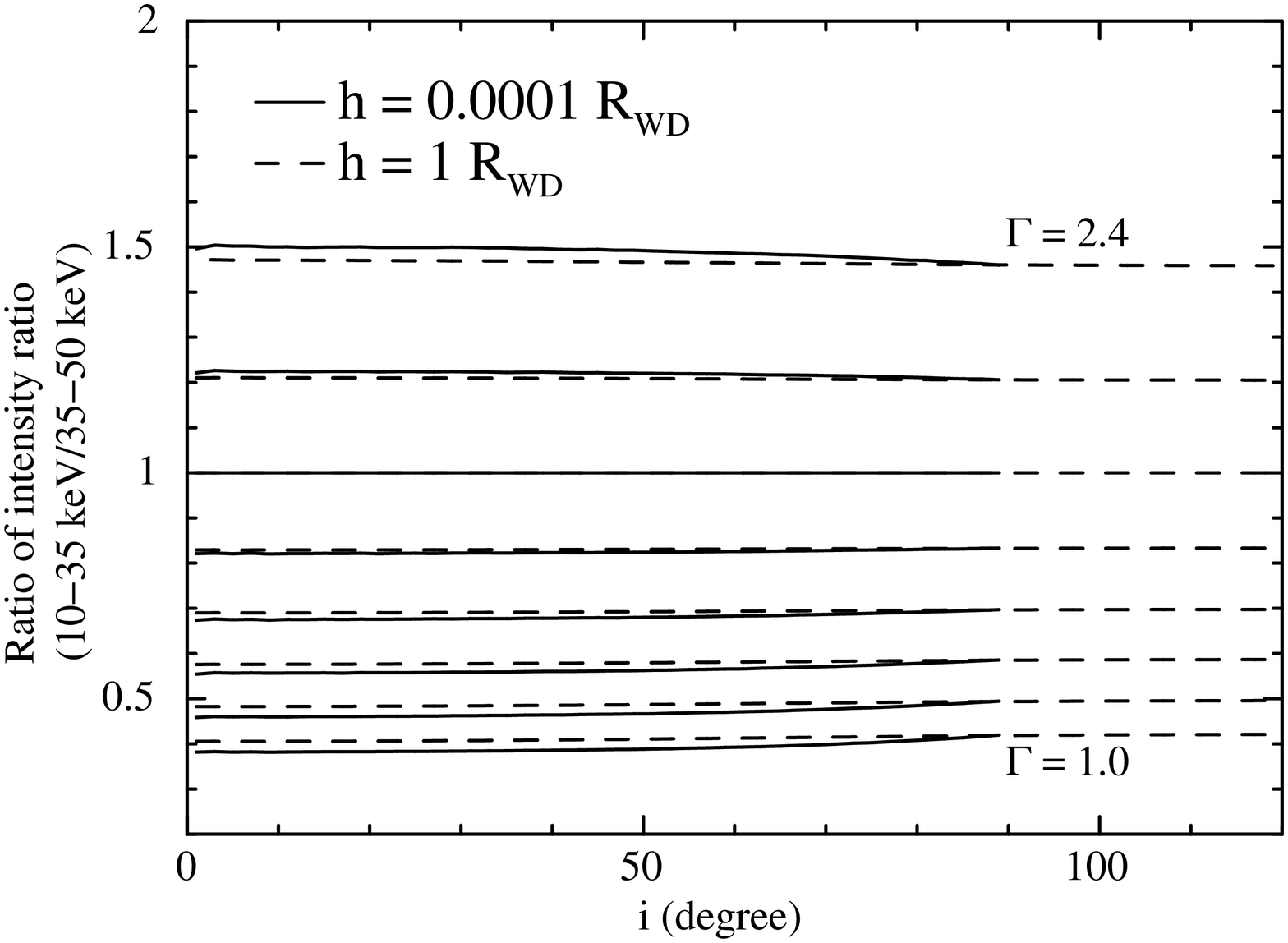}
\includegraphics[width=80mm]{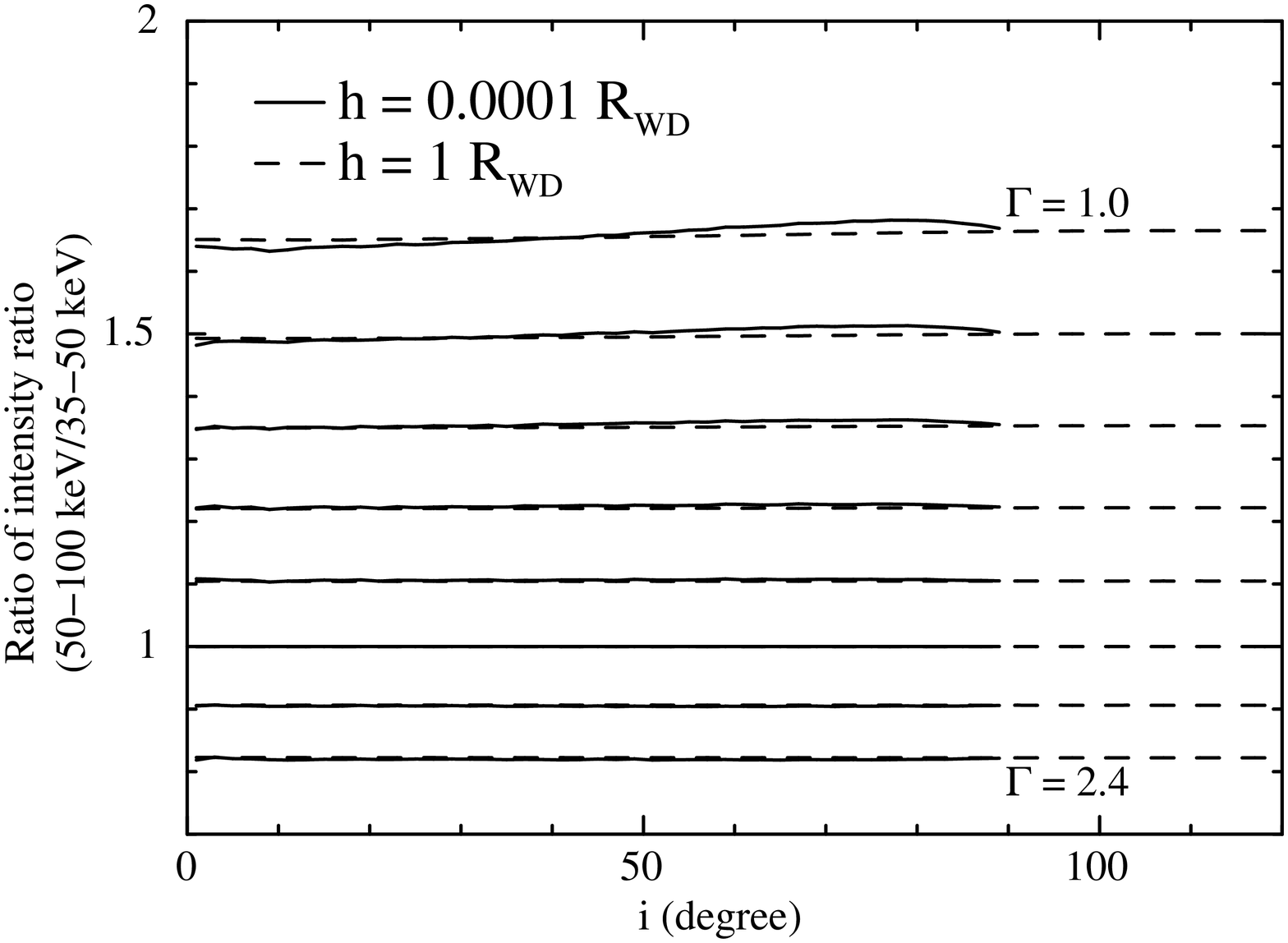}
\vspace{1mm}
 \caption{Ratio of %ratio of 
 intensity in 10--35\,keV (top) and 50--100\,keV (bottom) of various  ${\it \Gamma}$
 to that in 35--50\,keV of $\Gamma = 2.0$ (see Figure\,\ref{fig:albedo_ratio})
as a function of %in 
the viewing %reflecting 
angle $i$. 
 Solid and %, 
 dashed %, chain and dotted 
 lines are cases of % assume 
 $h$ of 0.0001 %, 0.01, 0.1 
 and 1\,$R_{\rm WD}$.
Calculation is made for these two cases at ${\it \Gamma}$ of 1.0 to 2.4 with an interval %at intervals 
 of 0.2 are used to the two-line groups from bottom to top 
in the top panel and from top to bottom in the bottom panel.
%\textcolor{red}{This figure will be replaced by results of 10$^9$ of the escapes.}
 \label{fig:intensity_ratio_ratio}}
\end{figure}

These intensity ratios depend on  ${\it \Gamma}$ %, an index of the intrinsic power law %component 
as shown in figure\,\ref{fig:intensity_ratio_ratio}.
We note that the difference in %by %the  
${\it \Gamma}$ appears mainly in offsets of the intensity ratios
with only small dependence on $i$, which
%The offset is simply caused by the intrinsic power law. %component.
%Small %Slight 
%$i$ dependence of each line in figure\,\ref{fig:intensity_ratio_ratio} 
is caused 
by the nature of the reflection %component 
whose spectral shape %spectrum 
approaches the intrinsic power law %in shape 
in larger $i$
because the number of interactions is suppressed.

%\begin{figure}
%\includegraphics[width=80mm]{ref_spe_icomp_hcomp_z1_emin1_emax500_PI-2_enum1e9_refEratio10-35_35-50.eps}
%\includegraphics[width=80mm]{ref_spe_icomp_hcomp_z1_emin1_emax500_PI-2_enum1e9_refEratio50-100_35-50.eps}
%\vspace{1mm}
% \caption{ \label{fig:albedo_ratio}}
%\end{figure}

\subsubsection{Influence of abundance}\label{sec:influ_abund}
\begin{figure*}
%\hspace{5mm}
\begin{center}
\begin{tabular}{cc}
\includegraphics[width=80mm]{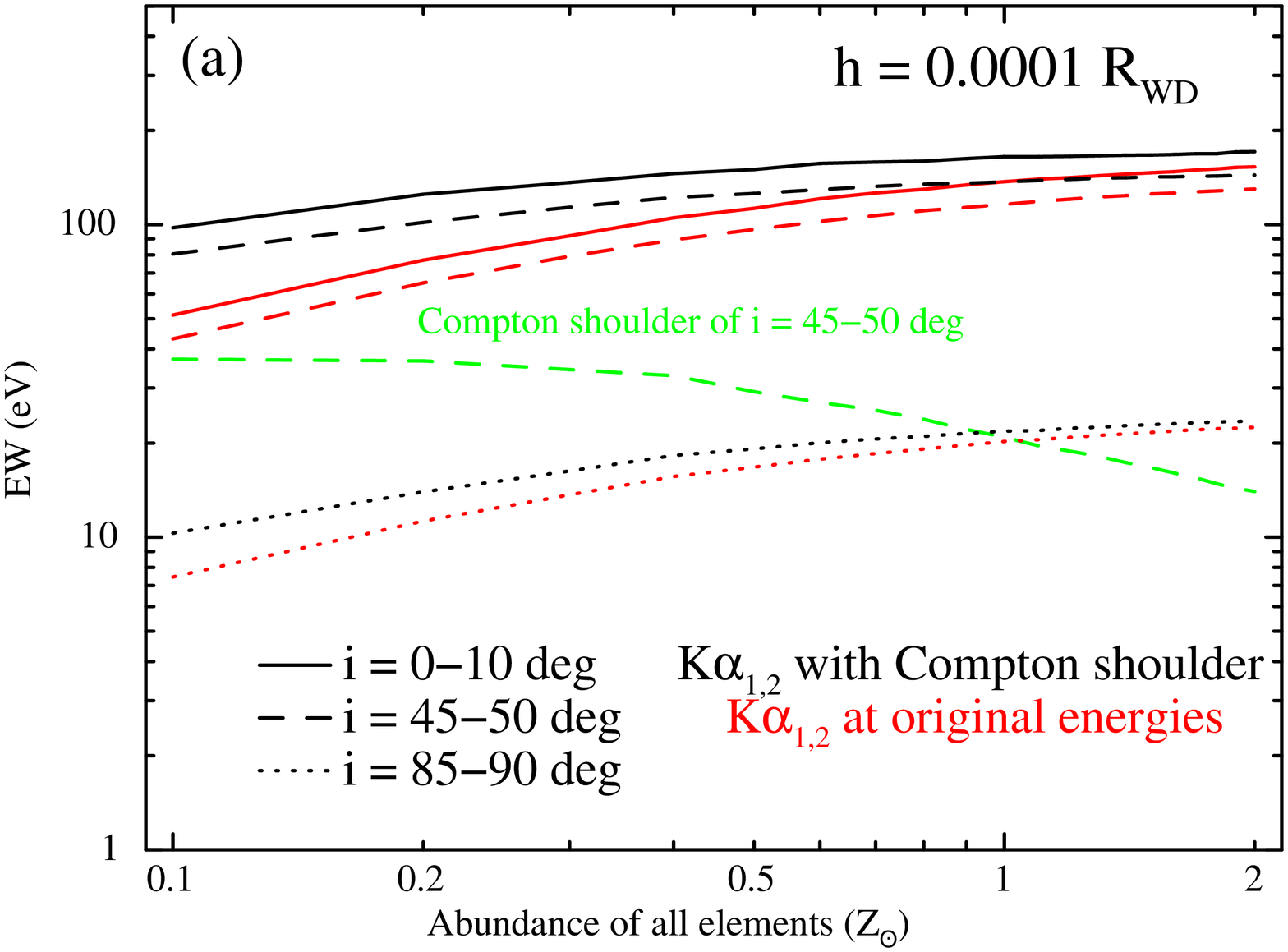}&
\includegraphics[width=80mm]{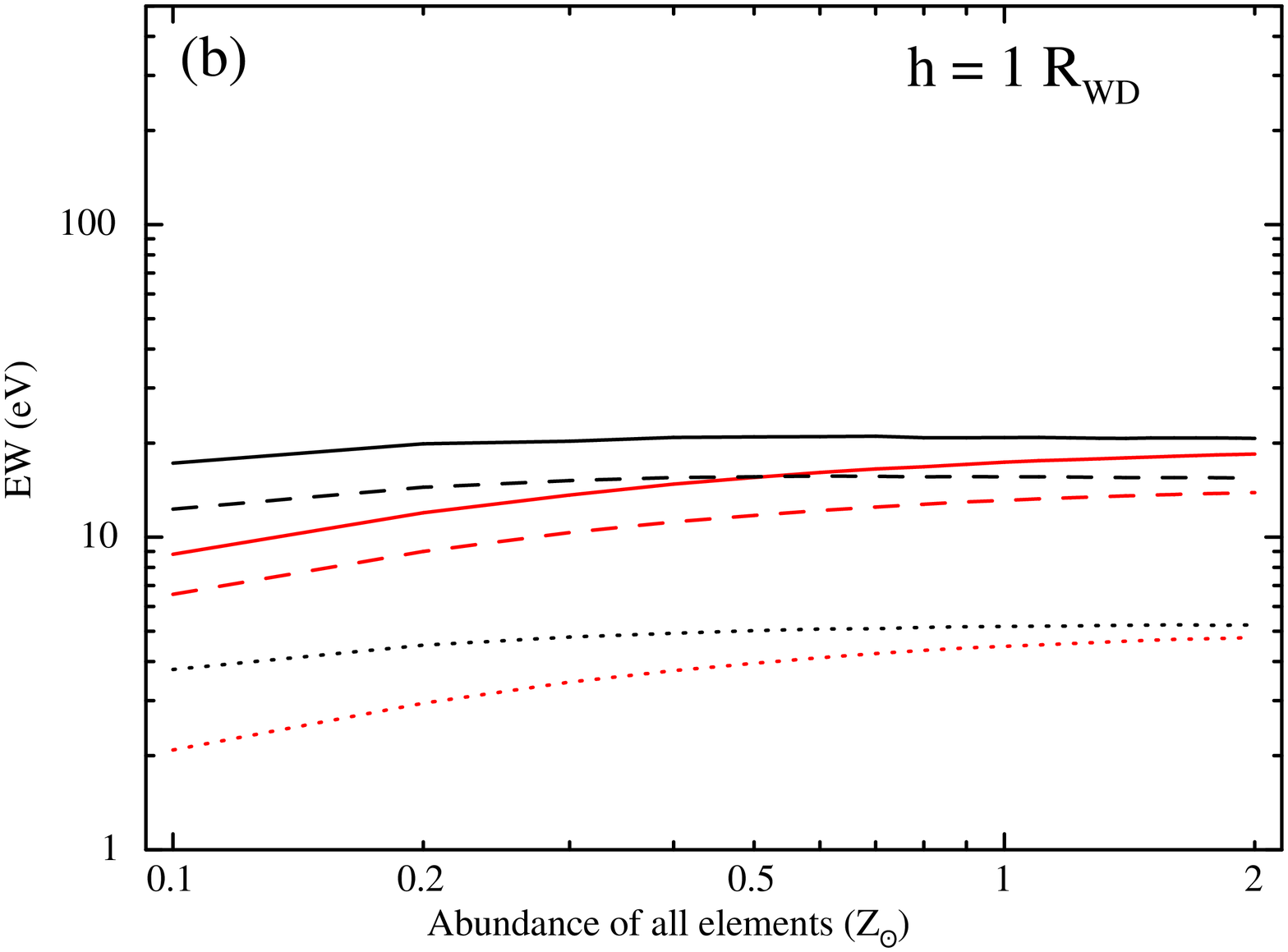}\\
\includegraphics[width=80mm]{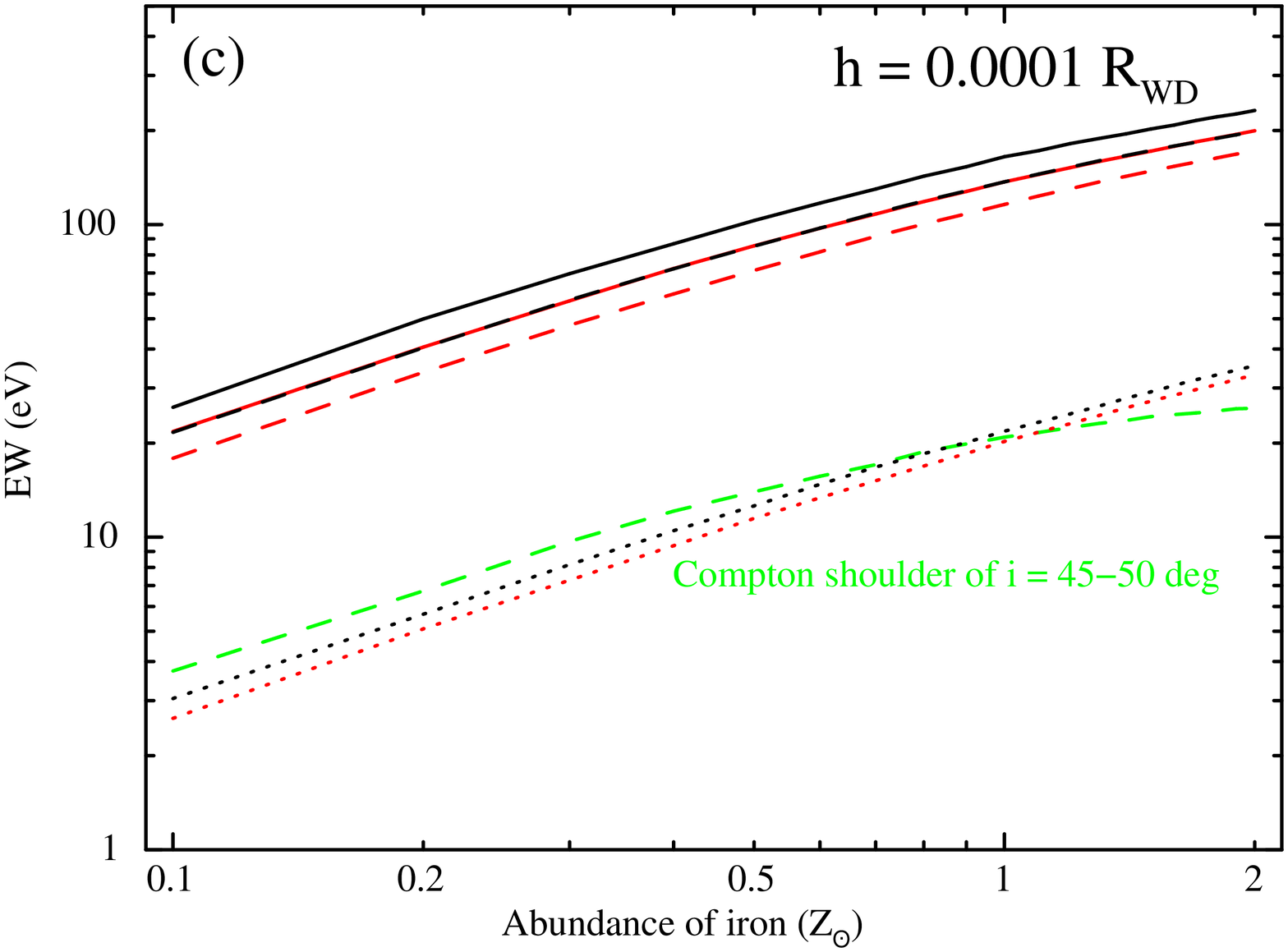}&
\includegraphics[width=80mm]{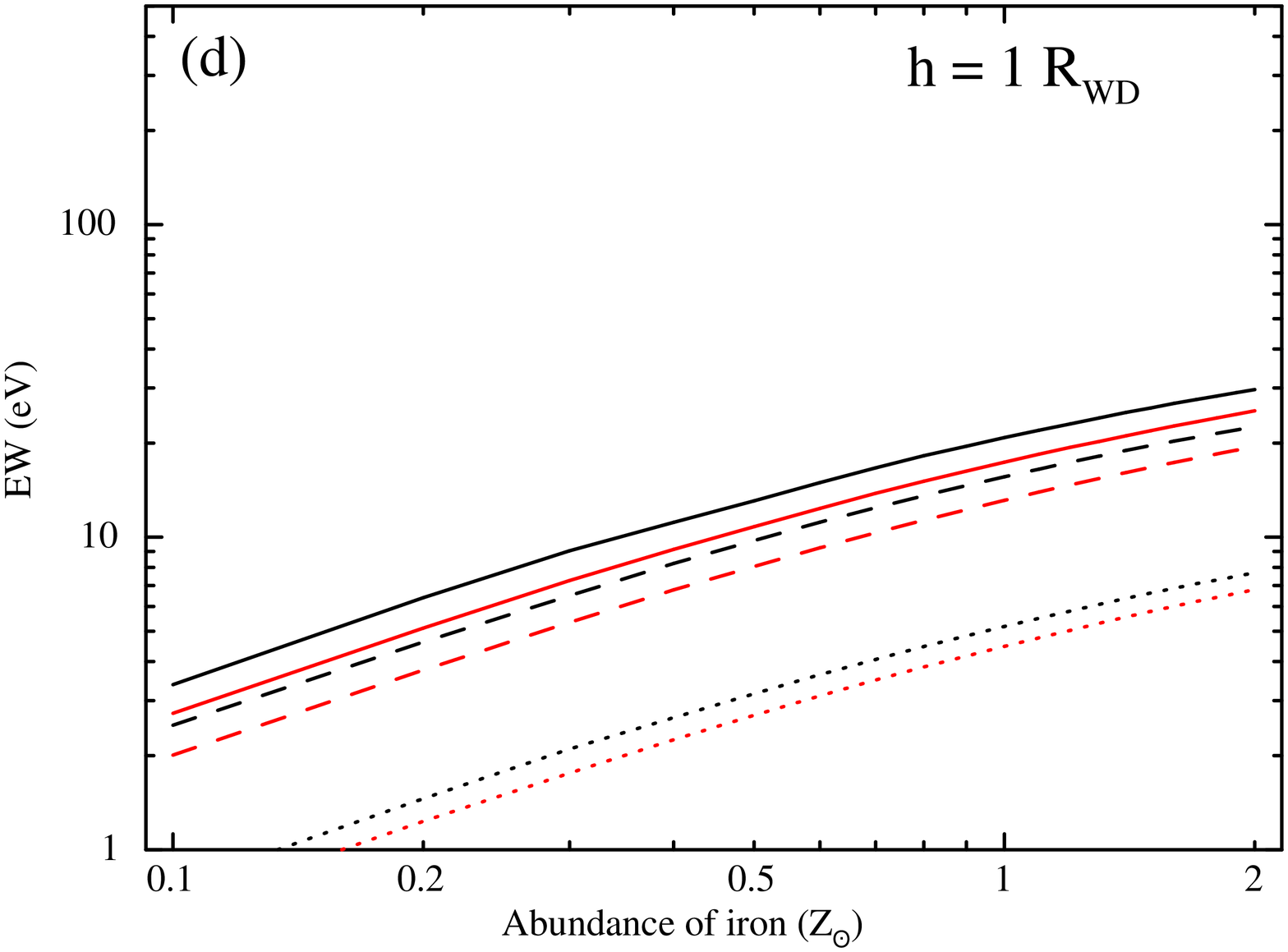}\\
\end{tabular}
\end{center}
 \caption{Relations between abundance and the total %sum of the 
 EW %s 
 of the %summed 
 fluorescent iron K$\alpha_{1,2}$ lines.
Abundances of all elements are %is simultaneously 
varied at the common rate relative to the solar composition
in the top two panels ((a), (b)). % as their ratio is fixed at that of the solar abundance.
In the bottom panels ((c), (d)), only iron abundance is varied and the others are fixed at one solar abundance.
$h$ = 0.0001\,$R_{\rm WD}$ and 1\,$R_{\rm WD}$ cases are shown %are assumed 
in the left ((a), (c)) 
and the right ((b), (d)) panels, respectively.
Solid, dashed and dotted lines are cases of $i$ = 5, 45 and 85\,deg, respectively.
The total EW of the %total of the 
fluorescent iron K$\alpha_{1,2}$ lines and their corresponding Compton shoulder is shown by black %thick
lines and 
that %sum 
of the %EWs %that of the summed 
%of the 
fluorescent iron K$\alpha_{1,2}$ lines is shown by red %thin 
lines.
%Red 
Green dashed line %s 
presents the EW of the Compton shoulder in the panels of $h = 0.0001\,R_{\rm WD}$ ((a), (c))
with $i$ = 45--50\,deg.}
%The ends are common to figure \ref{fig:density_geocomp}.}
 \label{fig:z_ew_dependency}
\end{figure*}

The fluorescent iron K$\alpha_{1,2}$ lines, % and 
their Compton shoulder and Compton hump %due to Compton scattering
%Compton shoulders and the Compton hump 
are influenced by the abundance.
Figure\,\ref{fig:z_ew_dependency} shows plots of the line EWs versus the abundance
%relations between the abundance and 
%their EWs %of %the total of 
%the K$\alpha$ lines %6.4\,keV line 
%and those with their %corresponding
%Compton shoulders
%% (black thick lines), and 
%%EW of sum of the K$\alpha$ lines line (black thin lines) 
with various viewing angle ($i$) and heights of the source ($h$). % parameters.
The EWs initially increase with %as 
the abundance, %increases, 
but
%and, on the other hand, 
peaks out at certain abundances in %the 
all cases, as reported in \cite{1991MNRAS.249..352G}.

When the abundances of all elements %(except hydrogen) 
are simultaneously increased from zero ((a) and (b) of figure\,\ref{fig:z_ew_dependency}),
%the effect of 
%the scattering is %processes are 
%suppressed and 
the probability of the absorption of all elements including iron %by any element 
is enhanced relative to %and 
the scattering %becomes relatively minor 
by degrees.
%This makes the line photons escape from the WD more efficiently without being disturbed by increase of the effective path length due to the scattering.
Consequently, %Therefore, 
the EW increases as the abundances %of the all elements 
increase.
%With the higher abundance, 
In the higher abundances where
%As the abundances increase further,
the scattering becomes %is 
negligible, %and 
the EW peaks out around 1 and 0.2 solar abundance
for h = 0.0001 and 1\,$\Rwd$, respectively, 
%becomes constant
because the %abundance fraction of 
iron abundance relative to the others %total 
is constant.
%fraction of the iron among the elements dose not vary.
By contrast, their Compton shoulders decrease in the %as drawn with a 
dashed %shown by
%red 
green line in panel (a) of %in 
figure\,\ref{fig:z_ew_dependency}
because the scattering is suppressed.
The decrease of the Compton shoulder counteracts 
the increase of the %total 
EW of the %summed 
iron K$\alpha_{1,2}$ lines. % with their Compton shoulders.

%With the higher abundance, the scatterings is almost negligible ($Z\geq$1\,Z$_\odot$) as shown with the red dashed line in the two top panels 
%and the possibility of absorption by iron is constant. 
%Therefore, the EW is independent from the abundance.
When only the iron abundance is varied and the other abundances are fixed at 1\,solar abundance
((c) and (d) of figure\,\ref{fig:z_ew_dependency}),
% at 1\,$Z\_\odot$,
the EWs are proportional to the iron abundance when the iron abundance is lower than %low and up to 
1 and 0.5 solar abundance %of iron
for $h = 0.0001$ and 1\,$\Rwd$, respectively.
With larger iron abundance, the increase of the EW becomes gentle
because the %cross section of the iron determines the total cross section 
total cross section is dominated by photoelectric absorption of iron
and the probability of absorption by iron asymptotically becomes %comes close to 
constant. 
The EW of Compton shoulders also increase as the iron abundance increases 
(dashed %red 
green line in panel (c) of %in 
figure\,\ref{fig:z_ew_dependency})
because the scattering is mainly caused by lighter elements than iron and is not suppressed by the low iron abundance.
%lighter elements than iron mainly cause the scattering and the scattering is not suppressed 
%when the iron abundance is low.
With larger iron abundance than the range of figure\,\ref{fig:z_ew_dependency}),
the EW of the Compton shoulders becomes constant as the EWs of the K$\alpha_{1,2}$ lines,
because the cross sections of the scatterings are also determined by %the 
iron \citep{1991MNRAS.249..352G}.

%We note that the EW of the Compton shoulder (red lines in $h=0$\,$R_{\rm WD}$ cases of figure\,\ref{fig:z_ew_dependency})
%influences on the total EW.
%Especially, in the case of all abundance varied,
%the EW of the Compton shoulder decrease as the abundance increase
%because the scatterings reduced.
%The decrease of EW of the Compton shoulder contributes to  peaking out of the total EW.

\begin{figure*}
%\hspace{5mm}
\begin{center}
\includegraphics[width=79mm]{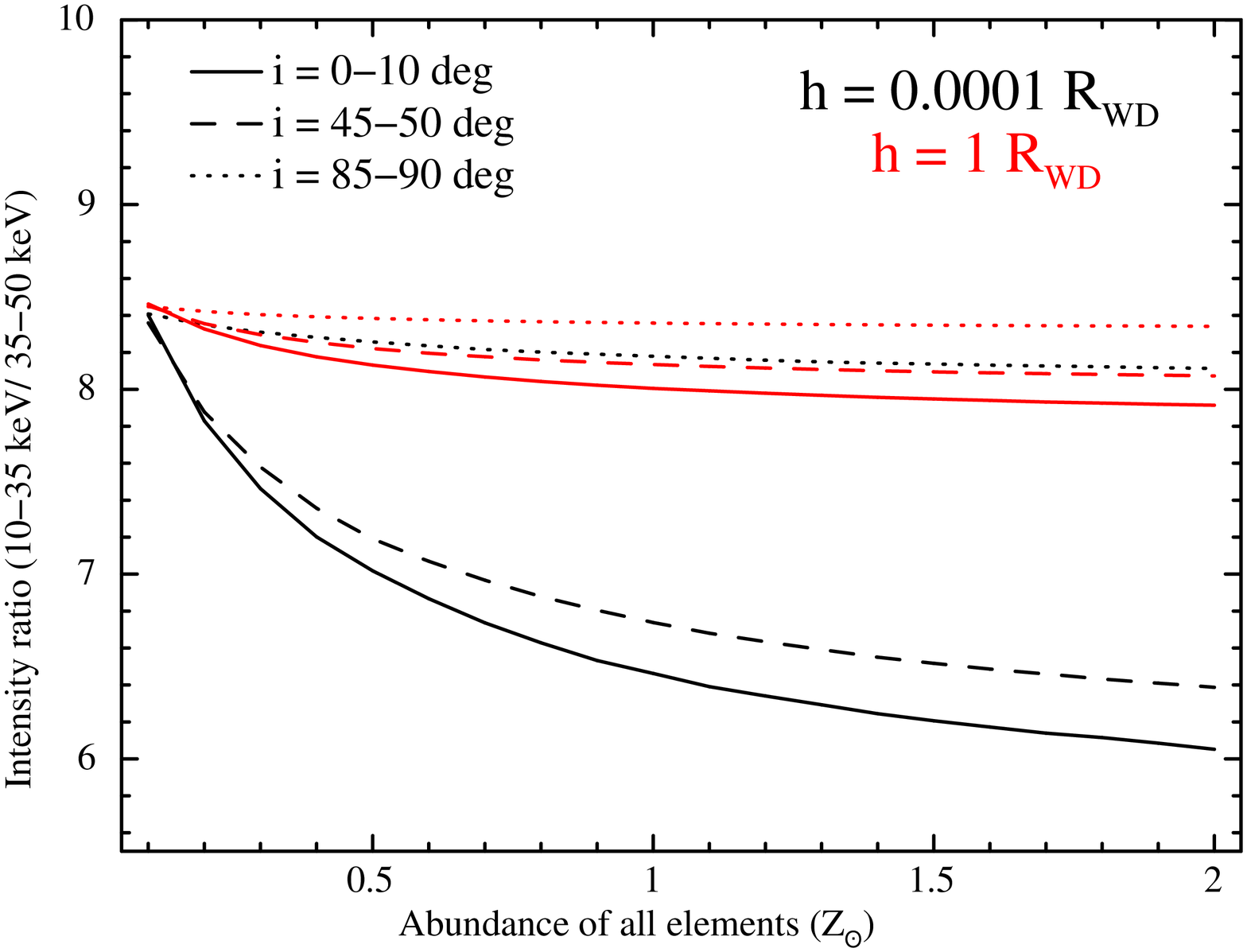}
\includegraphics[width=81mm]{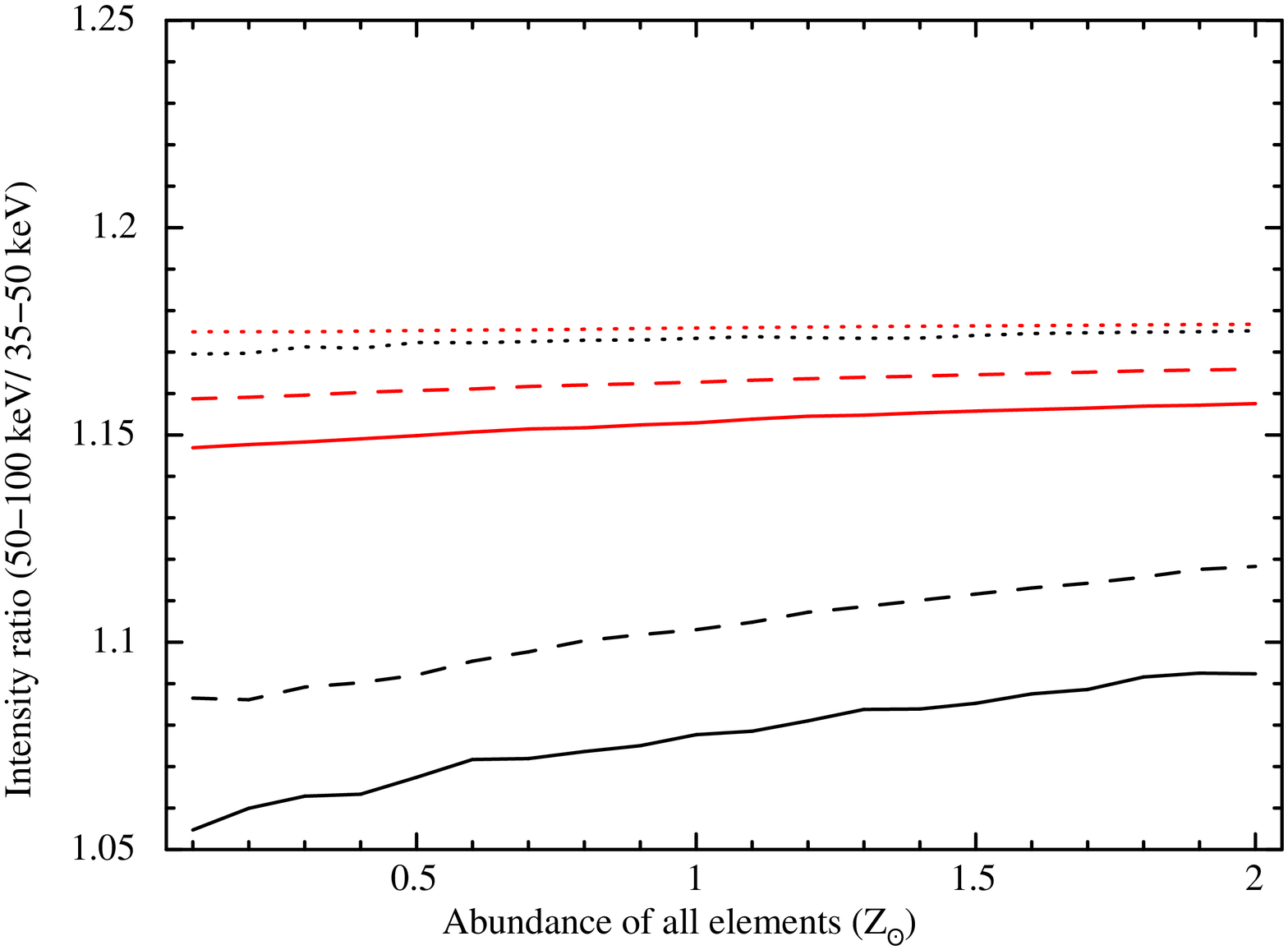}
\end{center}
 \caption{Relations between abundance and the ratio of intensities in 10-35\,keV (left) and 50-100\,keV (right) to that in 35-50\,keV.
Abundances of all elements are %simultaneously 
varied at the common rate relative to the solar composition.
%as their ratio is fixed at that of the solar abundance.
$h$ = 0.0001\,$R_{\rm WD}$ and 1\,$R_{\rm WD}$ are shown with black and red, respectively.
Solid, dashed and dotted lines are the cases of $i$ = 5, 45 and 85\,deg, respectively.}
%The ends are common to figure \ref{fig:density_geocomp}.}
 \label{fig:z_albedo_dependency}
\end{figure*}

The Compton hump depends on the abundance as well as the fluorescent iron K$\alpha_{1,2}$ lines.
Figure \ref{fig:z_albedo_dependency} shows ratios of intensity in 10--35\,keV % (left panel) 
and 50--100\,keV % (right panel)
to that in 35--50\,keV as functions of the abundance with various $i$ and $h$. %parameters.
%In this figure, $h$ of 0.0001\,$R_{\rm WD}$ (black) and 1\,$R_{\rm WD}$ (red), 
%and $i$ = 5, 45 and 85\,deg (solid, dashed and dotted lines) are adopted.
This figure indicates that the intensity in 10--35\,keV decreases and by contrast, that in 50--100\,keV increases against that in 35--50\,keV
in any cases, which means that cut-off energy %low energy bending %energy of low energy side 
of the Compton hump increases as the abundance increases.
%The shift of the cut-off energy is caused by vary of the photoelectric absorption dominated by the abundance.
The heavier photoelectric absorption caused by the higher abundance 
results in a larger absorption optical depth at any given energy, and, as a result, increases the cut-off energy
%The heavier photoelectric absorption caused by the higher %larger 
%abundance prevents the higher
% energy %etic
% photons from escaping 
%and increases the cut-off %bending 
%energy.

\section{PSAC source}\label{sec:result_psac}
In this section, we present results of the simulation %s 
using the PSAC source introduced in \S\ref{sec:source}.
The shape of reflection spectrum depends on %is varied by 
the WD mass, the specific accretion rate, the abundance and 
the viewing angle. %$i$.
The results with the PSAC source are more complex than those with a
%the 
point source of a power-law spectrum (\S\,\ref{sec:result_po}).
However, the reflection spectrum with the PSAC source 
can be basically represented by summation of the point sources. % with various parameters.
% except for 
%features caused by emission lines in the intrinsic spectrum.
%Therefore, we concentrate on describing the results and in most cases in this section.

\subsection{Spectrum}

\subsubsection{dependence on the viewing angle}%}%Influence of reflection angle}
\begin{figure}
%\hspace{5mm}
\begin{center}
\includegraphics[width=80mm]{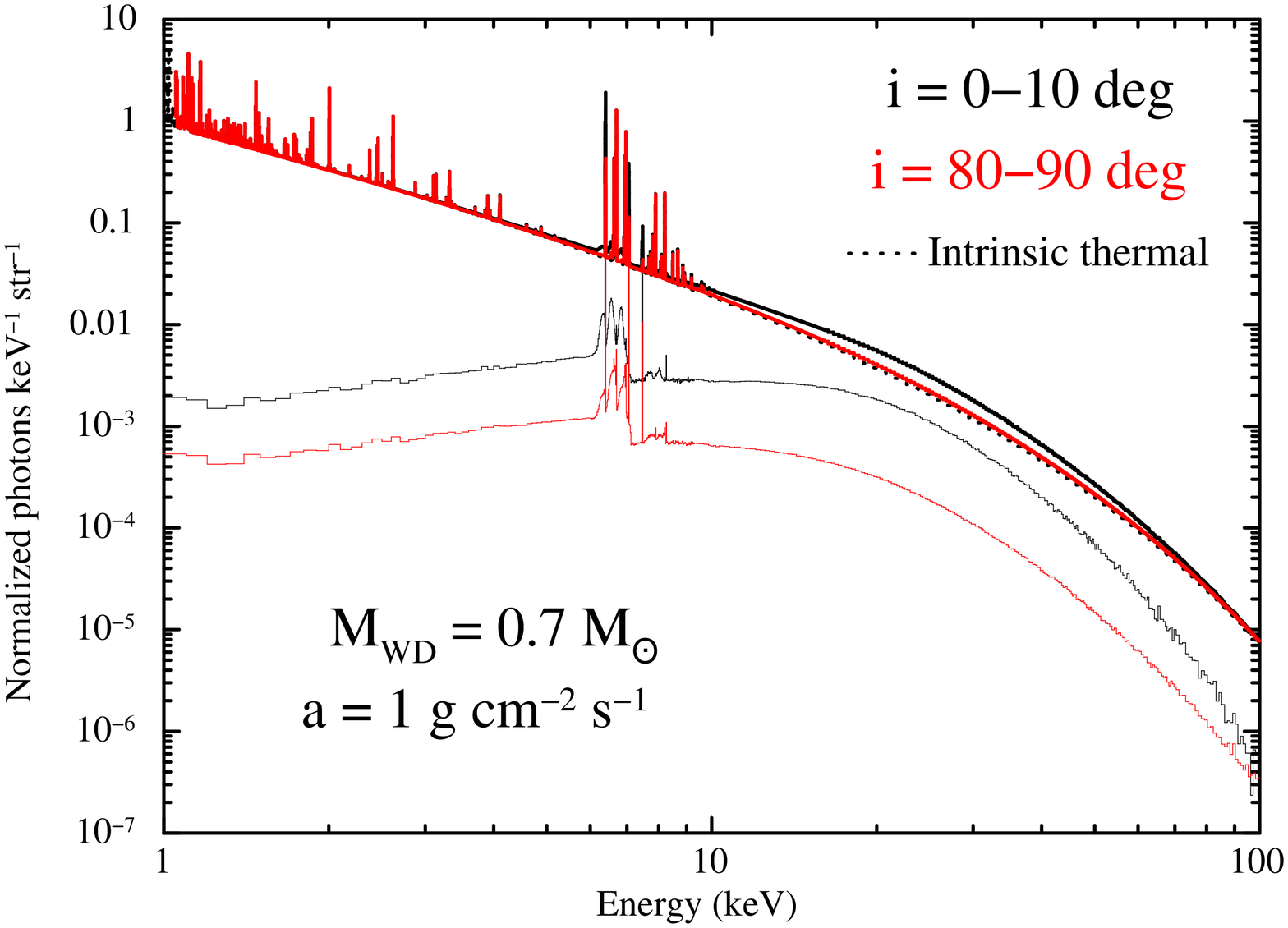}
\includegraphics[width=80mm]{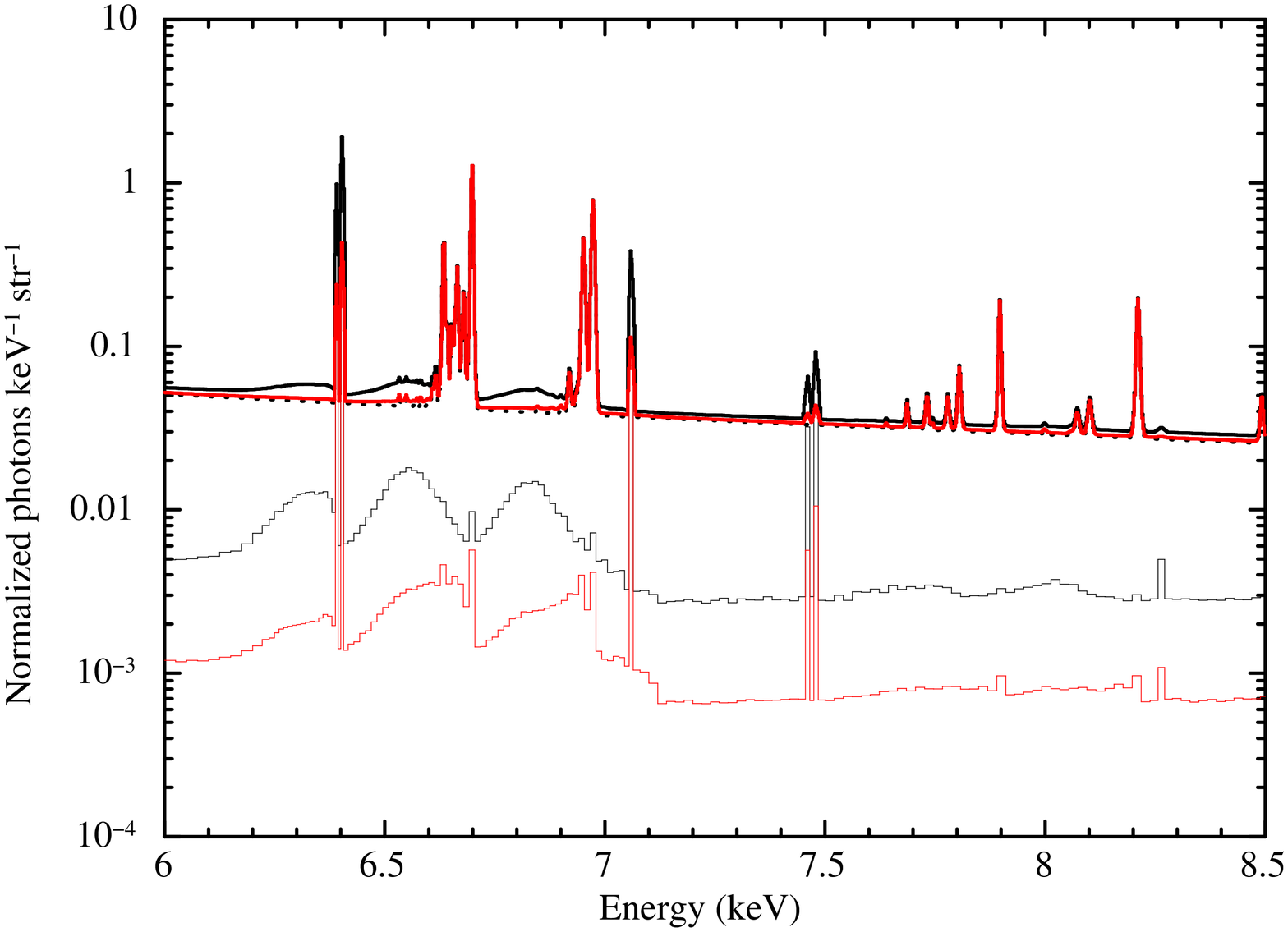}
\end{center}
 \caption{PSAC %Accretion column 
 spectra with the WD mass of 0.7\,$\Mo$, the specific accretion rate of 1\,g\,cm$^{-2}$\,s$^{-1}$
 and $i$ of 0-10 (black) and 80-90\,deg (red).
Solid, thin solid and dotted lines are the total, the reflection and %intrinsic 
the intrinsic thermal plasma components, respectively.
Bottom panel is a blowup in 6--8.5\,keV.}
%The ends are common to figure \ref{fig:density_geocomp}.}
 \label{fig:mwd0p7_psac_spe_i}
\end{figure}

\begin{figure}
\includegraphics[width=85mm]{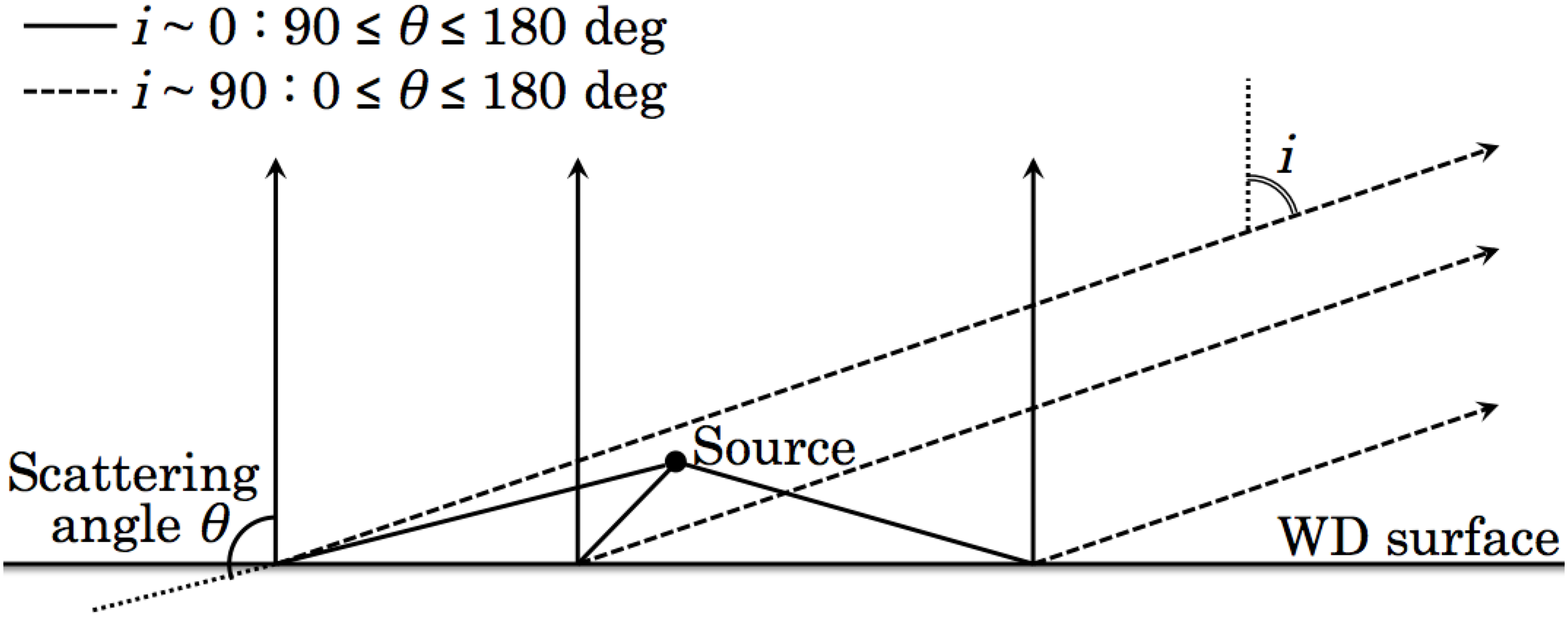}
%\vspace{1mm}
 \caption{Scattering paths of photons %a photon irradiated 
 from a source above %the outside of 
 the WD.
 When the reemitted photon is reflected to $i \sim$ 0\,deg (solid line),
 the scattering angle is between 90 and 180\,deg.
 By contrast, when the reflection angle $i \sim $90\,deg (dashed line),
 the scattering can occur at %with 
 any angle between 0 and 180\,deg.
 \label{fig:compton_shoulder_shape}}
\end{figure}

Figure \ref{fig:mwd0p7_psac_spe_i} shows  spectra of the intrinsic thermal plasma, the reflection and the total of them 
with $i$= 0--10\,deg %(black) 
and 80--90\,deg, %(red), 
$\Mwd$ = 0.7\,$\Mo$ and $a$ = 1\,g\,cm$^{-2}$\,s$^{-1}$. %g$^{-1}$\,cm$^{-2}$\,s$^{-1}$.
The reflection %component 
decreases as $i$ increases and the Compton shoulder shape of 
the fluorescent lines varies
in common with %the 
that %those 
of the point source of the power-law spectrum.
The reflection spectrum has 
%structures %similar to the Compton shoulder of the 6.4\,keV line
%constructed by %photons of 
%the scattering with %of 
%the intrinsic emission %K$\alpha$ 
%lines 
Compton shoulders of He-like and H-like iron K$\alpha$ emission lines at
%of highly ionized iron
%of the emission lines in the intrinsic spectrum 
around a few hundred eV down of their nominal line energies % intrinsic %lines around 
(see bottom of figure\,\ref{fig:mwd0p7_psac_spe_i}).
The scattered line profile is flatter in larger $i$.
%scattered intrinsic lines vary also in shape with relation to $i$.
%The Especially, the peaks of the scattered intrinsic lines shift toward higher energy, 
%that is, come close to the nominal energies of the intrinsic emission lines as the $i$ increases.
%The intrinsic emission irradiates the WD above its 
%surface, %from its outside 
%in contrast to fluorescent emissions 
%which is originated from the WD surface %re-emitted in the WD
%(see figure\,\ref{fig:compton_shoulder_shape}).
%Assuming a source of effectively low $h$ and that the photons in the spectral structure are scattered once, 
If we confine ourselves to consider a low enough PSAC ($h \ll \Rwd$) 
and a single scattering as shown in figure\,\ref{fig:compton_shoulder_shape}, 
around $i$ %of 
$\sim 0$\,deg, the scattering angle is limited in 90--180\,deg.
On the other hand%By contrast
, around $i$ %of 
$\sim 90$\,deg, the scattering angle is permitted to be 0--180\,deg.
%The PSAC source most strongly irradiates the WD surface around its footprint with nearly normal to the WD surface.
%The normal incident photons tend to be scattered at deeper position.
%For smaller $i$, the normal incident photons scattered backward can be easily escaped,
%on the other hand, for lager $i$, they are hardly escaped.
Consequently, %Therefore, 
the scattered line profile becomes flatter and wider %which is determined by the scattering angle 
as $i$ increases.
The fluorescent lines do not change by this effect
because they %the fluorescent lines 
are originated within the WD unlike the intrinsic lines.

\begin{figure}
%\hspace{5mm}
\begin{center}
\includegraphics[width=80mm]{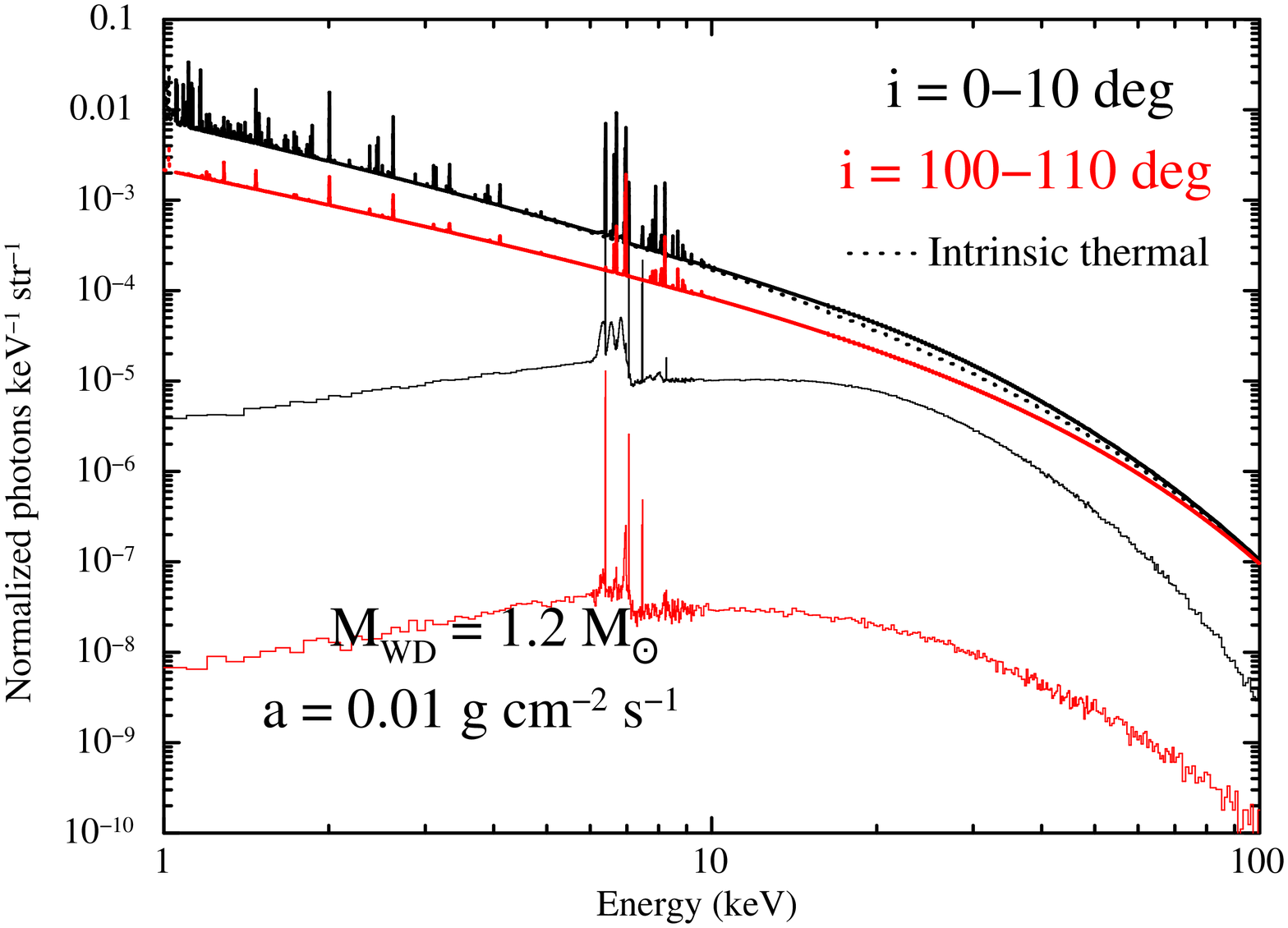}
\includegraphics[width=80mm]{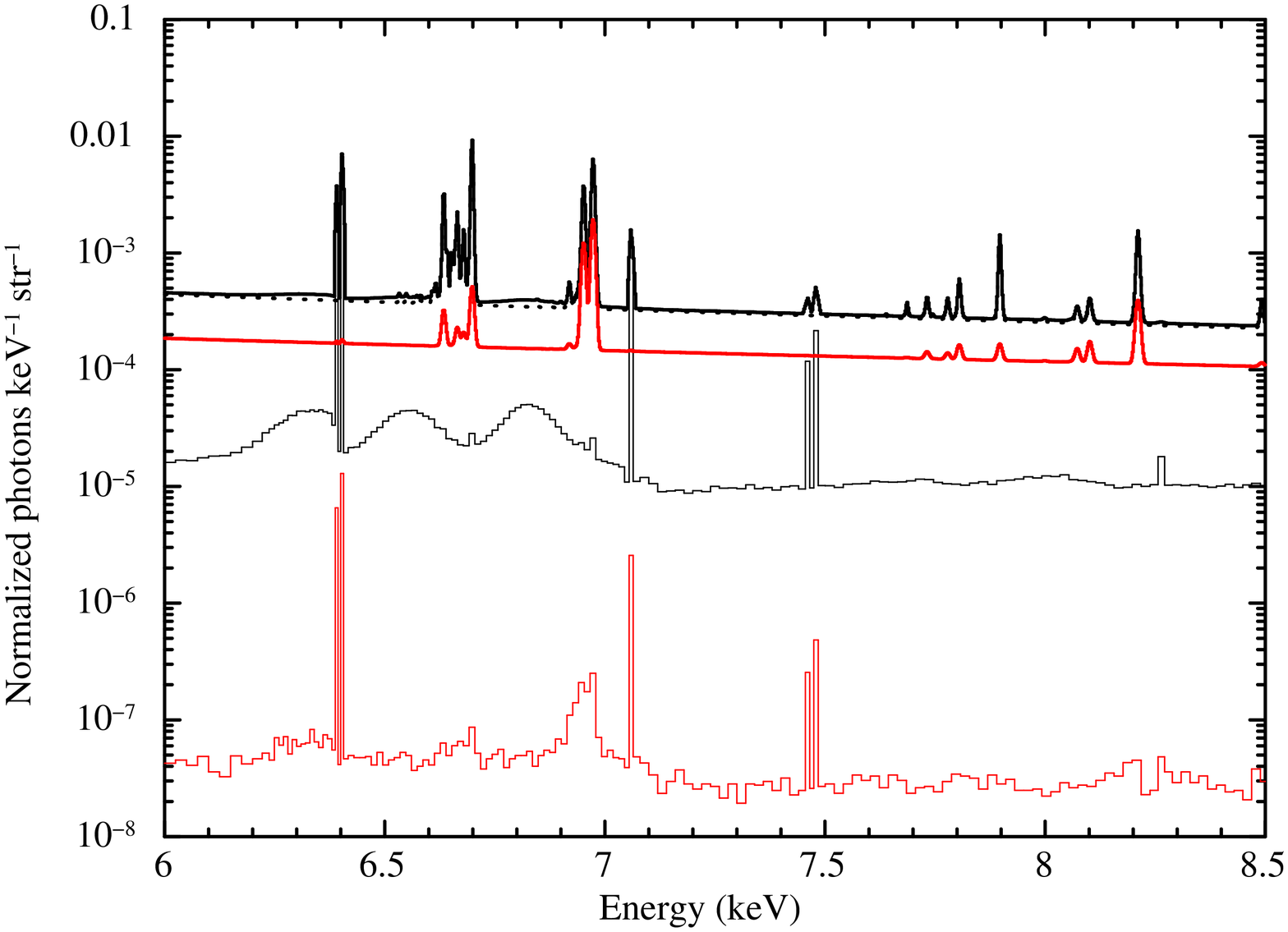}
\end{center}
 \caption{PSAC %Accretion column 
 spectra with the WD mass of 1.2\,$\Mo$, the specific accretion rate of 0.01\,g\,cm$^{-2}$\,s$^{-1}$,
 and $i$ of 0--10 (black) and 100--110\,deg (red).
 Solid, thin solid and dotted lines are the total, the reflection and %intrinsic 
the intrinsic thermal plasma components, respectively.
Bottom panel is a blowup in 6--8.5\,keV.}
%The ends are common to figure \ref{fig:density_geocomp}.}
 \label{fig:mwd1p2_psac_spe_i}
\end{figure}

Although it is weak, the %The 
reflection with the PSAC source may appear 
in the domain $i >$ %above %in $i >$ 
90\,deg as shown by red in figure\,\ref{fig:mwd1p2_psac_spe_i}
(where WD of 1.2\,$\Mo$ 
and $a$ = 0.01\,g\,cm$^{-2}$\,s$^{-1}$ are
%is 
adopted and the resultant
PSAC height is $\sim$ %about 
33\% of the WD radius)
because of the finite height of the PSAC source. % although it is weak.
In this case, only photons %of 
grazing %incidence %incident 
%to 
the WD surface %with nearly parallel %to the WD surface 
and then getting %hence 
scattered forward 
appear. %in the reflection 
The coherent scattering is dominant in this situation (see figure\,\ref{fig:dif_cross_section}). % where the coherent scattering is dominant.
Consequently, %Therefore, 
the energies of the scattered %intrinsic 
lines nearly %almost 
agree with %their %that of the 
those %the nominal energies 
of the intrinsic lines.
Note that the intrinsic thermal spectrum
which can be seen from the observer in the case of $i > 90$\,deg is harder than that of $i < 90$\,deg
because a lower PSAC part consisting of a low-temperature and high-density plasma
%part of the PSAC source 
is occulted by the WD
and only the hotter %emission from the 
higher part is visible. %observed, which is hotter than the lower part.

\subsubsection{Dependence on the WD mass}% Influence of WD mass}
\begin{figure}
%\hspace{5mm}
\begin{center}
\includegraphics[width=80mm]{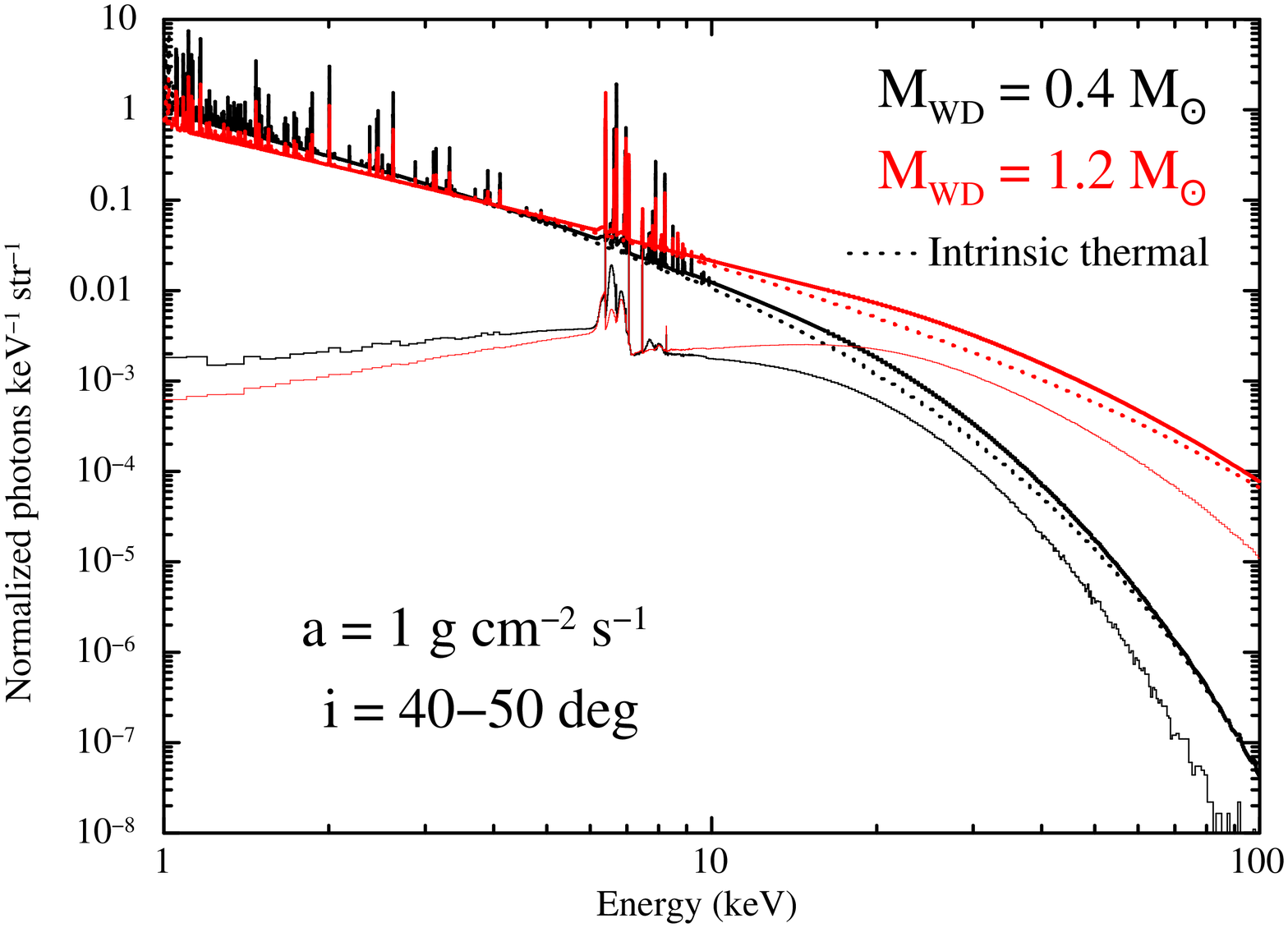}
\includegraphics[width=80mm]{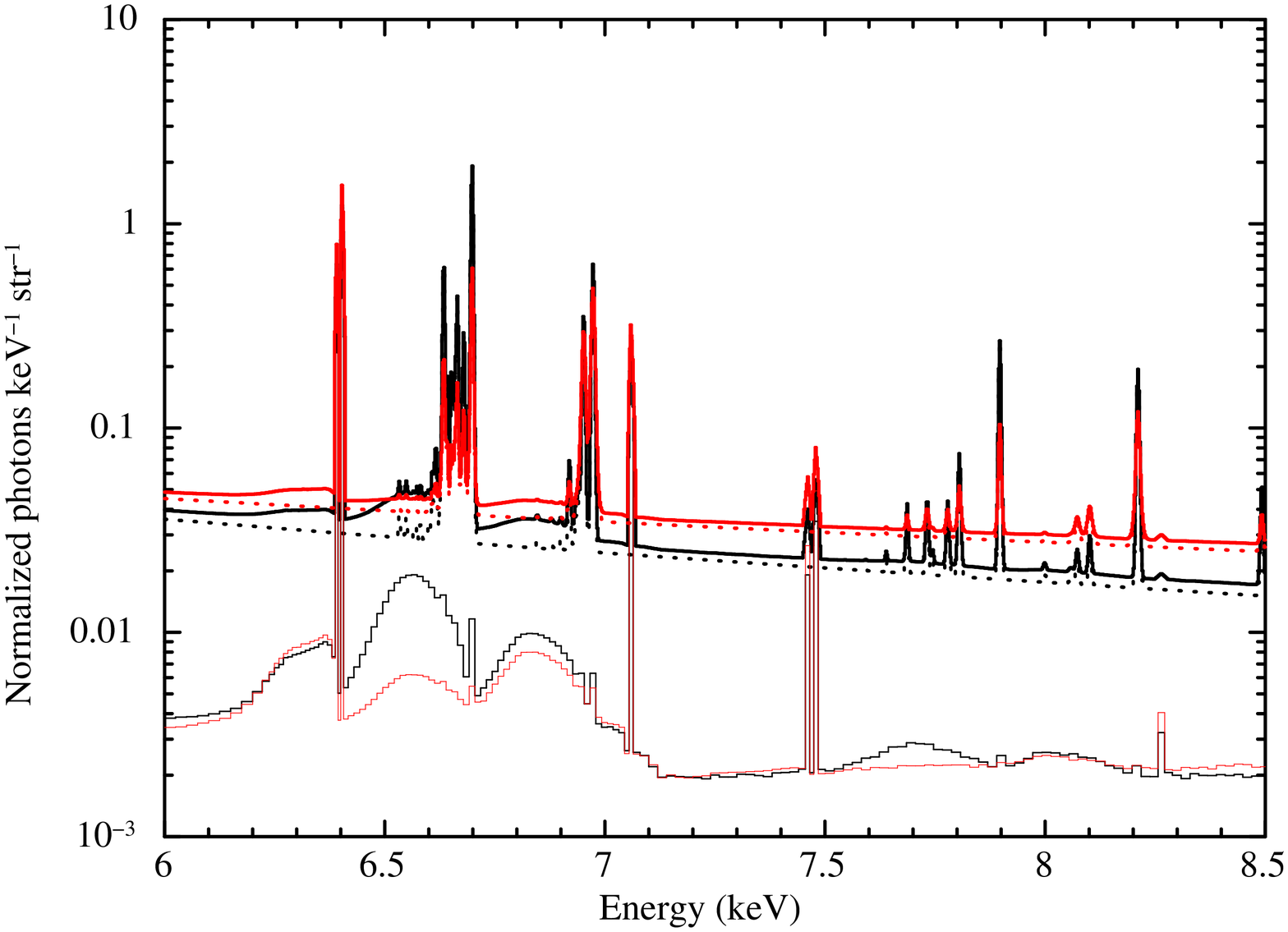}
\end{center}
\caption{PSAC %Accretion column 
spectra with the WD mass of 0.4\,$\Mo$ (black) and 1.2\,$\Mo$ (red),
 the specific accretion rate of 1\,g\,cm$^{-2}$\,s$^{-1}$ and $i$ of 40-50\,deg.
Solid, thin solid and dotted lines are the total, the reflection and %intrinsic 
the intrinsic thermal plasma components, respectively.
Bottom panel is a blowup in 6--8.5\,keV.}
%The ends are common to figure \ref{fig:density_geocomp}.}
\label{fig:mcomp_psac_spe_i}
\end{figure}

Figure\,\ref{fig:mcomp_psac_spe_i} shows the PSAC X-ray spectra 
with the WD masses of 0.4\,$\Mo$ %(black) 
and 1.2\,$\Mo$, %(red), 
the specific accretion rate of $a$ = 1\,g\,cm$^{-2}$\,s$^{-1}$
and the viewing angle of $i$ = 40--50\,deg.
The more massive WD makes a harder intrinsic
thermal %X-ray 
spectrum because of its deeper gravitational potential.
The harder intrinsic spectrum enhances the %EWs of 
the fluorescent lines and the Compton hump %reflection continuum above 10\,keV
as shown in \S\ref{subsec:po_spe}.
By contrast, the PSAC is taller relative to %against 
the WD radius for the more massive WD
because the more massive WD has a smaller %small %less 
radius 
and a taller PSAC due to longer cooling time
%the hotter plasma needs longer cooling time 
(e.g.\,\citealt{2014MNRAS.438.2267H}).
The taller PSAC preferentially reduces the contribution of its higher part to the reflection.
Consequently, %Therefore, 
the WD mass intricately influences the reflection.
For example, the EW of the fluorescent iron K$\alpha_{1,2}$ lines is not a 
monotonically increasing function of the WD mass
in a certain domain of %according to 
the specific accretion rate as described later (\S\ref{subsec:psac_ew}).

\subsubsection{Dependence on the %Influence of 
specific accretion rate}
\begin{figure}
%\hspace{5mm}
\begin{center}
\includegraphics[width=80mm]{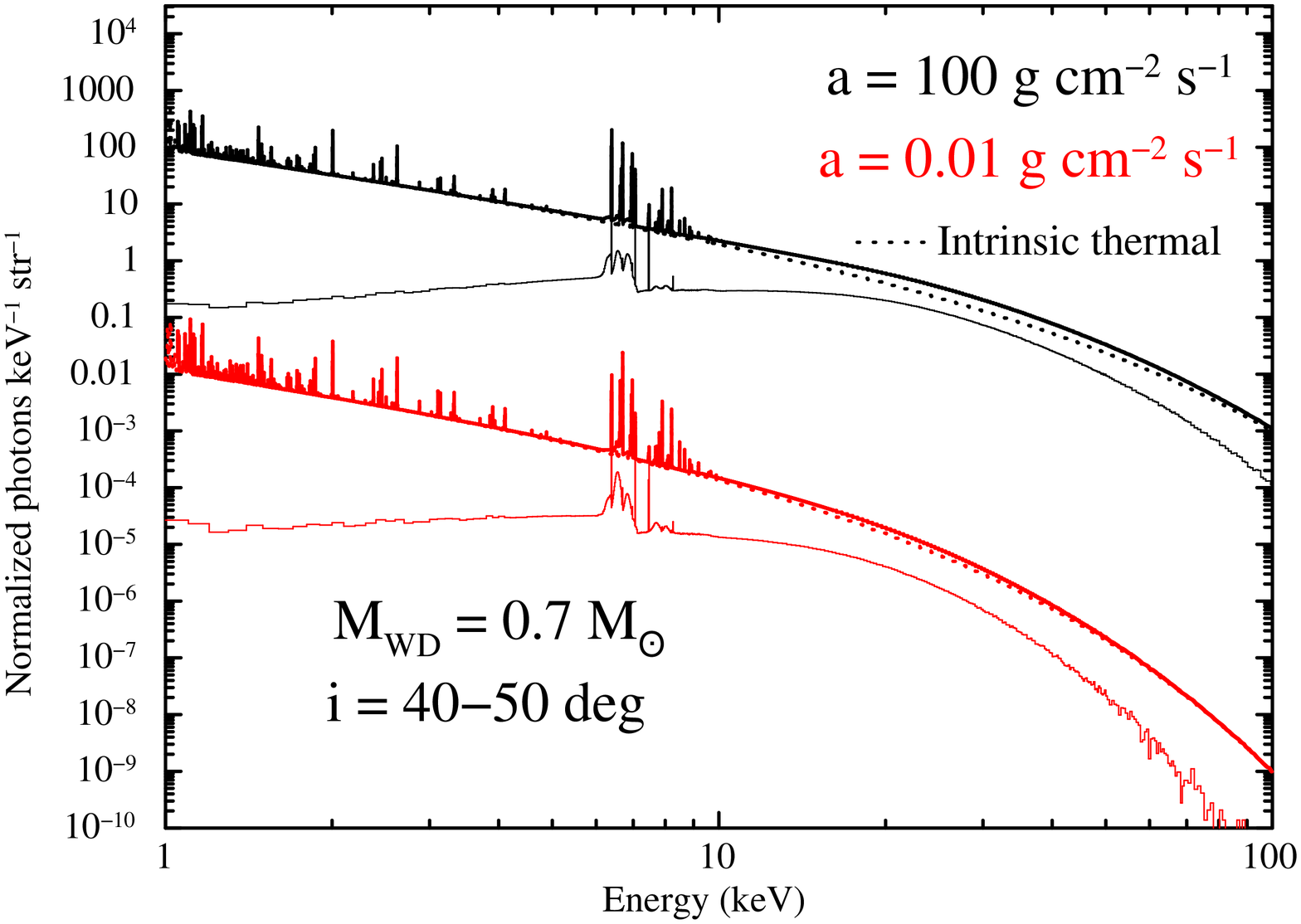}
\includegraphics[width=80mm]{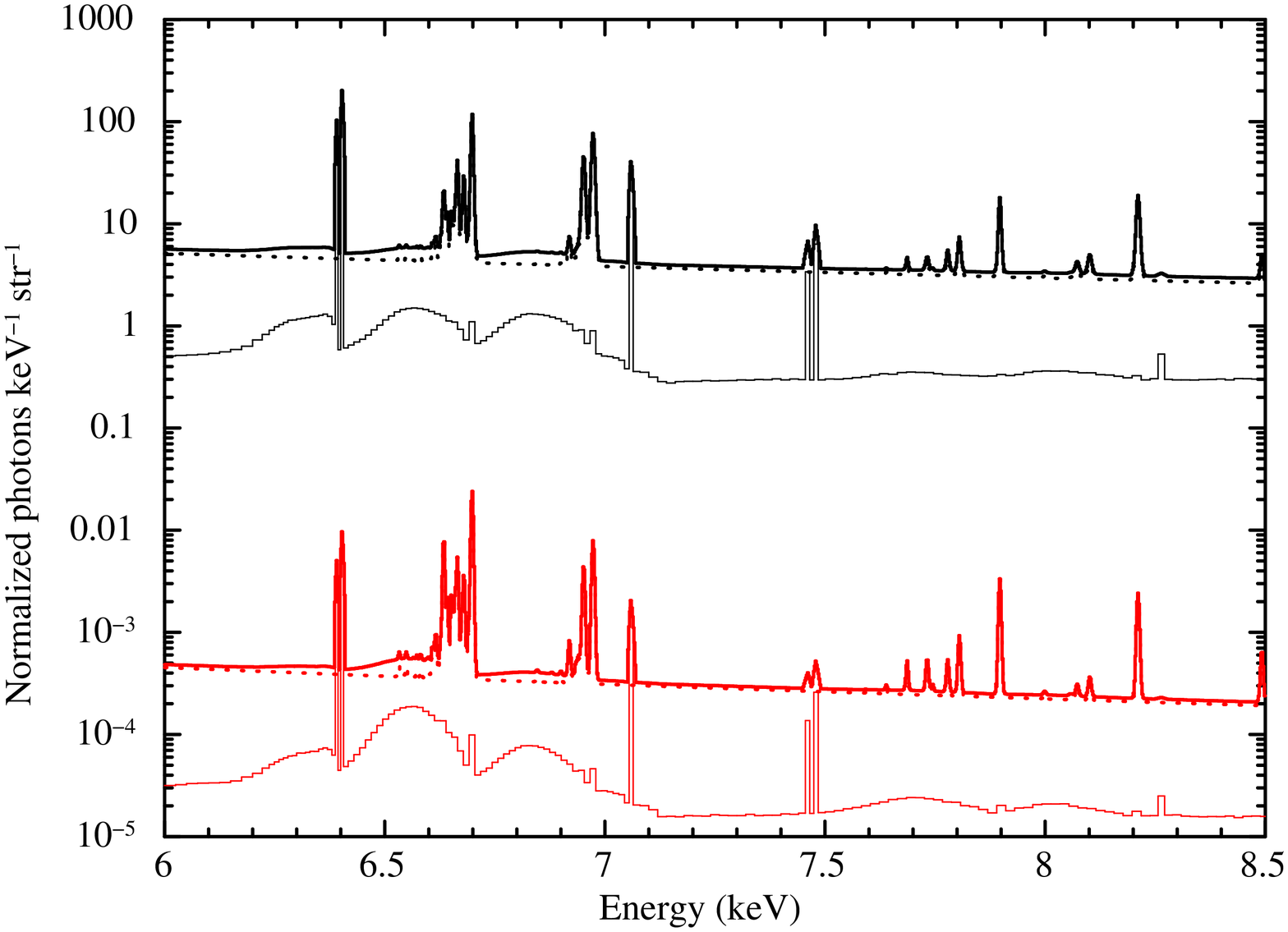}
\end{center}
\caption{PSAC %Accretion column 
spectra with the WD mass of 0.7\,$\Mo$, the specific accretion rate of 100 \,g\,cm$^{-2}$\,s (black) and 0.01 \,g\,cm$^{-2}$\,s$^{-1}$ (red) and $i$ of 40--50\,deg.
 Solid, thin solid and dotted lines are the total, the reflection and %intrinsic 
the intrinsic thermal plasma components, respectively.
Bottom panel is a blowup in 6--8.5\,keV.}
%The ends are common to figure \ref{fig:density_geocomp}.}
 \label{fig:acomp_psac_spe_a}
\end{figure}

The specific accretion rate also influences the intrinsic thermal and the reflection spectra.
Figure\,\ref{fig:acomp_psac_spe_a} shows the X-ray spectra 
with the specific accretion rates of $a$ = 100\,g\,cm$^{-2}$\,s$^{-1}$ %(black) 
and 0.01\,g\,cm$^{-2}$\,s$^{-1}$, % (red),
the WD mass of 0.7\,$\Mo$ and the viewing angle of $i$ = 40--50\,deg.
The smaller %less 
specific accretion rate increases %extends 
the cooling time and lengthens the PSAC.
With the taller PSAC, released gravitational energy %potential 
at the shock is less
which makes the PSAC cooler and the intrinsic
thermal spectrum softer.
The taller and cooler %softer 
PSAC reduces the reflection.
Consequently, %Therefore, 
the influence of the specific accretion rate on the reflection is simpler than that of the WD mass.
The EW of the fluorescent iron K$\alpha_{1,2}$ lines monotonically increases irrespective of the WD mass 
as the specific accretion rate increases  (\S\ref{subsec:psac_ew}).

\subsection{Sum of the EWs of fluorescent ion K$\alpha_{1,2}$ lines}\label{subsec:psac_ew}
\begin{figure}
%\hspace{5mm}
\begin{center}
\includegraphics[width=80mm]{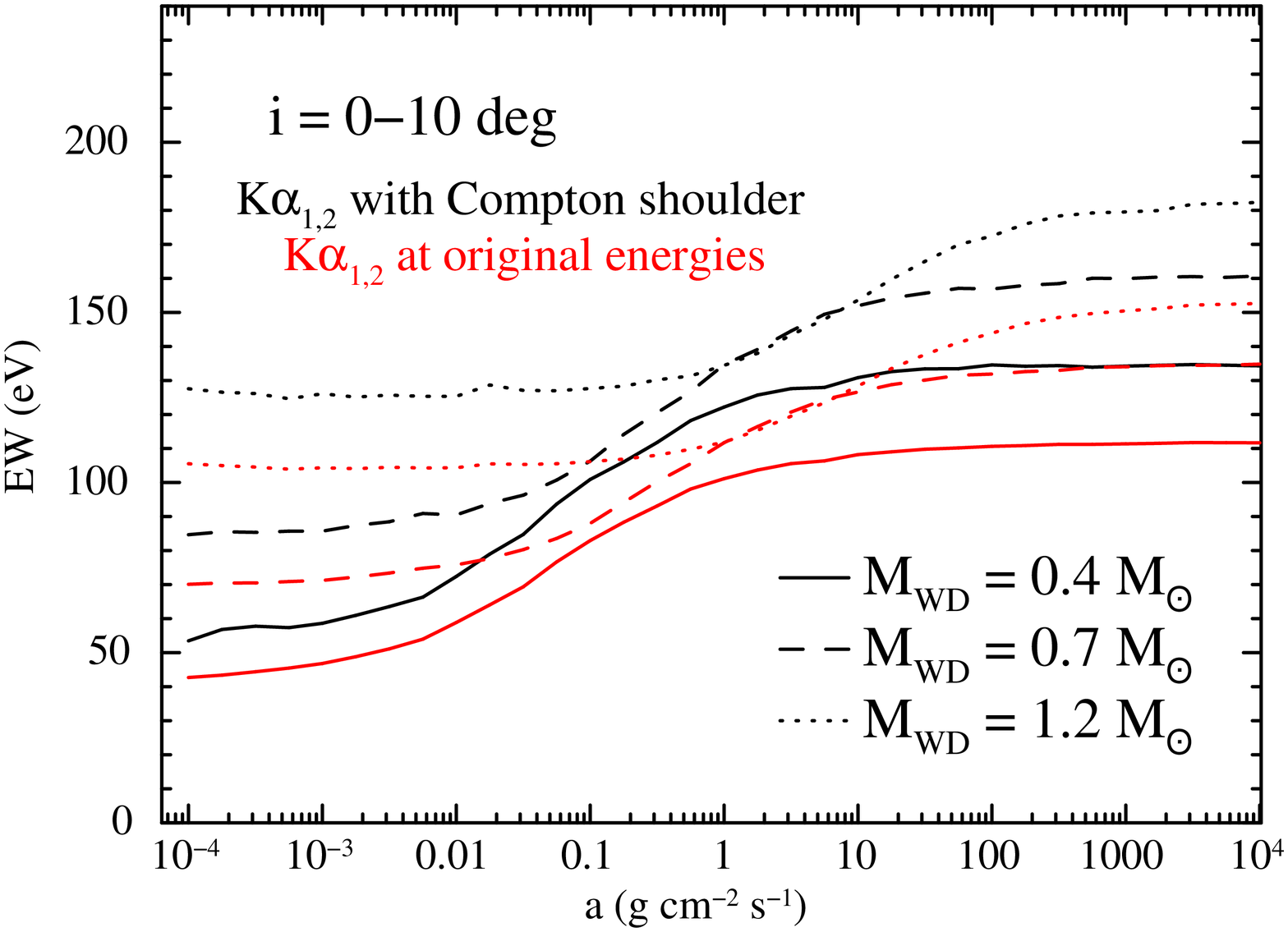}
\includegraphics[width=80mm]{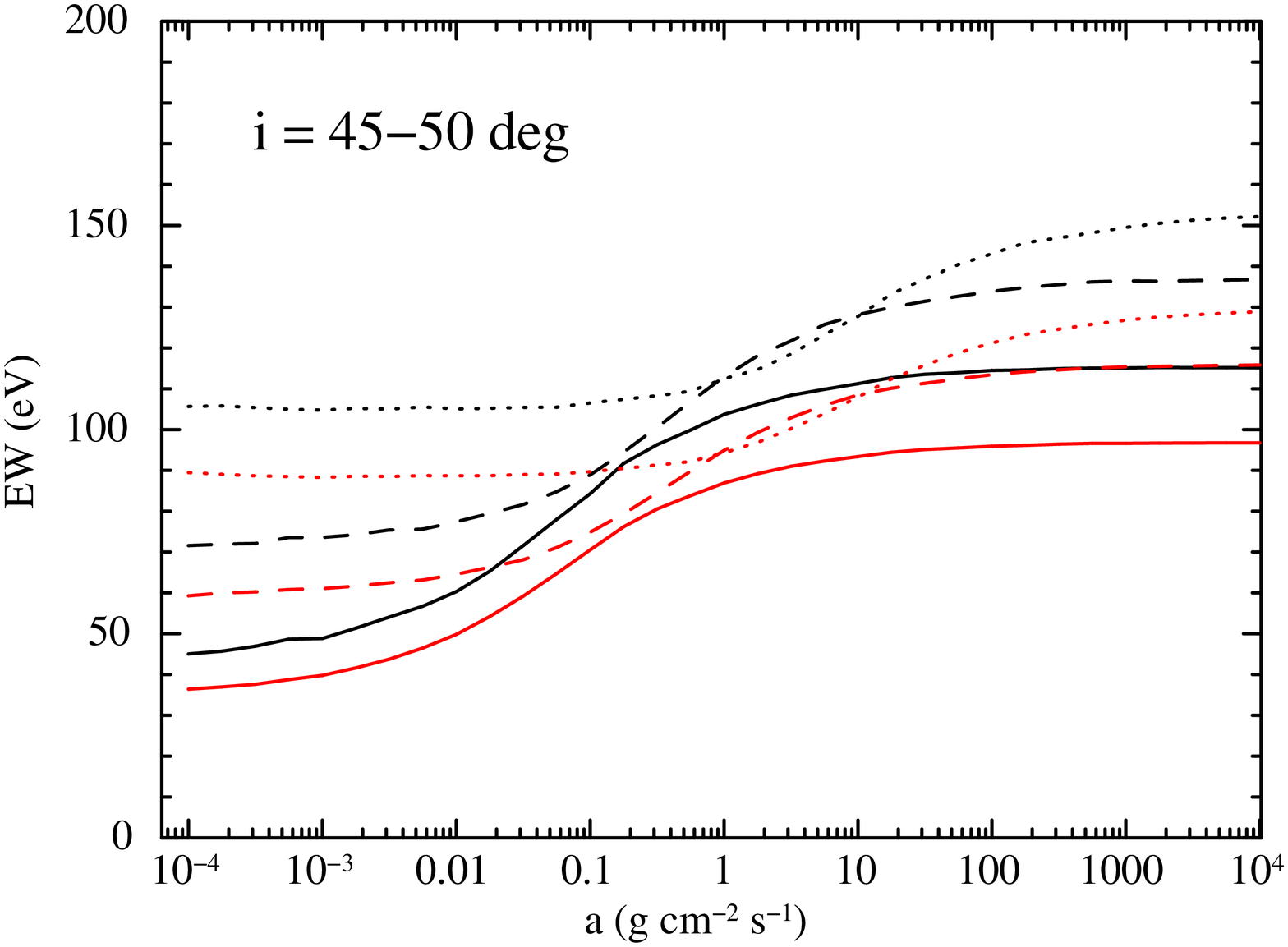}
\includegraphics[width=80mm]{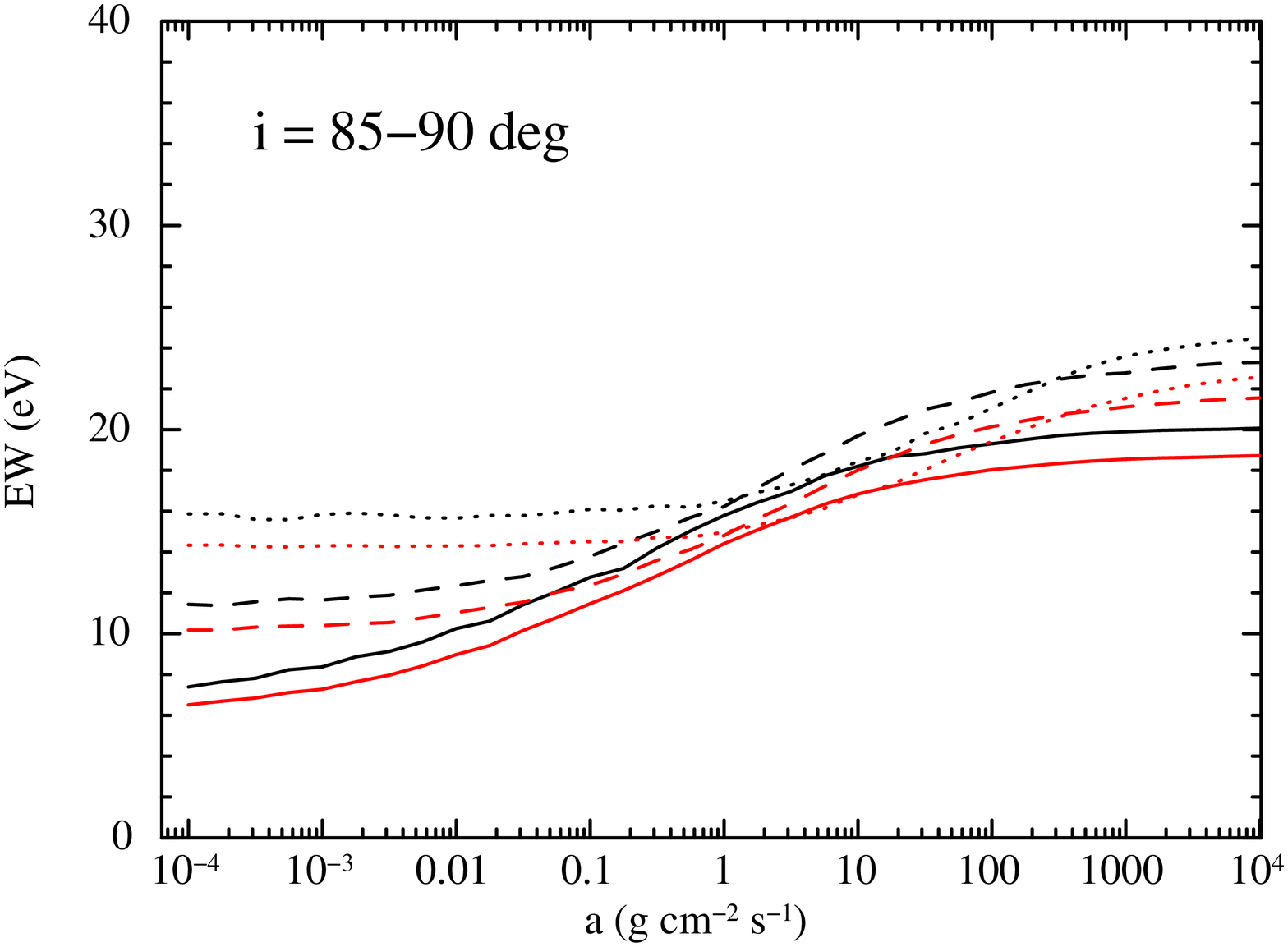}
\end{center}
 \caption{Relations between the specific accretion rate and the sum of the EWs of the fluorescent iron K$\alpha_{1,2}$ lines (red lines),
 and the total EW of the fluorescent iron K$\alpha_{1,2}$ lines and their Compton shoulders (black lines) 
 with the WD masses of 0.4, 0.7 and 1.2\,$\Mo$ (solid, dashed and dotted lines, respectively).
 $i$ of 0--10, 45--50 and 85--90\,deg are assumed in the top, middle and bottom panels, respectively.
 }
%The ends are common to figure \ref{fig:density_geocomp}.}
 \label{fig:a-ew}
\end{figure}

%The EWs of the fluorescent lines in the reflection are effected by the WD mass and the specific accretion rate.
Figure\,\ref{fig:a-ew} shows the EWs of the sum of the fluorescent iron K$\alpha_{1,2}$ lines,
and those %that 
with %the total of them and 
their Compton shoulders as a function of %against 
the specific accretion rate.
In this figure, the cases of the WD masses of $\Mwd$ = 0.4, %\,$\Mo$ %(solid), 
0.7 %\,$\Mo$ (dashed) 
and 1.2\,$\Mo$ %(dotted)
and the viewing angles of $i$ = 0--10, %\,deg (top), 
45--50\,deg %(middle) 
and 85--90\,deg %(bottom) 
are shown. %assumed.

The EWs monotonically increase with %as 
the specific accretion rate for %increases with 
any WD mass % and viewing angle
because the PSAC shortens and the its temperature rises (see \citealt{2014MNRAS.438.2267H}).
However, they become
%On the other hand, the EWs are 
almost constant 
at the highest end %ends 
of the specific accretion rate in %of 
this figure.
This is because,
in the larger specific accretion rate, %region,
the distributions of physical quantities normalized by the PSAC height does not vary,
and its height is negligible against the WD radius. % and the variation of its height is trivial.
Moreover, in the smaller specific accretion rate, %region,
the %peak and the 
maximum of the electron temperature in the PSAC 
is hardly affected %effected 
by the specific accretion rate.

The influence of the WD mass on the reflection is more complex
because a more massive WD raises the PSAC temperature and extends the PSAC, 
which %generally 
enhances and reduces the reflection, respectively.
%as referred in the previous subsection.
In fact, the EWs in %the 
figure\,\ref{fig:a-ew}
rise and fall as the WD mass increases in the specific accretion rate between 0.01--100\,g\,cm$^{-2}$\,s$^{-1}$.
The relations between the EWs and the WD masses also depend on the viewing angle.
Irrespective of the specific accretion rate and the WD mass, comparison of the three panels in figure 21 indicates that the EWs decrease with increasing viewing angle.

\subsection{Compton hump}

\begin{figure}
%\hspace{5mm}
\begin{center}
\includegraphics[width=80mm]{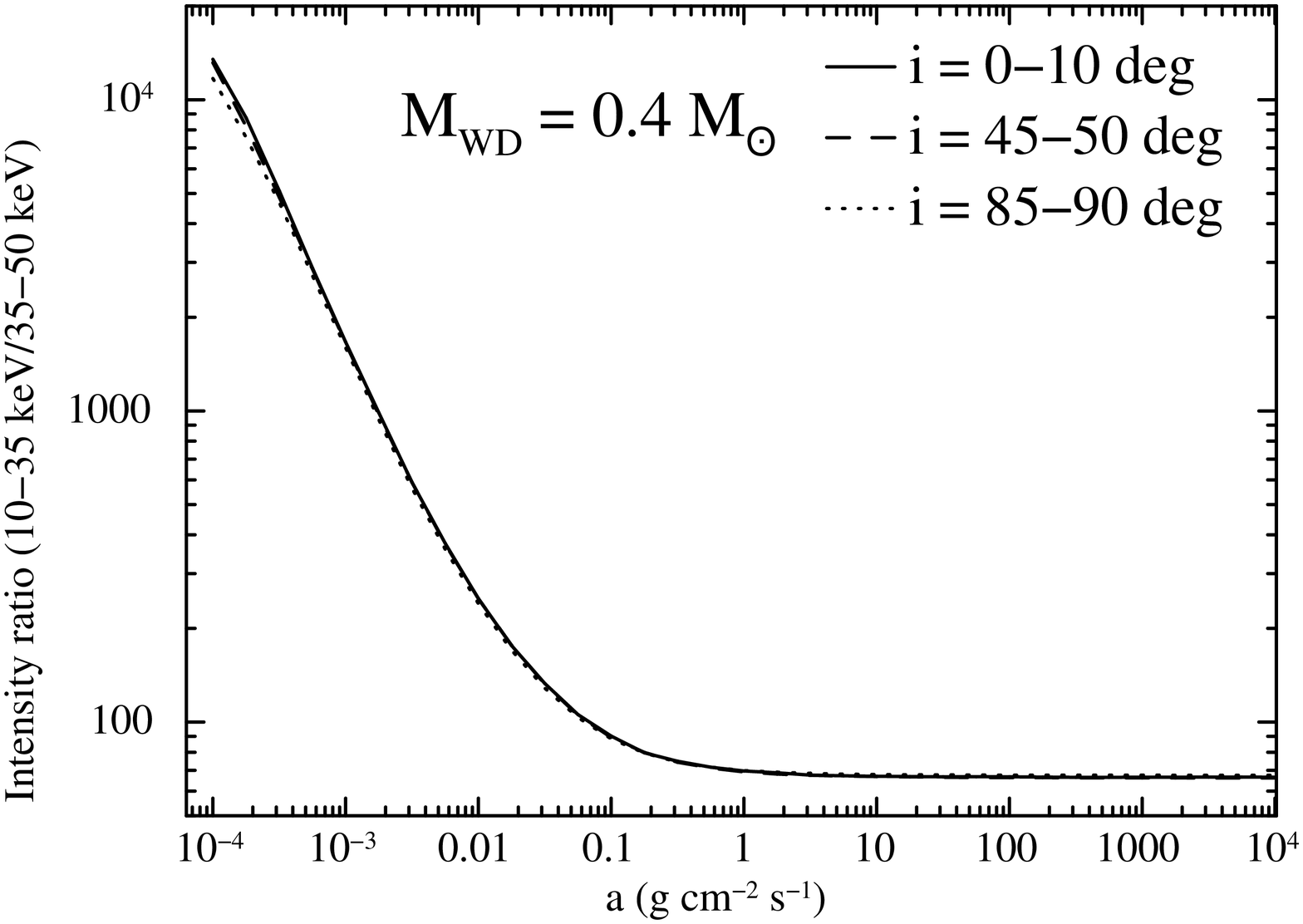}
\includegraphics[width=80mm]{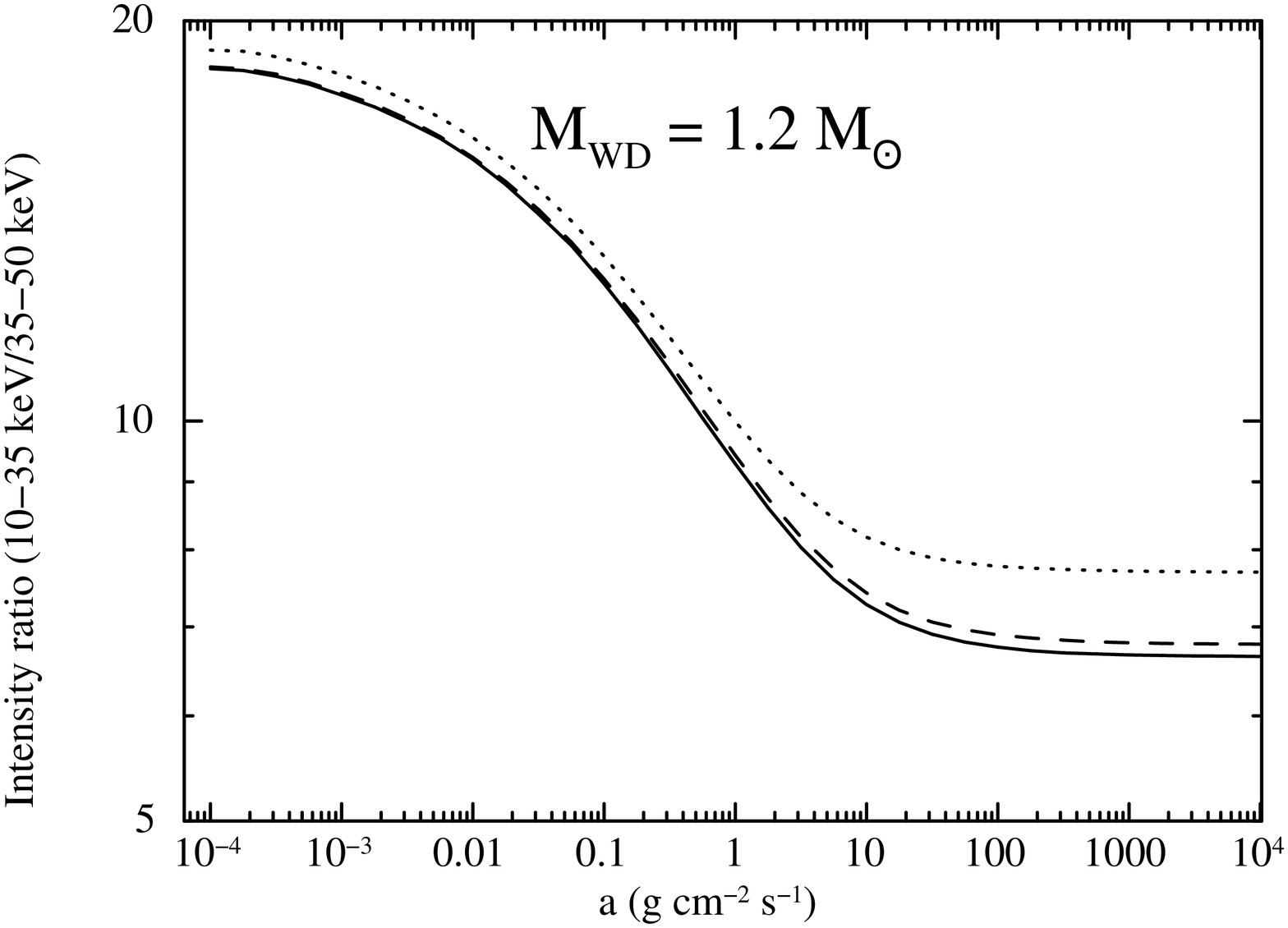}
\end{center}
 \caption{Relations between the specific accretion rate and the ratio of intensity in 10--35\,keV to that in 35--50\,keV
 with the WD masses of 0.4 %, 0.7 
 and 1.2\,$\Mo$ (top %, middle 
 and bottom panels, respectively) and 
 $i$ of 0--10\,(solid), 45--50\,(dashed) and 85--90\,(dotted)\,deg.
%{\Huge The inset of the upper panel is a blowup of the region of $a=0.1$--10\,g\,cm$^{-2}$\,s$^{-1}$.}
 } % are adopted in the top, middle and bottom panels, respectively.}
%The ends are common to figure \ref{fig:density_geocomp}.}
 \label{fig:a-hump}
\end{figure}

Since t%T
he Compton hump is the other important ingredient of the reflection spectrum, it also depends on the WD mass and the specific accretion rate %as the fluorescent lines
although its variation is somewhat minor %is 
compared to %than 
that of the intrinsic thermal spectrum.
Figure\,\ref{fig:a-hump} shows the ratio of the continuum intensity in 10--35\,keV to that in 35--50\,keV 
with the WD masses of $\Mwd$ = 0.4\,$\Mo$ %(top) %, 0.7\,$\Mo$ (middle) 
and 1.2\,$\Mo$, %(bottom)
and the viewing angles of $i$ = 0--10, %\,deg (solid), 
45--50 %\,deg (dashed)
and 85--90\,deg. % (dotted).
This figure indicates that the influence of the viewing angle on the intensity ratio is limited.
Especially, the intensity ratio of less massive WD (e.g. $\Mwd$ = 0.4\,$\Mo$) hardly depends on the viewing angle.
This fact means that the
shape of the reflection spectrum (the softness ratio) depends mainly on the intrinsic thermal spectrum (uniquely determined by the specific accretion rate at a given WD mass). %, irrespective of the viewing angle.
%relation between the specific accretion rate and the intensity ratio 
%is determined mainly by the variation of the intrinsic thermal spectrum. % regarding. %to the specific accretion rate.
On the other hand, in the case of more massive WD (e.g. $\Mwd$ = 1.2\,$\Mo$), 
the intensity ratio varies with %relating to 
the viewing angle and 
%the reflection is not trivial.
the reflection spectrum is more sensitive to the viewing angle. % than that of the $\Mwd = 0.4\Mo$ case.

\subsection{Spin profile}

In general, a %the 
reflection X-ray spectrum of %in 
the mCVs varies in association with the WD spin phase
because the PSAC is not coaligned %consistent 
with the WD spin axis. % in most cases.
%The occultation of the PSAC by the WD limb can vary the %intrinsic 
%thermal plasma component with the WD spin.
It has been well known from observations that
%We note that the X-ray spin modulation can be caused by 
the photoelectric absorption by the pre-shock accreting matter
can bring about spin modulation of the X-ray intensity.
However, here, we concentrate on %describe 
the spin modulations of the reflection and 
the intrinsic thermal plasma component caused by the occultation. % and the reflection.  %and %that of 

\begin{figure}
%\hspace{5mm}
\begin{center}
\includegraphics[width=80mm]{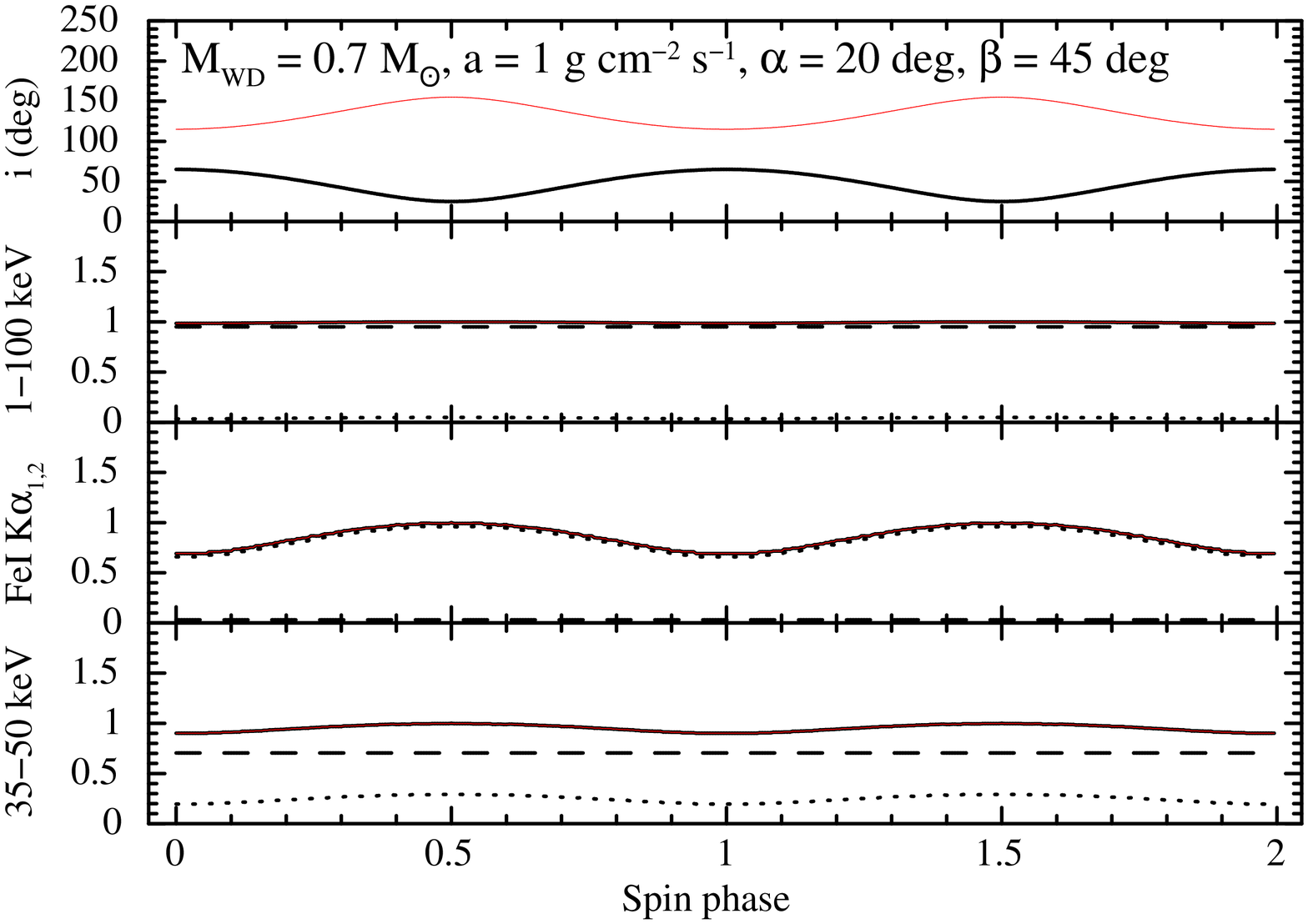}
\includegraphics[width=80mm]{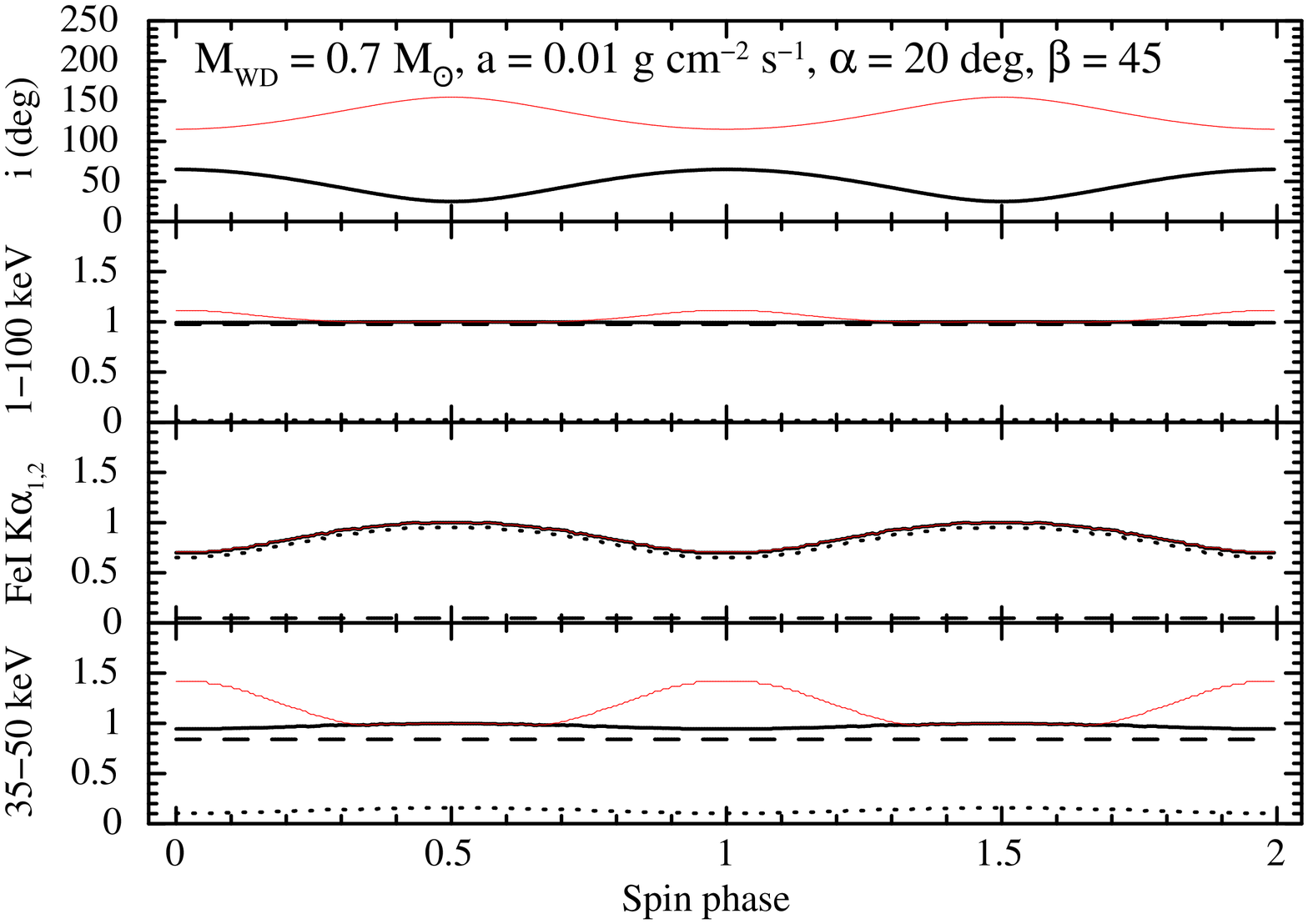}
\includegraphics[width=80mm]{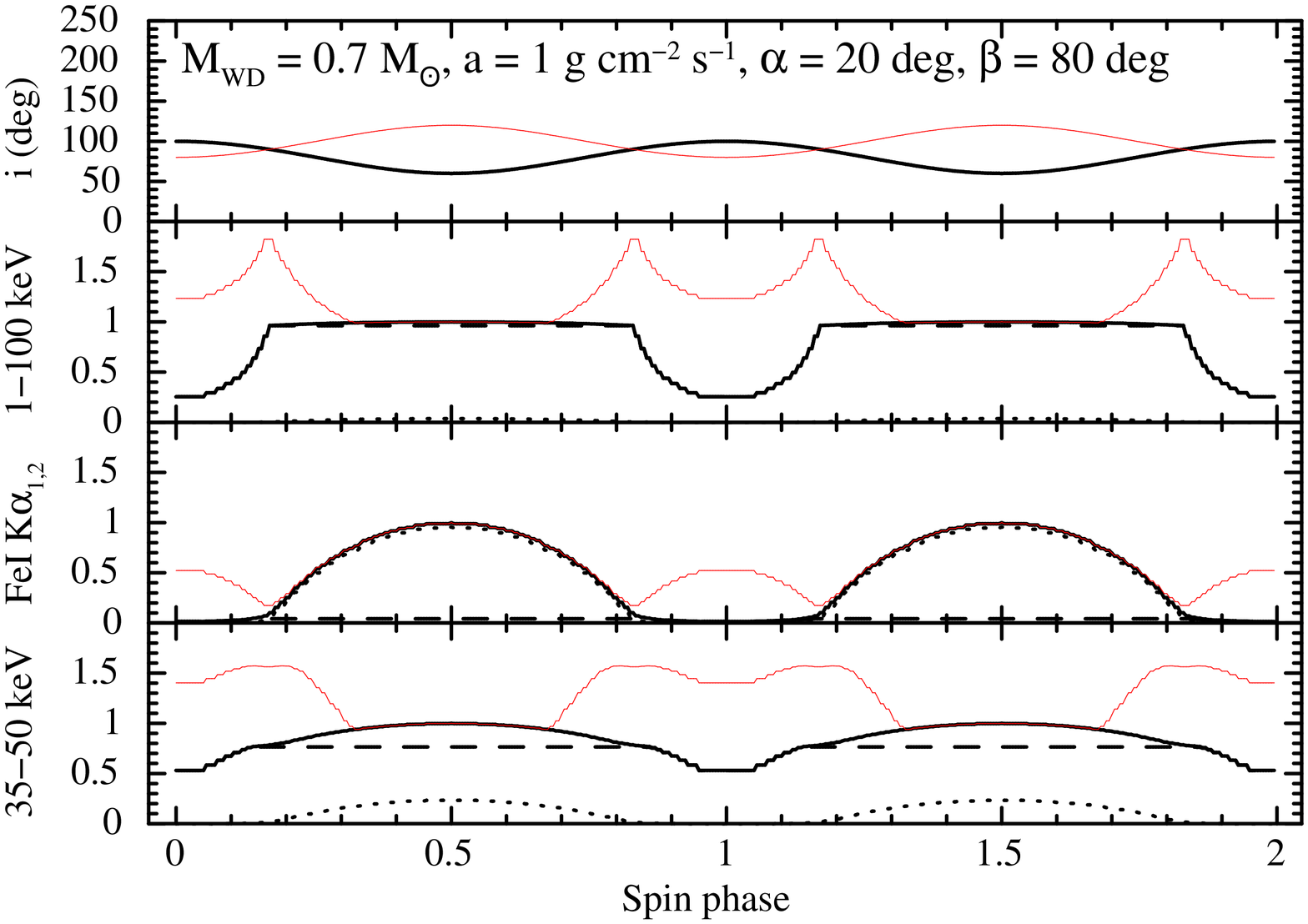}
\end{center}
 \caption{Spin profiles with the WD mass of 0.7\,$\Mo$
 and the angle of the spin axis from the line of sight ($\alpha$) of 20\,deg.
Sets of the specific accretion rate (g\,cm$^{-2}$\,s\,$^{-1}$) %, $\alpha$ (deg) 
and the angle of PSAC from the line of sight $\beta$ (deg) are 
assumed to be 1 %, 20 
and 45 in the top panel,
 0.01 %, 20 
 and 45 in the middle panel and 1 %, 20 
 and 85 in the bottom panel.
Top boxes in the individual panels show $i$ of the upper pole (black) and the possible lower pole (red).
Bottom three boxes in the individual panels are spin profiles 
of the intensities %corresponding to the $i$ in the top boxes
in the 1--100 keV band, of the 6.4 line
and in the 35--50\,keV band toward bottom.
Black dotted, dashed and solid lines are the intensities of the
reflection, the intrinsic thermal components and sum of them,
which are normalized by the maximum of the sum.
Red line is the total of the sum of the upper and the possible lower poles and 
normalized by the common value with black lines (the total of upper pole).
%its maximum. % of that total.
%{\Huge Deg}
}
%The ends are common to figure \ref{fig:density_geocomp}.}
 \label{fig:spin}
\end{figure}

The spin modulation of a single PSAC is determined by
angles of the spin axis from %between the spin axis and 
the line of sight ($\alpha$) and from %between the spin axis and 
the PSAC ($\beta$)
as well as the WD mass, the specific accretion rate and the abundance.
The spin modulation also depends on whether the two PSACs can be seen.
Figure\,\ref{fig:spin} shows the spin modulations in cases of %assuming 
$\Mwd = 0.7\,\Mo$ and $\alpha$ = 20\,deg
as common %constant 
parameters, and 
specific accretion of $a$ = 1.0 %(top and bottom panels) 
and 0.01%(middle panels), 
\,g\,cm$^{-2}$\,s$^{-1}$,
and $\beta$ = 45 and
80%(bottom panel)
\,deg.
Here, we assume that the upper and possible lower PSACs are 
separated by 180$^\circ$ on %symmetric about 
the WD surface. %center.

In the case of %the 
$a$ = 1.0\,g\,cm$^{-2}$\,s$^{-1}$ and $\beta = 45$\,deg, 
the upper PSAC can always be seen and the possible lower PSAC is invisible in %the 
any phase.
Consequently, %Therefore, 
the spin modulation is brought about
only by the reflection %component 
whose intensity
varies sinusoidally with %with sinusoidal curve and}
the pulse fraction %amplitude 
of %is 
about 30\% and 10\% in the fluorescent iron K$\alpha_{1,2}$ lines and 35--50\,keV band, respectively.
The intensity in 1--100\,keV is almost constant because it %this intensity 
is almost determined by the intrinsic thermal component.

%In the middle panel, 
When %the specific accretion rate is assumed to be 
$a$ = 0.01\,g\,cm$^{-2}$\,s$^{-1}$
and the other parameters are common with the previous case, %top panel.
%In this case, 
the PSAC becomes taller and the top of the possible lower PSAC is visible except for around phase of 0.5.
Since the %The 
top of the PSAC is hot ($>$ 10\,keV), % and therefore, 
the thermal component in the 35--50\,keV significantly rises
around the phase of 0. Consequently, %and 
the spin modulation is not sinusoidal.
% when the lower 
%PSAC emerges from the limb of the WD. %exists.
The pulse fraction %amplitude 
in the 35-50\,keV is about 10\% without the lower PSAC and 30\% with the lower PSAC.
Note, however, that the intensity of the fluorescent iron K$\alpha_{1,2}$ lines are not influenced by the lower PSAC because its base where the lines are emitted is mostly behind the limb of the WD.

%In the case of the bottom panel, 
When %the angle between the spin axis and the PSAC is 
$\beta$ = 80\,deg
and the other parameters are common with the first case, %top panel.
%the 
both %of 
the upper PSAC and the possible lower PSAC are seen and occulted in turn.
The spin profiles are complex and significantly change %s %varies 
with the X-ray %in 
energy.
The spin profile of the fluorescent iron K$\alpha_{1,2}$ lines has a
single-peak profile with 100\% amplitude
when the lower PSAC does not exist.
When the lower PSAC exists, its profile is a %has 
double peak. % whose amplitude is 80\%.
In the 35--50\.keV band, the single-peak with 50\% amplitude profile appears when the lower PSAC is not present. %considered.
When the lower PSAC is took into account, the spin profile has two flat regions %phases 
centered at the %around 
phases of 0 and 0.5
whose intensities differ by about 30\%.
The spin profile of %in 
the 1--100\,keV intensity has a dip 
around the phase %of 
0 with 70\% amplitude if we consider %with 
only the upper PSAC.
With the upper and the lower PSACs, the spin profile has two peaks
which correspond to the phases of switching the two PSACs.
%The spin profile in the 1--100\,keV anti-correlates with the spin profile of the fluorescent iron K$\alpha_{1,2}$ lines
%regardless of consideration of the lower PSAC.

\section{Discussion}\label{sec:discussion}

\subsection{Measurement of system parameters}
G%The g
eometrical parameters such as the viewing angle and the height of the intrinsic X-ray source
can be measured with our reflection model.
Since the observed spectrum is the sum of the intrinsic PSAC spectrum and its reflection,
%Moreover, 
the reflection model is of essential importance to evaluate the shape of the intrinsic PSAC spectrum, which enhances the measurement accuracy of %for 
the parameters of the thermal component
such as the WD mass and the specific accretion rate
 in cooperation with the intrinsic thermal model of %s for 
 the PSAC (e.g. \citealt{2014MNRAS.438.2267H}).

The EW of the fluorescent iron K$\alpha$ line is significantly different among the IPs.
%For instance, the EWs of EX Hya and V1223 Sgr 
%were measured to be 24\,eV and 98\,eV, respectively \citep{2014MNRAS.441.3718H}.
For instance, the EW of EX Hya and TV Col
were measured to be 24\,eV \citep{2014MNRAS.441.3718H} and 170\,eV \citep{1999ApJS..120..277E}, respectively.
The small EW such as that of EX Hya suggests that 
the system has a light WD, low specific accretion rate, low metal abundance and a large viewing angle (see figure\,\ref{fig:a-ew}).
The WD mass  and the specific accretion rate of EX Hya were measured in some works, for example : 
$\Mwd = 0.790\pm0.026\,\Mo$ \citep{2008A&A...480..199B}, 
$\Mwd = 0.4\pm0.02\,\Mo$ with $a$ fixed at 1\,g\,cm$^{-2}$\,s$^{-1}$ \citep{2010A&A...520A..25Y} and
$\Mwd = 0.63^{+0.17}_{-0.14}\,\Mo$ and 
$a$ = 0.049$^{+0.66}_{-0.04}$\,g\,cm$^{-2}$\,s$^{-1}$
%$a$ = 0.049$^{+0.66}_{-0.035}$\,g\,cm$^{-2}$\,s$^{-1}$ 
\citep{2014MNRAS.441.3718H}.
All of them %ese estimations 
need a large %larger 
viewing angle $i > 60$\,deg to explain the small EW 
with the measured iron abundance of 0.57$\pm0.03$ \citep{2014MNRAS.441.3718H}.
% (see figure\,\ref{fig:a-ew})
%even 
If the %reemission from absorbers around the system do 
line emission from the pre-shock absorption matter %dose not 
contributes to the EW, an even larger viewing angle is required.
By contrast, %with 
the large EW 170\,eV %such as that 
of TV Co implies that
%l,
%if it is assumed that the absorbers do not contribute,
the WD %mass 
should be massive as long as the entire fluorescent K$\alpha_{1,2}$ lines originate from the WD surface.
However, there is significant contribution 
from the pre-shock cold accreting matter
%of the absorbers 
to the EW \citep{1999ApJS..120..277E},
which makes difficult to put a %weakens the 
constraint from the EW
 on the WD mass. %reflecting angle.
%has to be taken into consideration.
However, since the central energies of the K$\alpha$ lines from the pre-shock accreting matter shift from the values in laboratories according to the radial flow velocity of the flow, they will be resolved from those from the WD surface with a high resolution detector, such as a micro-calorimeter, in the near future.

The Compton hump is another important spectral feature to constrain some system parameters such as the viewing angle and the height of the PSAC, like the fluorescent iron lines, yet it is difficult to resolve it from the observed continuum spectrum because its spectrum spreads widely in energy and somewhat similar to the intrinsic PSAC spectrum.
%can hardly disentangle the influences of the parameters by itself.
Recent observations of some IPs with the {\it NuSTAR} satellite undoubtedly 
verified existence of %need 
the Compton hump \citep{2015ApJ...807L..30M}.
However, it is difficult to segregate
%that 
the Compton hump %is distinguished 
from the intrinsic PSAC thermal 
spectrum with enough accuracy to evaluate the system parameters.
%unlike the emission lines,
%which complicates the estimation of the %intensity and the shape of the 
%Compton hump.
T%On the other hand, t
he Compton hump is useful %should contribute 
to constrain the parameters 
only if its accurate spectral model like ours presented in this paper is applied to the observed spectrum
in combination with the fluorescent iron K$\alpha$ line. % and the thermal emission
%such as fitting by the models for them.
Application of the reflection model with the intrinsic PSAC
thermal model \citep{2014MNRAS.438.2267H}
to observational data will be presented in a subsequent paper.

\subsection{Interpretation of the red wing of the fluorescent iron K$\alpha$ line}
The iron fluorescent K$\alpha$ line has a red wing in some IPs.
For GK Per in outburst, the red wing extending to 6.33\,keV was 
observed by High-Energy Transmission Grating (HETG) 
onboard the %the on 
{\it Chandra} satellite \citep{2004MNRAS.352.1037H}.
In addition, the {\it Suzaku} observation of V1223\,Sgr with CCD cameras
(X-ray Imaging Spectrometer: XIS) reveals that
%Although energy resolution is limited, %For V1223 Sgr, 
the energy centroid of the fluorescent iron K$\alpha$ line
synchronously
modulates %synchronizing
with the WD spin %according to
%the {\it Suzaku} CCD camera (X-ray Imaging Spectrometer: XIS) observation of V1223\,Sgr %X-ray charge-coupled devices
%named X-ray Imaging Spectrometers (XISs)
%on {\it Suzaku} satellite 
\citep{2011PASJ...63S.739H}.
The energy centroid is shifted from the rest frame value 6.40\,keV and
reaches the minimum of $\sim$ 6.38\,keV 
at a spin phase of the heaviest photoelectric absorption.

Our simulation shows that the Compton shoulder extends %makes red wing extending 
down to $\sim$ 6.2\,keV (figure\,\ref{fig:spectra}, \ref{fig:mwd0p7_psac_spe_i}, \ref{fig:mcomp_psac_spe_i})
and does not make the step around the
6.33\,keV found in the GK\,Per X-ray spectrum %found 
by the {\it Chandra} HETG
%spectrum of  
\citep{2004MNRAS.352.1037H}.
Consequently, %Therefore, 
the red wings should be caused not by %not 
the Compton shoulder
but by the fluorescent K$\alpha$ reemission from the pre-shock accreting matter
falling to the WD by a velocity of the %the 
order of thousands km\,s$^{-1}$ as the authors of the finding papers suggested.
The centroid energy shift %modulation of the energy centroid 
by the %$\sim$ 
20\,eV observed in V1223\,Sgr
can hardly be caused by the Compton shoulder (figure\,\ref
{fig:i_ecen}%{fig:i_ew}
).
%Moreover, even if the modulation by 10\,eV is caused by the Compton shoulder,
%the EW significantly modulates by about its maximum value in a WD spin period (see figure\,\ref{fig:i_intensity_ratio}).
%\textcolor{red}{
The amplitude of the EW modulation is $\pm$20\%, taking into account the errors (Fig. 10 of Hayashi et al. 2011). Such an EW modulation is possible with the reflection in the viewing angle range $0^\circ$--$70^\circ$. Although some contribution from the pre-shock accreting matter may exist, the 20~eV shift observed in V1223~Sgr can be attributed to the reflection from the WD surface.
Moreover, if the observed energy centroid modulation were attributed to the reflection from the WD, the EW would also significantly modulate.
%}
%\textcolor{blue}{
%Moreover, even if the modulation of the Compton shoulder is %significantly
%the main cause of %contributes 
%the observed energy centroid  modulation,
%%by 10\,eV is caused by the Compton shoulder,
%the EW should %also 
%significantly modulates
%(figure\,\ref{fig:i_ecen} and \ref{fig:i_ew}). 
%%by about its maximum value in a WD spin period (see figure\,\ref{fig:i_ecen} and \ref{fig:i_ew}).
%However, the EW modulation have not detected \citep{2011PASJ...63S.739H}.
%}

The {\it Chandra} HETG also found the fainter shoulder extending to 6.23\,keV in GK\,Per 
\citep{2004MNRAS.352.1037H}.
The extension is consistent with our simulation and its shape is similar to the simulated spectrum
(see the bottom panel of figure\,\ref{fig:spectra}).
O%Therefore, o
ur simulation therefore supports the scenario that the fainter shoulder is caused by the Compton shoulder
%which 
as suggested by \cite{2004MNRAS.352.1037H}. %the authors 
%suggested.

\subsection{Inspection of the WD with %the complex of 
the fluorescent iron K$\alpha$ line complex}
The reflection iron fluorescent line profile enables us to inspect the 
state of matter 
on the WD surface directly. %in direct.
In our simulation, the matter is assumed to be cool, that is, the scattering electrons 
are bound in atoms at the ground state.
%This assumption enhances the EWs of the fluorescent iron K$\alpha_{1,2}$ lines
%because of the coherent scattering,
%makes the valley between those lines and their Compton shoulder
%and makes the Compton shoulder blurred by the Compton profile.
This assumption enhances the EWs of the fluorescent iron K$\alpha_{1,2}$ lines
because the coherent scattering
releases the line photons from the WD without energy loss.
In other words,
the line photons coherently scattered accumulate on its original line.
Moreover, the incoherent scattering %with %by the bound electrons
makes the valley between the fluorescent iron K$\alpha_{1,2}$ lines %at their original energies 
and their Compton shoulder,
and makes the Compton shoulder blurred by the Compton profile.
The profiles of the valley and the Compton shoulder %profile 
show the fraction of bound %binding state of the scattering 
electrons that contribute to the scattering. % bound in atoms.
%The profiles of the valley and the Compton shoulder %profile 
%show the binding state of the scattering 
%electrons. % that contribute to the scattering. % bound in atoms.

On the other hand, if the scattering electrons are free,
any scattered reemitted photon does not contribute to the EWs of %the 
its original
%fluorescent iron K$\alpha_{1,2}$ 
line 
and the valley does not appear \citep{1991MNRAS.249..352G}.
%The Compton shoulder is blurred by the motion of the electrons.
If the electrons are almost at rest,
the Compton shoulder has double-peaked shape because of the Klein-Nishina differential cross section.
By contrast, if the electrons move with significant velocity,
the Compton shoulder blurred by the Doppler effect like figure\,2 of \cite{2003ApJ...597L..37W}.

The inspection of the state of the matter on the
WD with %the complex of 
the fluorescent iron K$\alpha$ line complex
requires high energy resolution and high photon statistics 
%high 
enough to resolve the Compton shoulder from
the fluorescent iron K$\alpha_{1,2}$ lines. %,
%the valley and Compton shoulder and, to investigate the shape of the Compton shoulder.
Although the {\it Chandra} HETG shows the hint of the Compton shoulder,
%however, 
it cannot be used to investigate its shape
due to limited energy resolution at the iron K$\alpha$ line energy. % (6--7\,keV).
%Moreover, the fluorescent iron K$\alpha_{1,2}$ lines can not be distinguished from the red wing.
Observations with higher energy resolution and higher photon statistics are desired.

\section{Summary}\label{sec:summary}
 We modeled the X-ray reflection spectrum
 from %by 
 the WD surface in the mCV using the Monte Carlo method. %simulation.
 The two types of source are simulated. %assumed.
 One %of them 
 is the point source emitting the power-law spectrum distant away by %finite distance of 
 $h$ from the WD surface.
 The other is the PSAC source which has %is 
the shape of pole with a finite length ($h$) and a
negligible width.
 The PSAC source has a %emits 
 thermal %spectra 
 %of temperatures corresponding its position based on 
 spectrum of a plasma with a temperature distribution %s 
 stratified according to
 the PSAC model calculated by \cite{2014MNRAS.438.2267H}.
 The sources irradiate the cool and spherical WD.
 The %arrived 
 photons from the sources interact with the electrons bound in the atoms.
 We took into account %\textcolor{red}{for the bound electrons} 
 the interactions of the photoelectric absorption, the coherent scattering and the incoherent scattering.
Moreover, %the reemission is considered 
for fluorescent K$\alpha_{1,2}$ and K$\beta$ lines of the iron and the nickel
are considered in the reflection spectrum. 
 Note that the modeled spectrum involves the fluorescent lines 
 as well as the Compton hump, which enhances the measurement accuracy of the parameters of 
the WD in the mCVs. % compared with the model.}
 
With the point source,
%we found that most reflected X-rays %come out %
%are emitted 
%from the circle of radius of 10$^{-3}\,\Rwd$ on the WD for the $h = 0.0001\,\Rwd$ case
%(figure\,\ref{fig:images_h0p0001}).
the larger viewing angle ($i$) makes 
a reflection %reemission %shining 
area %by the reflection 
elliptical because of different angular dependence of the
% the differential 
cross sections of the scattering.
By contrast, for $h$ = 1$\,\Rwd$, most of the %almost 
hemisphere of the WD centered on the illuminating source shines through %by 
the reflection %in the smaller $i$
(figure\,\ref{fig:images_h1p0}).
Accordingly, in larger $i$, the part of the shining area is occulted by the WD limb.
Since $h$ is large enough compared to $R_{\rm WD}$, part of the reflection is visible even in a configuration of $i > 90^\circ$.

We investigated %confirm 
the fluorescent lines, %and 
their Compton shoulder, and the Compton hump in 10--50\,keV.
The energy center of the fluorescent iron K$\alpha_{1,2}$ with their Compton shoulders slightly depends on %the 
$i$
and increases by $\sim$10\,eV as %the 
$i$ increases (figure\,\ref{fig:i_ecen}).
The total EW of the fluorescent iron K$\alpha_{1,2}$ with their Compton shoulders 
monotonically decreases as $i$ increases from $0^\circ$ (figure\,\ref{fig:i_ew}).
When $h$ increases, the solid angle of the WD viewed %ing 
from the point source reduces and the EW  decreases.
The significant EW appears even in the case of $i>$ %above 
90\,deg %of $i$ 
for non-negligible $h$.
%because %of the curvature 
%the point source with such $h$ can illuminate the area of the WD
%which is separate from the point source by more than 90\,deg in angle.
The EW ratio of the Compton shoulders %of the fluorescent iron K$\alpha_{1,2}$ 
to that of the sum of the K$\alpha_{1,2}$
decrease from 0.2 to 0.1 as the $i$ increases (figure\,\ref{fig:i_intensity_ratio}).
%This is because for larger $i$ only the photons interacted at shallow inside of the WD can escape
%and the number of interactions (especially %i.e. also 
%possibility of the incoherent scattering after reemission) 
%%of the photon 
%is less in the shallower inside.
The smaller photon index (${\it \Gamma}$) enhances the EW (figure\,\ref{fig:ew_gamma_ratio})
%of the fluorescent iron K$\alpha_{1,2}$ 
because the photon whose energy is higher than the K absorption edge increases. % in the intrinsic X-ray. 

The Compton hump reduces as $i$ and $h$ increase (figure\,\ref{fig:albedo_ratio}) %s 
like %as well as 
the EW of the fluorescent iron K$\alpha_{1,2}$.
%On the other hand, t
%The Compton hump is difficult to %can not 
%be separated from the intrinsic thermal %intrinsic 
%component solely with the continuum spectrum.
%Consequently, %Therefore, 
%w
We estimated the Compton hump with the intensity ratio 
of the sum of the intrinsic and reflected continuum
%total of the thermal %intrinsic 
%component and the reflection
among some energy bands (figure\,\ref{fig:intensity_ratio}).
%in the .
The intensity ratio varies by the photon index ${\it \Gamma}$ (figure\,\ref{fig:intensity_ratio_ratio}).
% mainly in its offset
%because of the intrinsic power law itself.
%
% in the 35-50\,keV relative to that around this energy band grows up when the $\Gamma$ decreases
%mainly because of the intrinsic power-low itself.
%%the higher energy photons increases which can result in that energy band owing to the down scattering.

The abundance also influences %on 
the fluorescent lines, %and 
their Compton shoulder, and the Compton hump.
When the abundances of %the 
all elements simultaneously increase, %s,
the total EW of the fluorescent iron K$\alpha_{1,2}$ with their Compton shoulders 
increases and peaks out in larger abundance
(figure\,\ref{fig:z_ew_dependency}). %around 1 and 0.2 solar abundance for h = 0.0001 and 1\,$\Rwd$, respectively.
On the other hand, the EW of the Compton shoulders decreases as the abundances %of the all elements 
increase
and partially counteracts the increase of the total EW.
%This is because the scattering is suppressed in the larger abundances. % of the all elements.
When the abundance only of the iron increases,
the total EW of the fluorescent iron K$\alpha_{1,2}$ with their Compton shoulders is more steeply enhanced.
%is almost proportional to the iron abundance
%up to 1 and 0.5 solar abundance of iron.
%Above these abundance of the iron, the increase of the EW become gentle 
%where the iron begins to determine the cross sections of the interactions in the WD.
The EW of the Compton shoulders increases with %as 
the iron abundance %s increases 
unlike the case of the fixed abundance %s 
ratio
because the most scatterings are attributed to elements except for the iron.
The cut-off energy %low-energy bending 
of the Compton hump rises %its cut-off energy of lower energy side 
as the abundance increases
because of the heavier photoelectric absorption (figure\,\ref{fig:z_albedo_dependency}).

We modeled the reflection spectrum of the PSAC source
 in the mCV with %parameterizing 
 the WD mass, the specific accretion rate and the abundance as the model parameters.
%using the PSAC source.
We confirmed that the reflection depends on these parameters %WD mass and the specific accretion rate
as well as the viewing angle.
The reflection spectrum includes a widespread %has scattered 
line structure constructed 
by the %incoherent 
scattering of the emission lines in the intrinsic thermal spectrum.
%unlike %the point-source emitting 
%the intrinsic power-law spectrum.
With the PSAC source, %As for the intrinsic thermal emission,
its lower temperature part %of the PSAC 
can be hidden and only the higher temperature spectrum can be observed
because of the occultation of the base of the PSAC by the WD limb.

The WD mass influences %on 
the reflection in a %with 
somewhat complex %.This is because the 
manner (figure\,\ref{fig:a-ew}). While a more massive WD makes the PSAC hotter and enhances
 the reflection,
%and,  on the other hand, 
it makes the PSAC taller and reduces the reflection.
In fact, the EW %s 
of the fluorescent iron K$\alpha_{1,2}$ lines does not monotonically increase with %as 
the WD mass %increases
%depending on %with some 
%the specific accretion rate and the reflecting angle
although the more massive WD makes more intense reflection. % in a basic way.
On the other hand,
%By contrast, 
the larger specific accretion rate enhances the reflection in a unilateral way
because the larger specific accretion rate makes the %cooler and the taller 
PSAC hotter and shorter. %cooler and taller.

The shape of the Compton hump also depends on the WD mass and the specific accretion rate
(figure\,\ref{fig:a-hump}).
%However, the thermal component varies %with 
%even more than the reflection. %Compton hump.
Especially, with the lighter WD, the shape %variation 
of the Compton hump is independent of $i$.
%trivial.

We calculated the X-ray modulation caused by %in relation to 
the WD spin with a few %some 
parameter sets (figure\,\ref{fig:spin}).
The modulation profile significantly depends on the angles %between 
of %the PSAC and 
the line of sight measured from
the spin axis and from the PSAC
%and between the spin axis and the line of sight 
and whether the two PSACs can be seen
as well as the WD mass and the specific accretion rate.
Moreover, the modulation profile is drastically changed,
depending on the energy band in question. % by energy.

The EW of the iron fluorescent line is a good measure to constrain the viewing angle.
For example, %With 
the %lower 
EW of 24\,eV %of the fluorescent iron K$\alpha$ line 
in the EX Hya leads to %,
the viewing angle %of the system is constrained as 
$i > 60$\,deg.
%When 
If our reflection spectral model %for the reflection 
that includes the iron fluorescent line is applied to data directly,  %in direct,
the system parameters will %should 
be measured with higher accuracy, which will be presented in a forthcoming paper.
%We will report the application in a subsequent paper.

Our simulation indicates that the fluorescent iron K$\alpha$ line at 6.4 keV can be extended down to $\sim$ 6.2 keV due to the Compton scattering. Hence,
the red wing of the fluorescent iron K$\alpha$ line extending down to 6.33 keV %with $\sim$ 70\,eV 
in GK Per
%and the centroid energy shift of the
%fluorescent iron K$\alpha$ line %modulating in its red side 
%by $\sim$ 20\,eV in V1223 Sgr
can not be reproduced by the reflection.
In contrast,
%On the other hand, 
the fainter shoulder of the fluorescent iron K$\alpha$ line extending to 6.23\,keV in GK Per
can be reproduce by the reflection,
which probably contains the information about the state of matter on the WD.
The shift of the line centroid to 6.38 keV
detected from V1223 Sgr, on the other hand, can be attributed to the reflection from the white dwarf, although some contribution from the pre-shock accreting matter is not completely ruled out.
Higher energy resolution and higher photon statistics in future 
will unveil the state of the matter on the WD with %the complex of 
%the parameters of 
the fluorescent iron K$\alpha$ line
by its %such as the 
EW, %of the fluorescent iron K$\alpha_{1,2}$ lines,
%their 
Compton shoulder and the valley between them.

%
%We can apply our spectral and modulation models to data in direct,
%which will avert the problem of contamination between the reflection and the intrinsic thermal component.
%The reflection constrains the WD mass, specific accretion rate and abundance in the mCV 
%as well as the thermal component.
%Moreover, the reflection gives us the geometrical information such as the PSAC height and the reflecting angle.
%These parameters will be measured with unprecedented accuracy by our reflection model.

\section*{ACKNOWLEDGEMENTS}
T. H. is financially supported by the Japan Society for the Promotion of Science
(JP15J10520 and JP26800113).

\end{document}